\documentclass[a4paper,11pt]{article}
\usepackage{caption}
\usepackage{mathrsfs}
\usepackage{jheppub} 
\usepackage{amsmath,amssymb,amsfonts}

\usepackage{
epsf,
epsfig,
epstopdf,
graphics,
graphicx,
mathtools,
subcaption,
physics,
verbatim
}

\usepackage[utf8]{inputenc}

\usepackage{array}
\newcolumntype{P}[1]{>{\centering\arraybackslash}p{#1}}

\usepackage[shortlabels]{enumitem}

\usepackage{dcolumn}
\newcolumntype{M}[1]{>{\centering\arraybackslash}m{#1}}
\newcolumntype{N}{@{}m{0pt}@{}}

\usepackage{amsmath}
\usepackage{ulem}

\newcommand{\be}{\begin{equation}}  
\newcommand{\ee}{\end{equation}}

\usepackage{tikz}

\title{\boldmath Can Horndeski Genesis be Nonpathological?}

\author[1]{Han Gil Choi,}
\author[1]{Pavel Petrov,}
\author[1,2]{and Masahide Yamaguchi}

\affiliation[1]{Cosmology, Gravity and Astroparticle Physics Group, Center for Theoretical Physics of the Universe, Institute for Basic Science (IBS), Daejeon, 34126, Korea}

\affiliation[2]{Department of Physics, Institute of Science Tokyo,
Tokyo, 152-8551, Japan}

\emailAdd{hgchoi1w@gmail.com}
\emailAdd{ p669590371@ibs.re.kr}
\emailAdd{gucci@ibs.re.kr}

\abstract
{
       We present a minimal setup within the framework of Horndeski gravity that can describe a nonpathological Genesis scenario. Our setup allows for a fully stable transition to the kination epoch, during which General Relativity (GR) is restored. This Genesis scenario circumvents the no-go theorem at the cost of encountering the risk of strong coupling in the past. Interestingly, our scenario admits two different regimes for the background solution for Hubble parameter at the Genesis stage: power-law behavior and manifestly non-power-law behavior. We explicitly show that, in both regimes, our model remains within unitarity bounds. In most cases, the tensor spectrum is blue-tilted. Then, we adopt a mechanism with a spectator field that allows for a red-tilted scalar power spectrum. We also suggest a deformation of the model that enables us to achieve sufficiently small values for the r -- ratio. Finally, we discuss the geodesic (in)completeness of the current model.

}

\makeatletter
\gdef\@fpheader{}
\makeatother
\begin{document}

\maketitle
\flushbottom

\section{Introduction}
\label{sec: Introduction}

Currently, inflation is the conventional paradigm for describing the primordial Universe; however, it suffers from the initial singularity problem (see Ref.~\cite{Borde:1996pt}). This issue has attracted attention to non-singular cosmological scenarios, such as Genesis and Bounce Universe models. These scenarios could serve as alternatives or completions to the standard inflation model. 

In order to construct a non-singular universe, one must find a way to avoid the Penrose singularity theorem~\cite{Penrose:1964wq}. One way to achieve this is by violating the Null Energy Condition (NEC); for a review, see Refs.~\cite{Rubakov:2014jja, Kobayashi:2019hrl}. Unfortunately, stable NEC violation is quite difficult to achieve. Nevertheless, it is possible to violate the NEC in a healthy way within the framework of Horndeski gravity~\cite{Horndeski:1974wa}. Another possibility is to utilize Beyond Horndeski theories~\cite{Zumalacarregui:2013pma, Gleyzes:2014dya} or DHOST theories~\cite{Langlois:2015cwa}, which are generalizations of Horndeski gravity. Therefore, Horndeski gravity remains one of the simplest choices for model building, with numerous examples of stable early Genesis~\cite{Creminelli:2010ba, Creminelli:2012my, Hinterbichler:2012fr, Elder:2013gya, Pirtskhalava:2014esa, Nishi:2015pta, Kobayashi:2015gga} and Bouncing Universe models~\cite{Qiu:2011cy, Easson:2011zy, Battarra:2014tga, Ijjas:2016tpn} developed within this framework.

Unfortunately, non-singular scenarios within Horndeski gravity typically suffer from instabilities at some point during their evolution. This is not a coincidence, as noted in Refs.~\cite{Kobayashi:2016xpl, Libanov:2016kfc}. The statement can be summarized as follows: if the two integrals below diverge,
\begin{align}
\label{Integrals for No-go}
     \int_{-\infty}^t a(t)(\mathcal{F}_T+\mathcal{F}_S) dt = \infty, \\
     \int_t^{+\infty} a(t)(\mathcal{F}_T+\mathcal{F}_S) dt = \infty\;,\nonumber
\end{align}
then the model would be plagued by gradient instabilities at least at some point during its evolution. This statement is known as a no-go theorem. Here, $\mathcal{F}_T$ and $\mathcal{F}_S$ are the gradient coefficients in the quadratic action for tensor and scalar perturbations, respectively.

One way to circumvent the no-go theorem is to invoke Beyond Horndeski or DHOST terms. Another approach, first proposed in Ref.~\cite{Kobayashi:2016xpl}, considers the case in which the coefficients $\mathcal{F}_T$ and $\mathcal{F}_S$ in the quadratic action for perturbations tend to zero in the asymptotic past (we refer to these as models with strong gravity in the past). This presents a unique opportunity to overcome the no-go theorem and construct non-singular cosmology. One example of such a scenario is provided in~\cite{Kobayashi:2016xpl}, namely the Genesis followed by never-ending inflation.

However, it was soon realized that scenarios where the coefficients in the quadratic action for perturbations vanish in the asymptotic past could be pathological as effective field theories. Nevertheless, Refs.~\cite{Ageeva:2018lko, Ageeva:2020gti} demonstrate that cosmological models with strong gravity in the past can still be applicable if the energy scale of classical evolution is much lower than the strong-coupling scale of the theory.

Subsequently, a more advanced method was employed -- the unitarity bound from the optical theorem~\cite{Ageeva:2022fyq, Cai:2022ori, Ageeva:2022asq} -- which reached the same conclusion: there could exist a region of parameter space where the model can be legitimately described by classical field theory and weakly coupled quantum field theory. For Genesis with strong gravity in the past, it has also been shown that this statement holds even at arbitrary orders of perturbation theory, provided that loops are not considered. Moreover, Ref.~\cite{Ageeva:2020buc} claims that the most stringent conditions arise from the cubic Lagrangian for scalar perturbations.

As noted in Refs.~\cite{Creminelli:2016zwa, Kobayashi:2016xpl}, the convergence of the integrals in the no-go theorem indicates that space-time is geodesically incomplete for the propagation of gravitons. Nevertheless, it is argued that the concept of geodesic (in)completeness is frame-dependent and requires a generalized notion of geodesic (in)completeness~\cite{Rubakov:2022fqk, Wetterich:2024ung}. 

In this context, we discuss how to apply different definitions of generalized geodesic (in)completeness to our model in Section~\ref{sec:NEC and Einstein frame}. In addition, we explicitly show that, for our Genesis model, there exists a clock system in which the Universe is eternal, i.e., complete in a generalized sense.

In Ref.~\cite{Wetterich:2014zta}, it is claimed that a singularity in one frame may arise from a singularity in the field transformation, while in another frame, everything appears regular. This is precisely the situation for our model: we observe a singularity in the Einstein frame but none in the Jordan frame. Furthermore, Ref.~\cite{Wetterich:2014zta} states that these ``field singularities" do not represent actual physical singularities; rather, they are analogous to ``coordinate singularities" that arise from choosing a specific coordinate system (they are singularities in ``field-coordinates"). Thus, the absence of physical singularities is ensured if there is at least one frame in which all relevant physical observables remain regular. In our model, this frame is the Jordan frame, where the Genesis scenario takes place. In the last part of Section~\ref{sec:NEC and Einstein frame}, we present our version of a generalized geodesic completeness.

In light of this discussion, it is natural to inquire which minimal setup could produce an experimentally viable non-singular cosmology. \textit{This is the main aim of the present paper.} Indeed, non-singular cosmologies can potentially be realized in Beyond Horndeski theories; for a review, see~\cite{Kobayashi:2016xpl, Mironov:2024pjt}. However, before moving to a more complicated theoretical framework, it is important to determine whether a similar construction is possible using simpler modifications of gravity. Thus, let us first identify which minimal Horndeski subclass could facilitate a non-singular cosmological model.

For definiteness, we consider the Genesis scenario. As it was shown in Ref.~\cite{Kobayashi:2011nu} Horndeski gravity is equivalent to generalized Galileons \cite{Deffayet:2011gz}. Thus, for our purposes it is sufficient to consider generalized Galileons action. This action  encompasses four arbitrary functions $G_2$–$G_5$ of the field $\phi$ and the kinetic term $X = -\frac{1}{2}(\partial_{\rho}\phi)^2$ :
\begin{align*}
    \mathcal{L} = G_2(\phi,X) - G_3(\phi,X)\Box\phi + G_4(\phi,X)R + G_{4X} \left[(\Box\phi)^2 - \phi^{\mu\nu}\phi_{\mu\nu}\right] \\
    + G_5(\phi,X)G^{\mu\nu}\phi_{\mu\nu} - \frac{G_{5X}}{6} \left[(\Box\phi)^3 - 3\Box\phi\phi^{\mu\nu}\phi_{\mu\nu} + 2\phi_{\mu\nu}\phi^{\nu\lambda}\phi_\lambda^\mu\right].
\end{align*}

The function $G_2(\phi, X)$ is necessary as it contains the canonical kinetic term for the field. Additionally, the function $G_3(\phi, X)$ must be included to allow for stable NEC violation. Finally, at least one field-dependent function $G_4(\phi)$ should remain. This condition provides an opportunity to circumvent the no-go theorem. It is crucial to note that the field-dependent function $G_4(\phi)$ is sufficient to bypass the no-go theorem; thus, we need not consider the more general case where $G_4$ depends on both the field $\phi$ and kinetic term $X$.

At first glance, one might conclude that the function $G_4(\phi)$ does not contribute anything novel to the physical behavior of the theory, as it is possible to perform a conformal transformation to the Einstein frame with $G_{4}^{E} = \frac{M_{Pl}^{2}}{2}$. However, this is not the case for the current model: to evade the no-go theorem, $G_4$ must tend to zero at least in the asymptotic past. This behavior of the function $ G_4 $ causes the conformal transformation between the Jordan and Einstein frames to be singular at negative infinity, thereby distinguishing the Jordan frame from the Einstein frame physically.

Considering all subsequent requirements, we can conclude that the minimal subclass of Horndeski gravity for constructing a non-singular universe is given by:
\begin{align*}
     \mathcal{L} = G_2(\phi, X) - G_3(\phi, X)\Box\phi + G_4(\phi)R.
\end{align*}

In this paper, we will utilize this subclass to construct a Genesis scenario that could serve as a viable non-singular alternative to conventional cosmological inflation. Our scenario begins from flat space and time, expands, and then terminates with the standard kination stage. We ensure that there are no ghost and/or gradient instabilities throughout the entire evolution. Furthermore, there is no superluminal behavior beyond the background cosmological solution. Additionally, we demonstrate that it is indeed possible to obtain the correct physical predictions for the power spectrum and the $ r $ ratio, respectively.

Finally, we note that, to the best of our knowledge, the current scenario with strong gravity in the past is novel and has not been previously investigated in the literature. The closest analogues were discussed in Refs.~\cite{Ageeva:2021yik, Akama:2022usl}. In~\cite{Ageeva:2021yik}, a Genesis scenario in the context of Horndeski gravity was proposed; however, it had two main drawbacks. Firstly, it ended with infinite inflation, and, as mentioned in the same reference~\cite{Ageeva:2021yik}, the transition between stages could, in most cases, be plagued by instabilities. Thus, establishing a fully stable transition phase between different stages is a nontrivial task. Secondly, there was no investigation of the strong coupling regime at later times. As we point out in Sec.~\ref{sec: Unitarity Bounds for the late times}, the Genesis stage could enter a non-power-law regime when the solution for the Hubble parameter does not exhibit power-law behavior; this case requires further investigation.

In Ref.~\cite{Akama:2022usl}, a Genesis scenario with strong gravity in the past was indeed constructed. This scenario allows for a parameter space that leads to a red-tilted scalar power spectrum. This spectral index was achieved at the cost of two modifications. The first is abandoning the no-go theorem and introducing Beyond Horndeski terms at certain points to ensure stability throughout the cosmological evolution. The second is utilizing functional freedom to set certain terms to zero in the cubic Lagrangian for perturbations. The latter relaxes the unitarity bounds and creates the possibility for a red-tilted scalar spectrum. It remains to be seen whether the model still fits within the unitarity bounds when considering higher-order Lagrangians for both tensor and scalar perturbations.

The paper is organized as follows. In Section~\ref{sec: General expressions}, we present the general framework and expressions that will be used in our work. Section~\ref{sec:early_gen_stage} is devoted to the early Genesis stage; specifically, we outline the background solution and stability requirements. In Section~\ref{sec:NEC and Einstein frame}, we discuss the behavior of our model in the Einstein frame and the violation of the NEC condition. Additionally, we discuss geodesic (in)completeness and present our definition of generalized completeness.

Section~\ref{sec: The setup} focuses on the model that produces a stable Genesis scenario, which ends with the reheating stage, as well as the general method for constructing such scenarios. In Section~\ref{sec:numerical_example_1}, we present two numerical solutions for different parameter ranges. Both solutions avoid gradient and ghost instabilities and ensure subluminal speeds of propagation for scalar perturbations, while tensor perturbations always propagate at the speed of light. The first solution exhibits significant non-power law corrections for the early Genesis stage, while the second maintains power-law behavior in the Hubble parameter.

In Section~\ref{sec: Scalar primordial power spectrum}, we derive the primordial power spectrum for scalar and tensor perturbations, respectively. We point out a tension between the unitarity bounds in the asymptotic past and the red-tilted value of the scalar spectral index. However, this tension does not necessarily apply if the mode freezes in a regime where the background solution has significant non-power law corrections.

Section~\ref{sec: Unitarity Bounds for the late times} analyzes the unitarity bounds when the Hubble parameter undergoes significant non-power law corrections. We demonstrate that there exists a parameter range in this case where the theory can be accurately described by classical field theory and weakly coupled quantum field theory.

In Section~\ref{sec:TimeDependTermsScalarSpectrum}, we calculate the scalar power spectrum numerically when the Hubble parameter cannot be described by the analytic power-law background solution. We investigate the parameter space and show that obtaining a red-tilted power spectrum is extremely challenging.

Section~\ref{sec: The spectator field} discusses mechanisms that could produce a red-tilted power spectrum. Additionally, we examine model deformations that allow for a significantly small value of $ r $ -- the ratio. We conclude in Section~\ref{sec:conclusion}. In Appendix~\ref{app:fromJToE}, we provide formulas to express Einstein frame functions $ G^{E}_{2-4} $ in terms of the Jordan frame functions $ G^{J}_{2-4} $. Finally, in Appendix~\ref{app: Eikonal approximation}, we comment on the details of the Eikonal approximation, which we utilized during our numerical simulations.


\section{Generalities}
\label{sec: General expressions}

As mentioned in Section~\ref{sec: Introduction}, we will work in the Jordan frame and consider the following subclass of Horndeski theories:
\begin{align}
    \mathcal{S} = & \int d^4x \sqrt{-g} \left\{ G_2(\phi, X) - G_3(\phi, X) \square \phi + G_4(\phi) R \right\}, \quad X =  -\frac{1}{2} g^{\mu\nu} \partial_{\mu} \phi \partial_{\nu} \phi\;,
    \label{modelAction}
\end{align}
where $\square \phi = g^{\mu\nu} \nabla_\mu \nabla_\nu \phi$ and $R$ is the Ricci scalar. We will use the following metric signature: $(-,+,+,+)$. 

It is convenient to perform all calculations in the ADM formalism~\cite{Kobayashi:2019hrl}. In this framework, we write the metric as 
\begin{equation*}
    ds^2 = -N^2 d t^2 + \gamma_{ij} \left( dx^i + N^i d t \right) \left( dx^j + N^j d t \right),
\end{equation*}
where $t$ -- is Jordan frame coordinate time and $\gamma_{ij}$ is a three-dimensional metric, $N$ is the lapse function, and $N_i = \gamma_{ij} N^j$ is the shift vector. Next, we choose the unitary gauge (in this gauge, the field $\phi$ depends only on $t$ and can be expressed in the form $\phi = \phi(t)$). In this gauge, we can rewrite the action as follows:
\begin{align*}
\label{adm_lagr}
    \mathcal{S} = \int d^4x \sqrt{-g} \left[ A_2(t, N) + A_3(t, N) K + A_4(t) (K^2 - K_{ij}^2) + B_4(t) R^{(3)} \right],
\end{align*}
where
\begin{equation*}
    A_4(t) = -B_4(t)~.
\end{equation*}
Here, $\sqrt{-g} = N\sqrt{\gamma}$, $K = \gamma^{ij} K_{ij}$, and $^{(3)}R = \gamma^{ij} \phantom{0}^{(3)}R_{ij}$, with $K_{ij}$ defined as:
\begin{align*}
    K_{ij} &\equiv \frac{1}{2N} \left( \frac{d\gamma_{ij}}{d t} -\,^{(3)}\nabla_{i} N_{j} - \;^{(3)}\nabla_{j} N_{i} \right),
\end{align*}
which represents the extrinsic curvature of the hypersurfaces $t = \text{const}$. The relationship between the Lagrangian functions in the covariant and ADM formalisms is given by~\cite{Gleyzes:2014dya, Gleyzes:2013ooa, Fasiello:2014aqa}:
\begin{equation}
    G_2 = A_2 - 2X F_{\phi}, \quad G_3 = -2X F_X - F, \quad G_4 = B_4,
    \label{FromADMToCov}
\end{equation}
where $N$ and $X$ are related by:
\begin{equation*}
    N^{-1} \frac{d\phi}{d t} = \sqrt{2X},
\end{equation*}
and 
\begin{equation*}
    F_X = -\frac{A_3}{(2X)^{3/2}} - \frac{B_{4\phi}}{X}.
\end{equation*}

Now, let us turn to the perturbations about the FLRW background. To this end, we introduce the following notation:
\begin{subequations}
\begin{align*}
    N &= N_0(t) (1 + \alpha)\;, \\
    N_{i} &= \partial_{i} \beta + N^T_i\;, \\
    \gamma_{ij} &= a^{2}(t) \left(\text{e}^{2\zeta} (\text{e}^{h})_{ij} + \partial_i \partial_j Y + \partial_i W^T_j + \partial_j W^T_i\right) \; ,
\end{align*}
\end{subequations}
where $ a(t) $ and $ N_0(t) $ are the background solutions, while $ N^T_i $ and $ W^T_i $ satisfy $ \partial_i N^{T\,i} = 0 $ and $ \partial_i W^{T\,i} = 0 $, respectively.

We fix the residual gauge freedom by setting $ Y = 0 $ and $ W^T_i = 0 $. Here, the variables $ \alpha $, $ \beta $, and $ N^T_i $ are constraints; thus, they are nondynamical and enter the action without time derivatives. Therefore, the dynamical degrees of freedom are $ \zeta $ and the transverse and traceless $ h_{ij} $, which represent scalar and tensor perturbations. Consequently, we have three degrees of freedom in our theory: one for the scalar field and two for gravity.

The background equations of motion for the spatially flat FLRW background read \cite{Kobayashi:2015gga}:
\begin{subequations}
\label{eom}
\begin{align}
    &(NA_2)_{N} + 3NA_{3N}H + 6N^2(N^{-1}A_4)_{N} H^2 = 0,\\
    &A_2 - 6A_4H^2 - \frac{1}{N} \frac{d}{d t}\left( A_3 + 4A_4H \right) = 0,
\end{align}
\end{subequations}
where the Hubble parameter is given by $ H(t) = \frac{d\text{ln}[a(t)]}{N dt}$, and $ N(t) $ is a background lapse function.

Upon integrating out the non-physical variables such as $ \alpha $, $ \beta $, and $ N^T_i $, we obtain the quadratic actions for both tensor and scalar perturbations:
\begin{subequations}
\label{quadraticAction}
\begin{align}
    \label{quadraticActionScalar}
    \mathcal{S}_{\zeta \zeta}^{(2)} &= \int d t \, d^{3}x \, N a^3 \left[ \frac{\mathcal{G}_S}{N^2} \left( \frac{\partial \zeta}{\partial t} \right)^{2} - \frac{\mathcal{F}_S}{a^2} \left( \vec{\nabla} \zeta \right)^{2} \right] \;, \\
    \mathcal{S}_{hh}^{(2)} &= \int d t  \, d^3x \, \frac{N a^3}{8} \left[ \frac{\mathcal{G}_T}{N^2} \left( \frac{\partial h_{ij}}{\partial t} \right)^2 - \frac{\mathcal{F}_T}{a^2} h_{ij,k} h_{ij,k} \right] \;,
\end{align}
\end{subequations}
where
\begin{subequations}
\label{FS_GS}
\begin{eqnarray}
    \mathcal{F}_S &=& \frac{1}{a N} \frac{d}{d t} \left( \frac{a}{\Theta} \mathcal{G}_T^2 \right) - \mathcal{F}_T\;, \\
    \mathcal{G}_S &=& \frac{\Sigma}{\Theta^2} \mathcal{G}_T^2 + 3\mathcal{G}_T\;,
\end{eqnarray}
\end{subequations}
and
\begin{subequations}
\label{FT_GT}
\begin{align}
    \mathcal{G}_T &= -2A_4\;, \\
    \mathcal{F}_T &= 2B_4\;,
\end{align}
\end{subequations}
with
\begin{subequations}
\label{Sigma_Theta}
 \begin{align}
      \Sigma &= N A_{2N} + \frac{1}{2} N^2 A_{2NN} + \frac{3}{2} N^2 A_{3NN} H + 6H^2 A_4\;, \\
      \Theta &= 2H\left(\frac{NA_{3N}}{4H} - A_4\right)\;.
 \end{align}
\end{subequations}
Here, the expression for $\mathcal{F}_S$ is valid only when $A_4 = A_4(t)$. For more general formulas, see Ref.~\cite{Kobayashi:2015gga}. We note that tensor perturbations in our model always propagate at the speed of light:
\begin{equation*}
    u_T^2 = \frac{\mathcal{F}_T}{\mathcal{G}_T} = 1\;,
\end{equation*}
while the sound speed in the scalar sector is given by:
\begin{equation*}
    u_S^2 = \frac{\mathcal{F}_S}{\mathcal{G}_S}\;,
\end{equation*}
which can be arbitrary. The scalar sound speed can be greater or smaller than unity, with the actual value being model-dependent.


\section{Genesis stage}

\label{sec:early_gen_stage}

In the framework discussed in the previous section, it is quite straightforward to construct Horndeski models that admit the Genesis solution with power-law asymptotic behavior at early times~\cite{Kobayashi:2016xpl, Ageeva:2021yik}. Therefore, in this work, we will consider such early-time behavior.

To this end, in order to build a Genesis model with this power-law solution, we choose the following form~\cite{Kobayashi:2016xpl, Ageeva:2021yik} for the Lagrangian functions as $t \to -\infty$:
\begin{subequations}
\label{LagrFunctions}
\begin{align}
    A_2(t, N) &= \frac{1}{2} (-c t)^{-2\mu - 2 - \delta} \cdot a_2(N) \;, \\
    A_3(t, N) &= \frac{1}{2} (-c t)^{-2\mu - 1 - \delta} \cdot a_3(N) \;, \\
    A_4(t) &= -B_4(t) = -\frac{1}{2} (-c t)^{-2\mu} \;,
\end{align}
\end{subequations}
where $c$ is a positive constant with dimension of mass, i.e., $[c] = 1$. Moreover, this choice of early-time behavior leads to the following asymptotics for the coefficients in the quadratic Lagrangians for perturbations~\cite{Kobayashi:2016xpl}:
\begin{equation*}
\label{second_order_coeff_asym}
    \mathcal{G}_{T} \propto (-t)^{-2\mu}, \;\; \mathcal{F}_{T} \propto (-t)^{-2\mu}, \;\; \mathcal{G}_{S} \propto (-t)^{-2\mu + \delta}, \;\; \mathcal{F}_{S} \propto (-t)^{-2\mu + \delta} \;.
\end{equation*}

We can circumvent the no-go theorem~\cite{Libanov:2016kfc, Kobayashi:2016xpl} and have a chance to build a fully stable cosmological scenario when the parameters $\mu$ and $\delta$ satisfy the following constraints:
\begin{align*}
    2\mu > 1 + \delta, \;\; \delta > 0, \;\; \mu > 0 \;.
\end{align*}
Thus, for this range of parameters, the coefficients in the quadratic actions for perturbations behave as:
\begin{equation*}
    \mathcal{G}_{T} \propto \mathcal{F}_{T} \to 0, \;\; \mathcal{G}_{S} \propto \mathcal{F}_{S} \to 0, \;\; t \to -\infty \;.
\end{equation*}
However, this behavior implies that one may encounter a strong-coupling regime in the asymptotic past, since the coefficients in the quadratic action for metric perturbations, which serve as effective Planck masses, tend to zero as $t \to -\infty$.

Nevertheless, it has been shown in Refs.~\cite{Ageeva:2018lko, Ageeva:2020gti} that the fact that $\mathcal{F}_T$, $\mathcal{F}_S$, $\mathcal{G}_T$, and $\mathcal{G}_S$ tend to zero at early times does not necessarily mean that classical field theory cannot legitimately describe the evolution of the background solution. Indeed, the classical theory is applicable~\cite{Ageeva:2018lko, Ageeva:2020gti} if the model parameters satisfy the following criteria:
\begin{align*}
    \mu + \frac{3}{2}\delta < 1.
\end{align*}
To summarize, the model parameters should be chosen within the range:
\begin{subequations}
\label{healthy_region}
\begin{align}
    2\mu > 1 + \delta > 1, \\
    \mu + \frac{3}{2}\delta < 1. \label{earlyUnitarityBound}
\end{align}
\end{subequations}

Next, it is more convenient to make a variable redefinition as follows:
\begin{subequations}
\label{redef}
\begin{align}
    u &\equiv (-c t)^{-\delta}, \; u \in (0, \infty), \; t \in (-\infty, 0),\\
    h(u) &\equiv H(t) \cdot N(t) \cdot (-c t)^{1 + \delta} \bigg|_{t = t(u)}\;.
\end{align}
\end{subequations}
Firstly, this variable redefinition is useful when transitioning to the numerical solution. Secondly, after this redefinition, the background equations of motion will have a more convenient form for further analysis. Thus, by using the redefinition \eqref{redef} in conjunction with substituting the Lagrangian functions \eqref{LagrFunctions} into the background equations of motion \eqref{eom}, we arrive at
\begin{subequations}
\begin{align}
\label{EomU1}
    &a_{2}(N) + \frac{6u h^2}{N^2} + N\cdot \frac{d}{dN} a_{2}(N) + 3u\cdot h \cdot \frac{d}{dN} a_{3}(N) = 0,   \\ 
    \label{EomU2}
    &\frac{1}{N^2}\cdot\Big(N (N^2 a_2(N)-c \delta  N u a_3^\prime(N) N^\prime(u)-c N a_3(N) (\delta +2 \mu +1) \nonumber \\
    &+4 c \delta  h u)+4 c h (N (\delta +2 \mu +1)-\delta  u N^\prime(u))+6 h^2 N u\Big) = 0.     
\end{align}
\end{subequations}
To clarify the equations above, we will write them order by order in terms of the variable $ u $. This expansion is valid since $ u $ is dimensionless and tends to zero in the asymptotic past; thus, it can serve as a small expansion parameter. Therefore, we decompose $ h $ and $ N $ as follows:
\begin{align*}
    h &= h_0 + u \cdot h_1 + \ldots, \\
    N &= N_0 + u \cdot N_1 + \ldots\;.
\end{align*}
After that, we arrive at:
\begin{itemize}   
\item From \eqref{EomU1} at order $ u^0 $:
\begin{align*}
   a_2(N_0) + N_0 a_{2}^{\prime}(N_0) = 0\;.
\end{align*}

\item From \eqref{EomU2} at order $ u^0 $:
\begin{align*}
   N_0 a_2(N_0) + \frac{c \cdot (1 + \delta + 2\mu) \cdot (4h_0 - N_0 a_3(N_0))}{N_0} = 0\;.
\end{align*}

\item From \eqref{EomU1} at order $ u^1 $:
\begin{align*}
3 h_0 a_3'\left(N_0\right) + 2 N_1 a_2'\left(N_0\right) + N_0 N_1 a_2''\left(N_0\right) + \frac{6 h_0^2}{N_0^2} = 0\;.
\end{align*}

\item From \eqref{EomU2} at order $ u^1 $:
\begin{align*}
   &N_0^{-2} \cdot \Big( N_0^2 N_1 \left(-c (2 \delta +2 \mu +1) a_3'\left(N_0\right)+N_0 a_2'\left(N_0\right) +a_2\left(N_0\right)\right)
   \\&-4 c h_0 N_1 (2 \delta +2 \mu +1)+4 c h_1 N_0 (2 \delta +2 \mu +1)+6 h_0^2 N_0 \Big) = 0\;.
\end{align*}

\end{itemize}
Here we provide the expansion only up to the first order; however, this expansion could theoretically be continued to arbitrary order if needed.

Now, let us solve the equations of motion in the limit as $ t \to -\infty $ (i.e., $ u \to 0 $), or in other words, in the leading order by the $ u $ variable. In this limit, we can immediately express the leading-order solutions $ (h_0, N_0) $ in terms of $ u $ as:
\begin{subequations}
    \label{solBackEoMInU}
\begin{align}
    &h_0 \equiv \frac{1}{4} N_0 \left( -\frac{N_0 a_2(N_0)}{c + c \delta + 2c \mu} + a_3(N_0) \right), \\
    &a_{2}(N_0) + N_0 \cdot \frac{d}{dN} a_{2}(N) \Big|_{N = N_0} = 0.
\end{align}
\end{subequations}
Returning to the original variables, we obtain the Hubble parameter and scale factor at leading order ($u^0$) by the variable $u$:
\begin{align}
    H &= \frac{h_0}{(-c t)^{1 + \delta}}, \nonumber\\
    a &= a_g\left( 1 + \frac{h_0}{c \delta (-c t)^\delta} \right), \;\; t \to -\infty,
    \label{background Solutions for early genesis}
\end{align}
where $a_g$ is an integration constant that can be determined from the boundary conditions. Additionally, the constant $h_0 > 0$ should be positive to ensure Genesis at early times. Thus, we see that the Universe starts from a constant value of the scale factor and subsequently undergoes expansion, as is expected during the early Genesis epoch. Furthermore, it should be noted that the initial value of the lapse function $N_0$ can be set to any value by rescaling time. Therefore, without loss of generality, we set $N_0 = 1$ and maintain this choice throughout the text. With this choose the Jordan frame coordinate time $t$ coincides with Jordan frame cosmic time, which we will denote as $t^J$\;.

Now, let us turn to the stability analysis of the solution. For the solution \eqref{solBackEoMInU} and the Lagrangian functions \eqref{LagrFunctions}, the coefficients in the quadratic action for perturbations are given by \eqref{quadraticAction} and are expressed as follows:
\begin{subequations}
    \begin{align*}
        \mathcal{G}_S &= \frac{4 (-c t)^{\delta - 2\mu} \left( 2 a_2^{\prime}(1) + a_2^{\prime\prime}(1) \right)}{\left( 4 h_0 + a_3^{\prime}(1) \right)^2},\\
        \mathcal{F}_S &= \frac{4 (-c t)^{\delta - 2\mu} c (1 + \delta - 2\mu)}{4h_0 + a_3^{\prime}(1)},\\
        \mathcal{G}_T &=  \mathcal{F}_T = (-c t)^{-2\mu}.
    \end{align*}
\end{subequations}
The stability requirements read as follows:
\begin{equation}
    \mathcal{G}_S > 0, \;\; \mathcal{F}_S > 0, \;\; \mathcal{G}_T > 0, \;\; \mathcal{F}_T > 0.
    \label{stability}
\end{equation}
In addition, we require the absence of superluminal propagation:
\begin{equation}
   u_S < 1.
   \label{noSL}
\end{equation}
It has been stated that the latter requirement is essential for the existence of the UV completion of the theory~\cite{Adams:2006sv, deRham:2013hsa}.

\section{NEC violation and geodesic (in)completeness}
\label{sec:NEC and Einstein frame}
\subsection{NEC violation}
At the early Genesis stage, we have 
$$\dot{H}^J = h_0 c (1 + \delta) \cdot (-c t^J)^{-2 - \delta} > 0,$$
where $t^J$ is Jordan frame cosmic time and dot means derivative with the respect to the $t^J$.
The equation above means that $\dot{H}^J$ is positive, thus the Universe expands as expected in the Genesis scenario. In the Einstein frame, a similar type of expansion requires a violation of the NEC. This violation is essential for constructing non-singular cosmological scenarios (for a review, see~\cite{Rubakov:2014jja}), and the NEC violation makes the Genesis scenario possible. However, it is not so obvious how to address the NEC condition in the Jordan frame. In the case of non-minimal coupling with gravity, the distinction between the gravitational and scalar-field portions of the Lagrangian becomes ambiguous.

Nevertheless, let us introduce the energy density $\rho$ and pressure $p$ as follows:
\begin{align*}
    \rho^{J} &\equiv - S_{0}^{0}\;,\\
    p^{J} &\equiv S_{1}^{1}\;,
\end{align*}
where we introduce the new tensor $S_{\mu}^{\nu}$, given by:
\begin{align*}
    S_{\mu}^{\nu} \equiv R_{\mu}^{\nu} - \frac{1}{2} \delta_{\mu}^{\nu} R\;.
\end{align*}

The variation of the action with respect to the metric leads to the following covariant equations of motion (one can find the covariant equations of motion in terms of the functions $G_2$, $G_3$, and $G_4$ in Ref.~\cite{Kobayashi:2011nu}). For the action \eqref{modelAction} and the FLRW metric, the tensor $S^{\nu}_{\mu}$ is given by:
\begin{align*}
   \rho^J &\equiv S_{0}^{0} = - \frac{1}{2 G_4} \left( 2X G_{2X} - G_{2} + 6X \dot{\phi} H^J G_{3X} - 2X G_{3\phi} - 6H^J \dot{\phi} G_{4\phi} \right)\;, \\
   p^J &\equiv S_{1}^{1} = S_{2}^{2} = S_{3}^{3} = \frac{1}{2 G_4} \left( G_{2} - 2X(G_{3\phi} + \ddot{\phi} G_{3X}) + 2(\ddot{\phi} + 2H^J \dot{\phi}) G_{4\phi} + 4X G_{4\phi\phi} \right)\;.
\end{align*}
In the above equations, the dot denotes a derivative with respect to Jordan frame cosmic time. Here, the superscript $J$ indicates that $\rho^{J}$ and $p^{J}$ are not the actual energy density and pressure, but rather they exhibit analogous properties similar to those in the Einstein frame.

Indeed, with these definitions, the equations of motion yield:
\begin{align*}
    \rho^{J} + p^{J} = -2\dot{H}^J,
\end{align*}
where again the dot denotes a derivative with respect to Jordan frame cosmic time. 
We can understand the NEC in the Jordan frame as follows:
\begin{align*}
    &\rho^{J} + p^{J} \geq 0 \;\; \text{(No NEC violation)},\\
    &\rho^{J} + p^{J} < 0 \;\; \text{(NEC violation)}.
\end{align*}

Now, for completeness, let us use the definitions provided above to estimate the energy density $\rho^{J}$ and pressure $p^{J}$ at early times. After some calculations, we find:
\begin{align*}
    \rho^{J} & = 3 (H^J)^2 \propto |t|^{-2-2\delta},\\ 
    p^{J} & \approx -2 \dot{H}^J \propto - |t|^{-2 - \delta}, \;\; \rho^{J} \ll |p^J|,\;\; t \to -\infty,\\
    \rho^{J}& + p^{J}  < 0 \;\; \text{(NEC violation)}.
\end{align*}

Here, for completeness, we clarify the behavior of our model at early times when transitioning to the Einstein frame. The conformal transformation from the Jordan frame to the Einstein frame takes the following form:
\begin{align}
    \label{eq: Na Einstein Frame}
    g_{\mu\nu}^{E} &= \Omega \cdot g_{\mu\nu}\;, \nonumber\\
    N^{E} &= \sqrt{\Omega} \cdot N\;, \\
    a^{E} &= \sqrt{\Omega} \cdot a\;, \nonumber
\end{align}
where $a$ and $N$ are Jordan frame scale factor and lapse function, respectively.
Here we choose
\begin{equation*}
    \Omega(\phi) \equiv \frac{2 G_{4}(\phi)}{M_{Pl}^2}~.
\end{equation*}
Here, we temporarily restore the Planck Mass to clarify the formula. The function $\Omega(\phi(t))$ exhibits the following asymptotic behavior:
\begin{align*}
    \Omega(\phi(t)) \to (-ct)^{-2\mu} \;\text{and} \;\; N \to 1, \;\; \text{as} \;\; t \to -\infty. 
\end{align*}
Thus, in the Einstein frame, the Lagrangian takes the form:
\begin{equation*}
    \mathcal{L}^{E} = G_{2}^{E}(\phi, X^{E}) - G_{3}^{E}(\phi, X^{E}) \square^{E} \phi + \frac{1}{2} R^{E}.
\end{equation*}

Next, we define the cosmic time in the Einstein frame as
\begin{align*}
    t^{E}_{c} \equiv \int N \sqrt{\Omega} \, dt \propto -\frac{1}{1 - \mu} (-t^J)^{1 - \mu}, \;\;\mu \neq 1\;,
\end{align*}
where $t$ is Jordan frame coordinate time and $t^J$ is Jordan frame cosmic time. At asymptotic past the connection between $t$ and $t^J$ reads
\begin{align}
    t^J = \int N(t) d t = t\;.
\end{align} 
The Jordan frame coordinate time $t$ coincides with the Jordan frame cosmic time $t^J$ , since we choose the asymptotic background value of lapse function $N$ equal to unity.  

Therefore, in the Einstein frame, the Hubble parameter and scale factor behave as follows:
\begin{align}
    \label{Einstein Frame Scale Factor}
    a^{E} &\propto \Big( -(1-\mu) t^{E}_{c} \Big)^{-\frac{\mu}{1 - \mu}}, \;\; \mu \neq 1\;, \\
    H^{E} &= -\frac{\mu}{(1 - \mu) \cdot t^{E}_{c}}\;, \nonumber \\
    \frac{d H^{E}}{d t^E_c} &= \frac{\mu}{1 - \mu} \cdot \frac{1}{|t^{E}_{c}|^{2}}\;.\nonumber
\end{align}
Here, we observe that for $\mu < 1$, $ \frac{d H^{E}}{d t^E_c} > 0$, indicating that the NEC is still violated in the Einstein frame. Conversely, for $\mu > 1$, there is no NEC violation in the Einstein frame, i.e., $\frac{d H^{E}}{d t^E_c} < 0$. Additionally, it is worth noting that the condition $\mu < 1$ corresponds precisely to the unitarity condition in the tensor sector of the current model. This raises the intriguing possibility of a deeper connection between unitarity bounds and NEC condition, which we leave as an open question for future work.

It is not a coincidence that the sign of $\mu - 1$ determines the sign of $\rho + p$ in the Einstein frame. Indeed, it can be proven that for any Genesis model with power-law suppression of the effective Planck mass in the asymptotic past, i.e., $M^{\text{eff}}_{Pl} \propto (-t)^{-2\mu} \to 0$ as $t \to -\infty$, the sign of $\frac{d H^{E}}{d t^E_c}$ coincides with the sign of $1 - \mu$.

For $H^E$ in the Einstein frame, we have:
\begin{align*}
    \frac{d H^{E}}{d t^E_c} = \frac{-3\dot{\mathcal{G}}_T^2 + 4 \mathcal{G}_T^2 \dot{H}^{J} + \mathcal{G}_T \cdot (-2 H^J \dot{\mathcal{G}}_T + 2 \ddot{\mathcal{G}}_{T})}{4 \mathcal{G}_T^3},
\end{align*}
where dot means derivative with respect to the Jordan frame cosmic time $t^J$.
Considering the power-law behavior for $\mathcal{G}_T = (-ct)^{-2\mu}$, this leads to:
\begin{align*}
    \frac{d H^{E}}{d t^E_c} = \frac{(-ct)^{-2\mu}}{t^2}\left(\mu(1 - \mu) + \mu t H^J + t^2 \dot{H}^J\right) \to \frac{(-ct)^{-2\mu}}{t^2}\left(\mu(1 - \mu)\right),
\end{align*}  
Here, we note that in the Genesis scenario, $H^J$ exhibits the following behavior as:
\begin{align*}
    t H^J \to 0,\; t^2 \dot{H}^J \to 0, \;\text{as } t \to -\infty.
\end{align*}

It is also noteworthy that while there is no singularity in the Jordan frame, we do have one in the Einstein frame. This behavior arises from the fact that the conformal factor $\Omega \to 0$ tends to zero in the asymptotic past; i.e., the conformal transformation becomes singular as $t \to -\infty$. \textit{This indicates that at negative infinity, the Jordan frame is physically distinct from the Einstein frame.}

Now, let us investigate how the quadratic action behaves in the Einstein frame. The scalar perturbation is invariant under the conformal transformation, i.e $\zeta^E = \zeta$ (here $\zeta$ is scalar perturbation in Jordan frame). Therefore, the scalar perturbation in Einstein has the same action as in the Jordan frame, thus in terms of
the Einstein frame variables the quadratic action for scalar perturbation reads
\begin{align*}
    \mathcal{S}^{E}_{\zeta\zeta} = \int N^{E} dt^E \big(a^{E}\big)^{3} d^{3} x \left[ \frac{\mathcal{G}^{E}_S}{\big(N^{E}\big)^2} \left(\frac{\partial{\zeta^{E}}}{\partial t^E}\right)^{2} - \frac{\mathcal{F}^{E}_S}{\big(a^{E}\big)^2} \left(\vec{\nabla} \zeta^{E}\right)^{2} \right]\;,
\end{align*}
where $t^E$ is Einstein frame coordinate time, 
$(a^E,\;N^E)$ are given by \eqref{eq: Na Einstein Frame} and  
\begin{align*}
   \mathcal{G}^{E}_{S} = \frac{\mathcal{G}_{S}}{\Omega}\;,\quad 
   \mathcal{F}^{E}_{S} = \frac{\mathcal{F}_{S}}{\Omega}\;, \quad
   \zeta^{E} = \zeta\;.
\end{align*}
Thus, $\mathcal{G}^{E}_{S}$ and $\mathcal{F}^{E}_{S}$ exhibit the following asymptotic power-law time behavior: 
\begin{equation*}
    \mathcal{G}^{E}_{S} \propto \mathcal{F}^{E}_{S} \propto (-t^{E}_{c})^{\frac{\delta}{1-\mu}} \to \infty,\;\; t_c^E \to -\infty,\;\mu<1,
\end{equation*}
here we consider the case $\mu<1,$ since values of $\mu$ greater than unity are restricted by the condition of strong coupling absence \eqref{earlyUnitarityBound}.

We would like to note two things here. First, in the Einstein frame, the coefficients in the quadratic action tend to infinity in the asymptotic past. This behavior is quite different from that in the Jordan frame, where the coefficients in the quadratic Lagrangian for scalar perturbations tend to zero. At first glance, this could imply that the model does not face the danger of a strong coupling regime at early times. However, the situation is different in practice. If we investigate the cubic Lagrangian, we will find that the risk of the strong coupling regime still persists because the interaction terms from the cubic Lagrangian for perturbations grow sufficiently fast compared to $\mathcal{F}_S$ and $\mathcal{G}_S$. This behavior was first mentioned in Ref.~\cite{Ageeva:2020gti} and further investigated in Ref.~\cite{Ageeva:2022fyq}.

Secondly, the Genesis model for $\mu < 1$ (in the Jordan frame), after transitioning to the Einstein frame, resembles the Modified Genesis model from Ref.~\cite{Libanov:2016kfc}. While the Lagrangian for our model is completely different from that of the Modified Genesis model, it gives rise to the same asymptotic behavior for the scale factor and Hubble parameter at an early stage. Moreover, another distinction of our model is the absence of the strong coupling regime at both early and late times. As mentioned in~\cite{Ageeva:2022fyq}, the Modified Genesis model could violate unitarity bounds at an early stage.

For $\mu > 1$ the theory violates unitarity bounds at the asymptotic past. In this case after transitioning to the Einstein frame, one arrives at power-law inflation:
\begin{align}
    \label{power law inflation}
    a^{E} \propto (t^E_{c})^{\frac{\mu}{\mu - 1}}, 
\end{align}
where for the case $\mu>1$ the Einstein frame cosmic time is always positive, i.e. $t^E_{c} \in (0,\ldots)$.
In the Einstein frame, it is clear that this particular case enters a strong-coupling regime. Indeed, the classical energy scale can be estimated as $E_{\mathrm{class}} \propto H \propto \frac{1}{t^E_{c}}$, while the strong-coupling scale (the cutoff) is roughly equal to $M_{\mathrm{Pl}}$. Therefore, there exist times at which the model becomes strongly coupled. For more details about strong coupling, see Section~\ref{sec: Unitarity Bounds for the late times}.


\subsection{Geodesic (in)completeness}

Now, let us discuss the geodesic (in)completeness of our model. Below, we will briefly describe various definitions of geodesic (in)completeness. Afterward, we will comment on which definitions make our model geodesically complete and which lead to geodesic incompleteness:

\textbf{1. Geodesic completeness in the Jordan frame.}

The condition for geodesic past-completeness in FRW cosmology in the Jordan frame reads as follows:
\begin{align*}
    \int_{-\infty}^{T_0} dt^J \; a^{J}(t^J) = \infty,
\end{align*}
where $t^J$ is Jordan frame cosmic time.
This condition is purely geometric and frame-dependent; it is defined only in the Jordan frame. For our model, the Jordan scale factor tends toward a constant in the asymptotic past (see \eqref{background Solutions for early genesis}). Thus, in this sense, our model is geodesically complete.

\textbf{2. Geodesic completeness in the Einstein frame.}

Similarly, the geometric notion of geodesic completeness in the Einstein frame can be expressed as follows:
\begin{align*}
    \int_{-\infty}^{T_0} dt^{E}_{c}\; a^{E}(t^{E}_{c}) = \infty.
\end{align*}
For our Genesis model, the Einstein scale factor tends to zero in the asymptotic past (see \eqref{Einstein Frame Scale Factor}). For $\mu < 1$, the cosmic Einstein-frame time $t^{E}_{c}$ is defined on the interval $(- \infty, \ldots)$, so it is legitimate to integrate from $-\infty$. 

Then, one can find that the Genesis model in the Einstein frame satisfies geodesic past-completeness:
\begin{equation*}
    \int_{-\infty}^{T_0} dt^{E}_{c}\; a^{E}(t^{E}_{c}) = \infty,\quad \text{if } \mu \leq \frac{1}{2}~,
\end{equation*}
and geodesic past-incompleteness:
\begin{equation*}
    \int_{-\infty}^{T_0} dt^{E}_{c}\; a^{E}(t^{E}_{c}) < \infty,\quad \text{if } \mu > \frac{1}{2}.
\end{equation*}
Note that inequality $\mu > \frac{1}{2}$ is the condition required to circumvent the no-go theorem in the tensor sector.

Therefore, if one bypasses the no-go theorem by sending the effective Planck mass to zero in the asymptotic past, then space-time becomes geodesically incomplete in the Einstein frame. This property could be interpreted as geodesic incompleteness for gravitons~\cite{Creminelli:2016zwa, Kobayashi:2019hrl}. Indeed:
\begin{align*}
    \int_{-\infty} dt^{E}_{c}\; a^{E}(t^{E}_{c}) 
    = \int_{-\infty} d t^J \; a^{J}(t^J) \,\mathcal{F}_{T}\;.
\end{align*}
Thus, requiring geodesic completeness for gravitons makes it challenging to realize the Bounce and/or Genesis scenarios within the framework of pure Horndeski gravity~\cite{Kobayashi:2016xpl, Kobayashi:2019hrl}. Nevertheless, it is argued that the purely geometric notion of geodesic (in)completeness should be replaced by more general conditions. Below, we will examine how to apply these conditions~\cite{Rubakov:2022fqk, Wetterich:2024ung} to our model.

\textbf{3. The Photon (Oscillation) Time}

One can define the photon time (the oscillation time for massless particles) in the spirit of Refs.~\cite{Wetterich:2014zta, Wetterich:2024ung}. Let us briefly describe the idea. A dimensionless oscillation time is defined by the number of zeros of the wave function. This quantity is coordinate- and frame-invariant.

As an example, consider the wave function of a photon (or the associated electromagnetic field) in the cosmic rest frame. This function satisfies:
\begin{align}
    (\partial^2_{\eta} + [2\partial_{\eta}\ln a]\partial_{\eta} + k^2)\psi(\vec{k}) = 0,
\end{align}
where $\eta$ is the conformal time, defined by $d\eta = dt/a$. For a chosen momentum $k$, the solution reads:
\begin{align*}
    \psi(\eta, \vec{x}) = B(\eta) \exp\big[i(\vec{k}\cdot\vec{x} - k\eta)\big],
\end{align*}
where the function $B(\eta)$ arises from Hubble damping. For momenta much higher than the Hubble parameter, $B(\eta)$ is nearly constant. In the high-momentum regime, the number of oscillations can be calculated as:
\begin{align*}
    n_{k} = \frac{k\eta}{2\pi}.
\end{align*}
Note that plane waves with different values of $k$ define different clocks, and the number of oscillations varies. However, the oscillation numbers for different momenta are related by an invertible function $f$, i.e., $n_{k_2} = f(n_{k_1})$. Therefore, these different clocks can be synchronized to define a unique time. Following Ref.~\cite{Wetterich:2024ung}, we refer to this set of modes as the clock system.

Thus, we deduce that if conformal time is unbounded toward minus infinity, the model is complete in the sense of photon time. Indeed, for our model, the conformal time tends to minus infinity as we go to the past:
\begin{align*}
    \eta(T) = \int_{-T}^{t_0} d t^J \, \frac{1}{a}, \quad \eta \to -\infty \text{ as } T \to \infty,
\end{align*}
where we note that in the Jordan frame, $a \to a_g = \text{const}$ as the Jordan frame cosmic time approaches minus infinity.

Here, we would like to make one remark: in principle, photons (or other massless particles) can couple non-minimally with the Horndeski field $\phi$. In this case, the photon time could be finite for some couplings, which are not necessarily artificial. For example, if one considers a 5-dimensional Horndeski theory and performs Kaluza-Klein compactification on a circle~\cite{Mironov:2024idn, Mironov:2024zzk}, then in 4 dimensions, terms like 
\begin{align*}
    G_4\!\left(\phi,-\frac{1}{2}(\partial_\rho \phi)^2\right) \left(R - \frac{\theta^2}{4} F_{\mu\nu}^2\right)
\end{align*}
naturally arise. Here, $\phi$ is the Horndeski field, $\theta$ is a dilaton, and $F_{\mu\nu}$ is the electromagnetic tensor. Therefore, such non-minimal couplings could emerge naturally after Kaluza-Klein compactification in higher-dimensional theories. In principle, for some non-minimal couplings, the photon (oscillation) time could be finite in the past. Nevertheless, in this paper, we consider only minimal couplings between photons and the field $\phi$. We leave the question of which non-minimal couplings between photons and $\phi$ are allowed by unitarity bounds, as well as which clock systems would be defined by non-minimally coupled photons, for future work.

To summarize, if we consider photons that couple minimally with the Horndeski field $\phi$ and use the photon clock system to define time, then the Universe exists for eternity.


\textbf{4. The Physical Time for Massive Particles}

Following Ref.~\cite{Rubakov:2022fqk}, we can define a frame-independent physical time for massive particles. Let us consider a free massive scalar field $\pi$:
\begin{align*}
    S = \int d^4 x\, Z(x) \sqrt{-g} \left(-\frac{1}{2}(\partial_{\mu}\pi)^2 - \frac{m^2(x)}{2} \pi^2 \right),    
\end{align*}
where $Z(x)$ and $m(x)$ are functions of the background Horndeski field $\phi$. The physical time can be defined by counting the number of oscillations of $\pi$.  Now we consider some Jordan frame $J_p$ in which massive particles couple minimally, i.e. $m^{J_p} \propto const$ and $Z^{J_p} = const.$
For the FLRW metric and constant $Z$, the physical time $t_{\text{ph}}$ is given by~\cite{Rubakov:2022fqk}:
\begin{align*}
    t_{\text{ph}} = \int_{-T}^{t_0} d t \,\frac{a^{J_p}\,N^{J_p}\,(m^{J_p})^2}{\sqrt{(a^{J_p})^2 m^2 + k^2}}.
\end{align*}
The interval is given by:
\begin{align*}
    ds^2 = -(N^{J_p})^2\,d t^2 + (a^{J_p})^2\,d\vec{x}^2.
\end{align*}

It is straightforward to see that the physical time $t_{\text{ph}}$ is a frame-invariant quantity. Under the Weyl scaling, we have:
\begin{align*}
    g^{J_p}_{\mu\nu} = \Omega^{2}(x)\,\tilde{g}_{\mu\nu}, 
    \quad m^{J_p} = \Omega^{-1}(x)\,\tilde{m}(x), 
    \quad Z^{J_p} = \Omega^{-2}(x)\,\tilde{Z}(x).
\end{align*}
Thus, the definition of $t_{\text{ph}}$ remains unchanged.

We can now consider two interesting cases for our model. In the first case, we couple matter (massive particles) minimally in the Jordan frame (in which genesis takes place) and use these massive particles to define the physical time $t_{\text{phJ}}$ and, consequently, the clock system. In the second case, we couple the massive particles minimally in the Einstein frame and use them to define a different clock system, with physical time $t_{\text{phE}}$. 

Both $t_{\text{phJ}}$ and $t_{\text{phE}}$ are frame-invariant (they do not change under Weyl scaling). However, they are defined by different particles that interact with the Horndeski field $\phi$ in distinct ways.

\textbf{4.1 Minimally Coupled Massive Particles in the Jordan Frame}

For our model in the Jordan frame with minimally coupled massive particles, we have:
\begin{align*}
    N &\to 1, \\
    a &\to a_g, \\
    m &= m_0 \propto const. 
\end{align*}
Therefore, the physical time $t_{\text{phJ}}$ is unbounded as the cosmic Jordan frame time tends to minus infinity. Hence, in this case, the Universe is eternal, and there are no signs of singularity or geodesic incompleteness. 

\textbf{4.2 Minimally Coupled Massive Particles in the Einstein Frame}

If we consider a minimally coupled massive particle in the Einstein frame (for expressions for the Einstein frame scale factor, see equation \eqref{Einstein Frame Scale Factor}), we have:
\begin{align*}
    m &= m_0 \propto const,  \\
    N^{E} & = (-ct)^{-\mu}, \\
    a^{E} & = a_g \cdot (-ct)^{-\mu}.
\end{align*}
The physical time is then given by:
\begin{align*}
    t_{\text{phE}} \propto  \frac{1}{1 - 2\mu}\cdot (T)^{1 - 2\mu}, \;\text{as } T \to \infty.
\end{align*}

We see that if we circumvent the no-go theorem by requiring the convergence of the integrals \eqref{Integrals for No-go}, then the Universe is geodesically incomplete. In other words, it is incomplete when one defines time using minimally coupled massive particles in the Einstein frame. This is expected, as there is a singularity in the Einstein frame, and particles that couple minimally to the metric in this frame experience this singularity.


\textbf{5. Coupling in the Arbitrary Frame}

Now, let us determine in which frames $J_{p}$ we should couple the additional massive field $\pi$ minimally to define a clock system for which the Universe is eternal. In other words, in which frame $J_{p}$ should we couple the scalar field so that the resulting clock system is identical to the photon clock system?

To this end, we make the transition to the frame $J_{p}$ using the following conformal transformation:
\begin{align*}
    g^{J_p}_{\mu\nu} = e^{2 j_p\phi}\,g_{\mu\nu},
\end{align*}
where $j_p$ is a real number. If we couple particles minimally in the frame $J_p$, the physical time is given by
\begin{align*}
    t_{phJ_{p}} = \int_{-T}^{t_0} dt\,\frac{e^{2 j_p\phi} a_g m_0^2}{\sqrt{e^{2 j_p\phi} a_g^2 m_0^2 + k^2}},
\end{align*}
where $t$ is a coordinate time in the frame $J$, where Genesis takes place. We remind here that at asymptotic past in the frame $J$, where Genesis takes place the cosmic Jordan frame time $t^J$ coincides with the coordinate Jordan frame time $t$. 

For $j_p \leq 0$, the physical time $t_{phJ_{p}}$ is unbounded as one approaches minus infinity. Thus, in this case, the Universe shows no signs of past incompleteness.

For the case $j_p > 0$, we find:
\begin{align*}
    t_{phJ_{p}} \to \infty \quad \text{if } j_p \leq \frac{1}{2},
\end{align*}
while for $j_p > \frac{1}{2}$, $t_{phJ_{p}}$ remains finite as one moves toward the asymptotic past. Therefore, if we couple massive particles minimally in the frame with metric $g^{J_p}_{\mu\nu} = e^{2j_p\phi} g_{\mu\nu}$ and $j_p \leq \frac{1}{2}$, these particles will define the same clock system as the photon system. If one measures time with this clock system, the Universe is complete and exists for eternity.

\textbf{6. Cosmic Physical Time}

Next, following Ref.~\cite{Rubakov:2022fqk}, we define the cosmic physical time (defined in the ``cosmic reference frame" where $x^{0}$ is identified with conformal time) as
\begin{align*}
    t_{cph} = \int_{\eta_{init}}^{\eta_0} d\eta \,\sqrt{k^2 + a^2 m^2}.
\end{align*}
For our model, this time diverges as we approach the asymptotic past for particles that couple minimally in both the Jordan frame and the Einstein frame. Therefore, if we define the lifetime of the Universe through $t_{cph}$, then the Universe exists for eternity. It is noteworthy that both the ``cosmic physical time" and the photon time belong to the same clock system.


Finally, we would like to point out that it is not possible to define a physical time in which the Universe is complete for every model. As a simple example, let us consider a massless scalar field with a rolling solution, i.e.,
\begin{align*}
    \mathcal{L}^{E} = X + \frac{R}{2}, \; H^{E}_{\text{bg}} = \frac{1}{3t^E_c}, \; \phi_{\text{bg}} = \sqrt{\frac{2}{3}} \ln(c\cdot t^E_c), \quad t^E_c > 0,
\end{align*}
where $ c $ is a constant. Here, $ t^E_c $ is the Einstein-frame cosmic time. Hence, the Einstein-frame metric is given by
\begin{align*}
   g^E_{\mu\nu} = \mathrm{diag}\bigl(-1,\, a_{\mathrm{init}} \cdot |t^E_c|^{1/3} \cdot \delta_{ij}\bigr),
\end{align*}
where $ a_{\mathrm{init}} $ is a constant. After that, we can perform the following conformal transformation:
\begin{align*}
    g^{J}_{\mu\nu} = e^{-2K(\phi)} \cdot g^{E}_{\mu\nu},
\end{align*}
with
\begin{align*}
    K(\phi) = \frac{\phi}{\sqrt{6}} + \ln\!\left(\frac{a_{\mathrm{init}}}{c^{1/3}}\right).
\end{align*}
This conformal transformation leads to the following Lagrangian:
\begin{align}
    \mathcal{L}^{J} = \frac{a^{2}_{\mathrm{init}}}{c^{2/3}} \cdot \chi^2 \cdot \frac{R}{2},
    \label{JordanFrameMasslessFieldLagrangian}
\end{align}
where we introduce the new field $\chi = e^{\sqrt{\tfrac{1}{6}}\,\phi}$ for convenience. 
Solving the equations of motion for the above Lagrangian, we obtain:
\begin{align}
    H^{J} = 0,\quad \chi = \sqrt{\bigl|\tilde{c} \cdot t^J \bigr|},\quad \tilde{c} = \text{const},
    \label{backgroundSolutionForJordanFrameMasslesField}
\end{align}
where $t^J$ is the Jordan frame cosmic time. 

It can be seen that the metric in the Jordan frame is flat. Thus, if we couple the massive field minimally in this frame and then define the clock system using this field, the physical time will be unbounded as it approaches the past, indicating no signs of incompleteness in the Jordan frame. In other words, in the Jordan frame, the metric is Minkowski, allowing all particles to propagate toward an infinite past.

However, this conclusion is \textit{totally incorrect}! First, the derivative of the field $\chi$ diverges as $t^J$ approaches zero. Second, when $t^J \to 0$, the classical description of the theory becomes invalid, and the background solution \eqref{backgroundSolutionForJordanFrameMasslesField} no longer describes the evolution (this does not happen in our Genesis model, for which the classical solution remains valid throughout the entire evolution!).


Now, let us explicitly show that for the model \eqref{JordanFrameMasslessFieldLagrangian}, there exists a point during the evolution where the classical solution becomes illegitimate. The quadratic and cubic actions in the Jordan frame for tensor perturbations are given by:
\begin{align*}
    \mathcal{S}_{hh} &= \int dt^J d^3x \,\frac{1}{8}\Bigl[
    \mathcal{G}_T\Bigl(\frac{\partial h_{ij}}{\partial t^J}\Bigr)^2 - \mathcal{F}_T\, h_{ij,k}\, h_{ij,k}
    \Bigr], \\
    \mathcal{S}_{hhh} &= \int d^3 x\, dt^J \,\frac{\mathcal{F}_T}{4}\Bigl(
    h_{ik}\,h_{jl} - \tfrac{1}{2}\,h_{ij}\,h_{kl}
    \Bigr) h_{ij,kl},
\end{align*}
where
\begin{align*}
    \mathcal{G}_T &= \mathcal{F}_T = \tilde{c}\,\bigl|t^J\bigr|,\quad [\tilde{c}] = 3.
\end{align*}

We will describe the analysis of the strong coupling scale in detail in Section~\ref{sec: Unitarity Bounds for the late times}, but we provide only a brief overview here. We consider the regime in which the scattering time $t_{\mathrm{scatter}}$ is much smaller than the timescale of evolution of the coefficients $\mathcal{G}_T$ and $\mathcal{F}_T$:
\begin{align*}
t_{\mathrm{scatter}}^{-1} \gg \frac{\dot{\mathcal{G}}_T}{\mathcal{G}_T},
\end{align*}
where dot means derivative with the respect to the Jordan frame cosmic time $t^J$. 

Next, we canonically normalize the field as $h_{ij}^c = \sqrt{\mathcal{G}_T} h_{ij}$ and estimate the matrix element for two-to-two graviton scattering as
\begin{align*}
    M \;\propto\; \frac{E^2}{\tilde{c}\,\lvert t^J \rvert}.
\end{align*}
The unitarity bound is saturated when $M \propto 1$. Thus, the strong-coupling energy scale is roughly given by:
\begin{align*}
    E_{\mathrm{strong}} = \sqrt{\tilde{c}\,\lvert t^J \rvert} \;\to\; 0\quad \text{as } t^J \to 0,
\end{align*}
while the classical energy scale is
\begin{align*}
    E_{\mathrm{class}} = \max\Bigl[H,\;\frac{\dot{H}}{H},\;\chi,\;\frac{\dot{\chi}}{\chi},\ldots\Bigr] \;\propto\; \frac{1}{t^J} \;\to\; \infty\quad \text{as } t^J \to 0,
\end{align*}
where again dot means derivative with the respect to the Jordan frame cosmic time.
The condition for the validity of the classical description is $E_{\mathrm{strong}} \gg E_{\mathrm{class}}$. 
From this, we immediately see that the classical description becomes invalid as $t^J \to 0$. This concludes the proof.

We would like to note that the situation analyzed above is reminiscent of models describing the crossing of the Big Bang singularity; for example, see Refs.~\cite{Kamenshchik:2016gcy, Wetterich:2020oyy}. In some of these models, the behavior of the field and metric is regular in the Jordan frame, whereas the Einstein frame exhibits singularities. It would be interesting to investigate whether the classical description of those theories remains valid in cases where the crossing of the Big Bang singularity is described. We leave this intriguing question for future work.

In summary, we have shown that for our Genesis model, there exists a clock system in which the Universe is complete, i.e., it exists for eternity. Thus, if one is looking for a model that is complete in the generalized sense~\cite{Rubakov:2022fqk}, it is indeed possible to construct such a model within the framework of Horndeski gravity. However, if one imposes a more restrictive condition—namely, geometric geodesic completeness in the Einstein frame (geodesic completeness for gravitons)—then one must consider Beyond Horndeski or DHOST gravity~\cite{Kobayashi:2019hrl}.


Now, based on the previous discussion, we can formulate two distinct conditions for completeness:

\textbf{A. Geodesic Completeness for Gravitons (Completeness in the Einstein Frame):}
\begin{itemize}
    \item In the Einstein frame, the integral diverges:
    \begin{align*}
        \int_{-T} a^{E}(t^E_c)\,dt^E_c = \infty,\quad T \to \infty.
    \end{align*}
\end{itemize}

\textbf{B. Generalized Completeness:}
\begin{itemize}
    \item The conformal time diverges toward the asymptotic past:
    \begin{align*}
        \int_{-T}^{t_0} \frac{N(t)}{a(t)}\,dt = \infty,\quad T \to \infty.
    \end{align*}
    \item The classical description is valid throughout the entire evolution:
    \begin{align*}
        \frac{E_{\text{strong}}}{E_{\text{class}}} \gg 1.
    \end{align*}
    \item There exist frames $J_p$ in which minimally coupled massive particles propagate to minus infinity:
    \begin{align*}
        \int_{-T}^{t_0} dt \,\frac{a^{J_{p}}(t) N^{J_{p}}(t) m_{0}^2}{\sqrt{a^{J_{p}}(t)^2 m_{0}^2 + k^2}} = \infty, \quad \forall k, \quad T \to \infty,
    \end{align*}
    where $t$ -- is a coordinate time and $m_0$ is a constant.
    \item The conformal or disformal transformation to frames $J_p$ is regular and invertible for every point except $\eta = -\infty$.
    \item In the asymptotic future, the frames $J_p$ coincide with the Einstein frame.
\end{itemize}

In other words, we require that in the late-time Universe (when General Relativity is restored), there may exist minimally coupled massive particles that have never been created, i.e., that exist for eternity with no beginning. Thus, our generalized completeness condition has a rather natural interpretation.

It is evident that power-law inflation does not satisfy condition B, as it violates unitarity at early times (see the paragraph below equation \eqref{power law inflation}). The question of whether other early Universe models are complete or incomplete in the sense of definition B is beyond the scope of this paper. Therefore, we leave this broad area of investigation for future work.

\begin{figure}[!ht]
\centering 
\includegraphics[width=7.2cm]{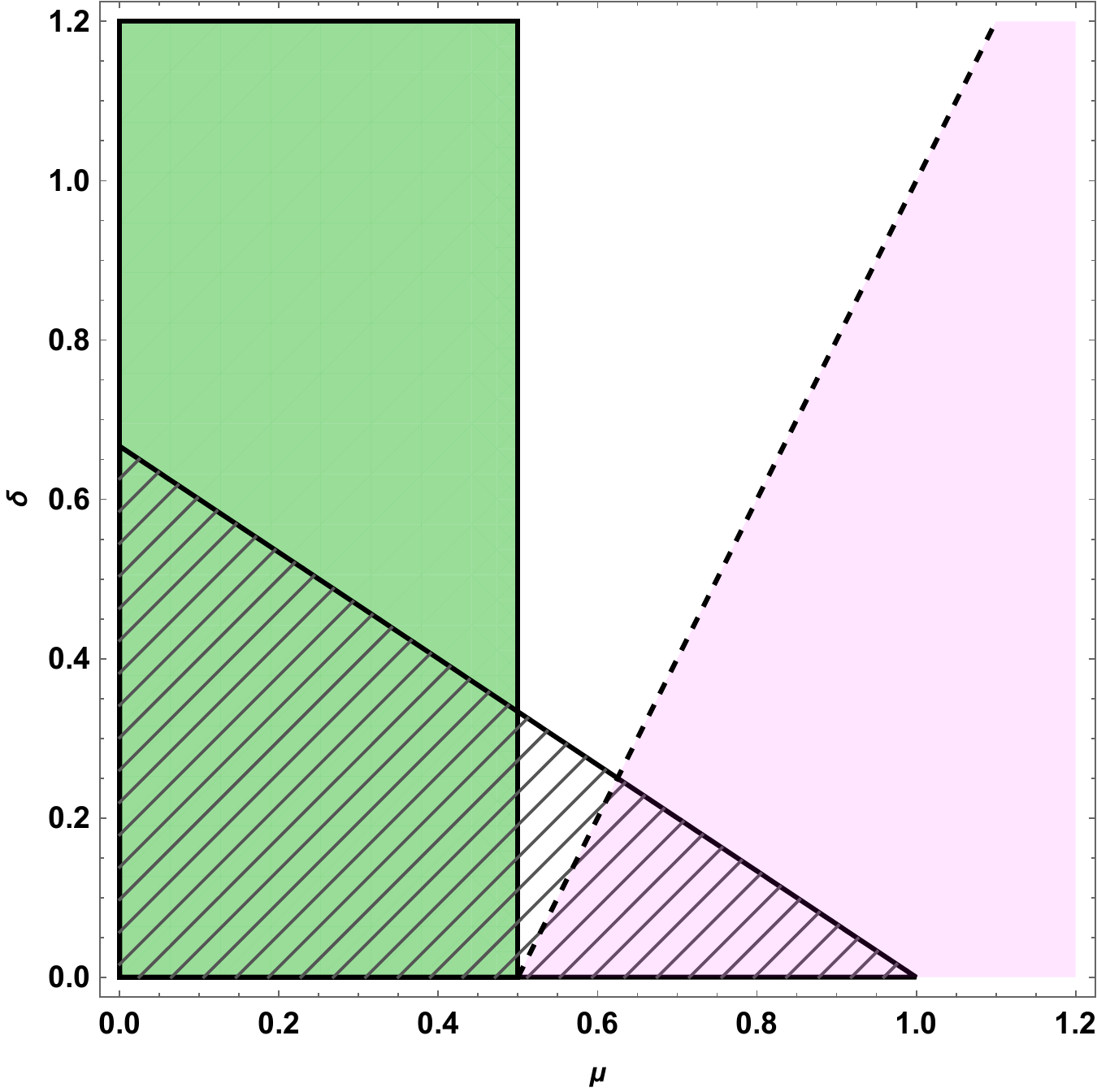}
 \caption{The green area corresponds to geodesic completeness for gravitons (condition A), while the shaded area represents the parameter range for which the model is complete in a generalized sense (condition B). The light magenta area indicates the condition for circumventing the No-Go theorem.
 The combination of magenta and shaded areas forms the parameter space suitable for constructing the nonpathological Genesis scenario.
 }
 \label{fig: Conditions A and B}
\end{figure}

It is worth noting that our Genesis model is complete according to definition B but incomplete according to definition A. However, condition A does not implicitly assume condition B. To illustrate this point, let us consider the initial Genesis stage from Section~\ref{sec:early_gen_stage} and identify the parameters $(\mu,\delta)$ for which this model satisfies conditions A and/or B. We present this in Fig.~\ref{fig: Conditions A and B}. The combination of magenta and shaded areas forms the parameter space in which the Genesis scenario circumvents the no-go theorem and is complete in the sense of condition B. Therefore, this area is suitable for constructing the nonpathological Genesis scenario.  

Additionally, it would be interesting to check whether the Genesis models in Refs.~\cite{Mironov:2019qjt, Ilyas:2020zcb, Zhu:2021ggm}, constructed within the framework of Beyond Horndeski gravity or DHOST, are complete in the sense of definition B. We leave this intriguing question for future work.

\newpage

In the next section, we will construct the stable Genesis solution throughout the entire evolution, demonstrating that it is relatively straightforward to satisfy all the requirements listed above, namely those in \eqref{stability} and \eqref{noSL}.

\section{The nonpathological Genesis scenario}

\subsection{The setup}
\label{sec: The setup}
Here, we will construct a stable Genesis solution that begins with flat space and time, undergoes a relatively slow expansion (in comparison with inflation), and finally transitions to a kination stage. During this stage, the evolution is described by a massless scalar field within the framework of General Relativity. We assume that this kination phase ends with reheating through one of the mechanisms discussed, for example, in Refs.~\cite{Armendariz-Picon:1999hyi, BazrafshanMoghaddam:2016tdk}.

To construct such a solution, we will adopt the general method suggested in Ref.~\cite{Ageeva:2021yik}. To this end, we will utilize a modification of the general Ansatz from Ref.~\cite{Ageeva:2021yik}:
\begin{align*}
    A_2(t, N) &= \frac{1}{2}f^{-2\mu - 2 - \delta}\left(\frac{x(t)}{N^2} + \frac{v(t)}{N^4}\right),\\
    A_3(t, N) &= \frac{1}{2}f^{-2\mu - 1 - \delta}y(t),\\
    A_4(t, N) &= -\frac{1}{2}f^{-2\mu}.
\end{align*}

To achieve the desired behavior, we require the following conditions from the functions $f(t)$, $x(t)$, $v(t)$, and $y(t)$:
\begin{enumerate}
    \item The Lagrangian must have the asymptotic behavior defined in \eqref{LagrFunctions} as $t \to -\infty$:
    \begin{subequations}
    \label{pastAsympForL}
    \begin{align}
        f(t) &\to -c t, \\
        y(t) &\to a_{30} = \text{const}, \\
        x(t) &\to a_{22} = \text{const}, \\
        v(t) &\to a_{24} = \text{const}, 
    \end{align}
    \end{subequations}
    where $a_{30}$, $a_{22}$, and $a_{24}$ are constants that satisfy both the stability requirement \eqref{stability} and the condition for the absence of superluminal propagation \eqref{noSL}, leading to a positive value of $h_0$ (ensuring the expansion of the Universe) and the unit lapse function $N_0 = 1$.
    
    \item We want to achieve a kination epoch in the future ($t \to +\infty$), so we set:
    \begin{subequations}
    \label{futureAsympForL}
    \begin{align}
        A_2 & = \frac{1}{3 t^2 N^2}, \\
        A_3 & = 0, \\
        A_4 & = -\frac{1}{2}.
    \end{align}
    \end{subequations}
    These Lagrangian functions will lead to the kination epoch, i.e.,
    \begin{equation*}
        H = \frac{1}{3tN_f},\; N_f = \text{const},
    \end{equation*}
    where $N_f$ is a constant equal to the value of the lapse function during the kination stage. This constant can be determined from the numerical solution.  Here the $t$ is a coordinate time, therefore one defines Hubble parameter as $H \equiv \frac{1}{N(t) a(t)}\cdot \frac{d a(t)}{dt}$\;. 

    To clarify why this form of $A_2$ corresponds to the kination stage, we note that, in covariant formalism, the evolution is governed by the massless scalar field with the following Lagrangian:
    \begin{align*}
         \mathcal{L} = X + \frac{R}{2}.
    \end{align*}
    Then, we can use the freedom of field redefinition and choose the background field $\phi$ as follows:
    \begin{align*}
         \phi = \sqrt{\frac{2}{3}} \ln(t).
    \end{align*}
    Using the formulas~\eqref{FromADMToCov}, we arrive at the expressions in \eqref{futureAsympForL}.

    \item We will also consider exponential damping in the asymptotic past and future. The Lagrangian functions exhibit power-law damping during the kination and Genesis stages, and we aim to achieve a rapid transition between these two epochs. Therefore, all parts of the Lagrangian functions that correspond to this transition must have much higher damping factors than those governing the evolution during the Genesis stage and the kination phase, respectively. Thus, exponential damping is perfectly suitable.

\end{enumerate}

Later in this section, we present a concrete example that satisfies all the requirements outlined above; see~\eqref{fullGenesisLagrangian}.


The condition (1) from above, together with \eqref{healthy_region}, leads to the following constraints on the parameters $a_{22}$, $a_{24}$, and $a_{30}$:
\begin{subequations}
\label{boundsFor_a}
\begin{align}
    &N_0 = 1: \quad a_{24} = -\frac{a_{22}}{3}, \\
    &h_0 > 0: \quad \frac{a_{30}}{4} - \frac{a_{22}}{6 c (1 + \delta + 2 \mu)} > 0, \\
    &\mathcal{G}_S > 0: \quad -a_{22} > 0, \\
    &\mathcal{F}_S > 0: \quad \text{holds automatically}, \\
    &u_S < 1: \quad \frac{2 c h_0 (\delta - 2 \mu + 1)}{a_{22}} < 1, \\
    &\text{No-go:} \quad 2\mu > 1 + \delta > 0, \\
    &\text{No strong-coupling:} \quad \mu + \frac{3}{2}\delta < 1.
\end{align}
\end{subequations}
Thus, it is relatively straightforward to choose concrete values for the model parameters and construct an example of the Genesis scenario. In the next section, we will present such an example and analyze its stability numerically throughout the entire evolution.

Next, let us choose the following model parameters $a_{22}$, $a_{24}$, and $a_{30}$:
\begin{align*}
    a_{22} \equiv -g < 0, \; a_{24} = \frac{g}{3}, \; a_{30} = 0.
\end{align*}
This set of parameters is fully consistent with the requirements in \eqref{boundsFor_a}.
This concrete choice leads to the following covariant Lagrangian at the Genesis stage:
\begin{align}
    \label{JordanFrameLagrangian}
    G_2 &= \frac{g X \left(-3 c^2 e^{2 \phi} + 2 X\right) e^{\phi (\delta + 2 \mu - 2)}}{3 c^4} + 4 \mu^2 X e^{2 \mu \phi} \ln\left(\frac{X}{X_0}\right) \;, \nonumber\\
    G_3 &= \mu e^{2 \mu \phi} \left(\ln\left(\frac{X}{X_0}\right) + 2\right) \;,\\
    G_4 &= \frac{1}{2} e^{2 \mu \phi}\;\nonumber,
\end{align}
where $X_0$ is an integration constant that appears during the transition from ADM to covariant formalism. The actual value of $X_0$ is physically irrelevant, since rescaling $X_0$ only adds a total derivative to the Lagrangian. For the Jordan frame Lagrangian, we have:
\begin{align*}
    \mathcal{L}(\lambda \cdot X_0) - \mathcal{L}(X_0) = -4 \mu^2 X \ln (\lambda) e^{2 \mu \phi} + \mu \ln (\lambda) e^{2 \mu \phi} \square \phi \Rightarrow 0.
\end{align*}
The situation is similar for the Einstein frame Lagrangian. One can define the field $\phi$, without loss of generality, such that $\phi(t) = -\ln(-t \cdot c)$ and then use the formulas \eqref{FromADMToCov}.


This Jordan frame Lagrangian admits the following early-time asymptotic solutions:
\begin{align}
    \label{jordanFrameSolutionsPhiH}
    \phi &\to -\ln(-ct)\;, \\
    H &\to \frac{g (-c t)^{-\delta-1}}{6 c (\delta + 2 \mu + 1)}\;,\nonumber
\end{align}
which are consistent with the analysis in the ADM formalism. These solutions lead to the following asymptotic behavior for the functions $G_2$, $G_3$, and $G_4$:
\begin{align*}
    G_2 &\to \frac{1}{3} (-c t)^{-\delta - 2 (\mu + 1)} \left(6 c^2 \mu ^2 (-c t)^{\delta} \ln \left(\frac{1}{2 t^2 X_0}\right) - g\right)\;, \\
    G_3 &\to \mu (-c t)^{-2 \mu} \left(\ln \left(\frac{1}{2 t^2 X_0}\right) + 2\right)\;, \\
    G_4 &\to \frac{1}{2} (-c t)^{-2 \mu}\;, \quad \text{as } t \to -\infty.
\end{align*}
The Einstein frame functions $G_{2}^{E}$ and $G_{3}^{E}$ are given by:
\begin{align*}
    G_{2}^{E}(\phi, X^{E}) &= \frac{1}{3} X^{E} \left(\frac{2 g X^{E} e^{\phi (\delta + 2 \mu - 2)}}{c^4} - \frac{3 g e^{\delta \phi}}{c^2} - 6 \mu^2\right)\;, \\
    G_{3}^{E}(\phi, X^{E}) &= \mu \log \left(\frac{X^{E} e^{2 \mu \phi}}{X_0}\right)\;.
\end{align*}
In the above expressions, we used the definitions from Appendix~\ref{app:fromJToE}. The dimension of $g$ is $4$, while $X_0$ is a constant with dimension $[X_0] = 2$.

We observe that as the time $t$ approaches minus infinity, the field $\phi$ also tends to minus infinity. This suggests that the effective Planck mass $M_{Pl}^{\text{eff}} \equiv e^{\mu \phi}$ tends to zero in the asymptotic past. This observation supports the statement made in Section~\ref{sec:NEC and Einstein frame} regarding the possibility of a strong-coupling regime at early times in our model.


Now, the most challenging part is approaching: we are going to find the transition functions that will describe the stable transition from the early Genesis stage in the past to the kination stage in the future. Through trial and error, we arrive at the following setup:
\begin{align*}
    c &> 0, \;\; s > 0\;, \\
    f(t) &= \frac{c}{2} \left( -t + \frac{\ln[2 \cosh(st)]}{s} \right) + 1, \\
    U(t) &= \frac{e^{st}}{1 + e^{st}}, \\
    x(t) &= a_{22} \cdot \left( 1 - U(t) \right) + \frac{2 \cdot U(t) \cdot f^{2\mu + \delta + 2}}{3 \cdot \left( \frac{2f(t)}{c} + t \right)^2}, \\
    v(t) &= a_{24} \cdot \left( 1 - U(t) \right).
\end{align*}

With these functions, the Lagrangian has the correct asymptotic behavior both in the distant past~\eqref{pastAsympForL} and in the future~\eqref{futureAsympForL}. This particular choice leads to the following ADM functions $A_2$, $A_3$, and $A_4$:
\begin{align}
\label{fullGenesisLagrangian}
    A_2 &= \frac{1}{2} f^{-2\mu - 2 - \delta} \left( -\frac{g}{N^2} + \frac{g}{3 N^4} \right) \cdot (1 - U) + \frac{U}{3 N^2 \left( \frac{2f}{c} + t \right)^2}\;, \nonumber\\
    A_3 &= 0\;, \\
    A_4 &= -\frac{1}{2} f^{-2\mu}\;.\nonumber
\end{align}
Additionally, let us emphasize that the choice $A_3 = 0$ does not necessarily imply the absence of the $G_3(\phi, X)$ term in the covariant Lagrangian, as seen in \eqref{JordanFrameLagrangian}.


\subsection{The numerical example}

\label{sec:numerical_example_1}

Now we will present a numerical example of the Genesis scenario that is fully stable throughout its entire evolution. To investigate small $\delta$ values numerically, we will split our solution into two stages: in the first stage, we obtain the solution in terms of the variables $(h,u)$, while in the second stage, we perform a numerical simulation in terms of the variables $(H,t)$.

Before moving to the explicit numerical examples, let us discuss the behavior of the solution during the early-time stage. To this end, we can write $h(u)$ up to second order in terms of the $u$ variable:
\begin{align}
\label{eq: series for h}
    h(u) &= \frac{g}{6 c (\delta +2 \mu +1)}
     -\frac{u g^2 (5 \delta +8 \mu +4)}{72 c^3 (\delta +2 \mu +1)^3 (2 \delta +2 \mu +1)} 
     \\&+
     \frac{u^2 g^3 (3 \delta +4 \mu +2) (23 \delta +32 \mu +16)}{864 c^5 (\delta +2 \mu +1)^5 (2 \delta +2 \mu +1) (3 \delta +2 \mu +1)} + O(u^3)\;.\nonumber
\end{align}
Thus, the time scale at which higher-order corrections by $u$ must be considered is approximately given by (for clarity, we restore the Planck mass):
\begin{align*}
    t_{nl} = \frac{1}{c} \left( \frac{g}{c^2 M_{Pl}^2} \right)^{1/\delta}\;,
\end{align*}
where $[c] = 1$ and $[g] = 4$, respectively. The corrections due to the $u$ variable are negligible if
\begin{equation*}
    -t \gg t_{nl}\;.
\end{equation*}
Therefore, it is evident that the solution can indeed acquire sufficient non-power law corrections. In this case, the early-time condition for the absence of strong coupling \eqref{healthy_region} could become invalid. We discuss this point further in Sec.~\ref{sec: Unitarity Bounds for the late times}.

To construct a realistic Genesis scenario, one must choose a sufficiently small value of $g$ to ensure a small Hubble parameter in comparison with the Planck mass. The inverse value of $c$ defines the characteristic evolution time, and we want this time to be much larger than the Planck time; therefore, it is natural to choose small values for $c$. Finally, we select the parameter $s$ to be of a similar order as $c$. This latter choice is reasonable, as it suggests that all timescales in the theory are of the same order.


Now we are ready to choose two sets of parameters. The first set will exhibit significant non-power-law behavior, while the second set will lead to a marginally power-law solution during the early Genesis stage. In this section, we demonstrate that it is indeed possible to achieve a fully stable Genesis scenario in both regimes, i.e., for both parameter sets.

\subsubsection{Solution with non-power-law behavior}

Here, by power-law behavior, we mean the following. As can be seen from Eqs. \eqref{eq: series for h} and \eqref{redef}, the Hubble parameter for times $-t \gg t_{nl}$ exhibits manifest power-law behavior $H \propto \frac{1}{(-t)^{1 + \delta}}$. While, for $-t \leq t_{nl}$, the Hubble parameter 
cannot be well described by a simple monomial power-law function of time.
However, in both cases, the background solution for the Hubble parameter cannot be well described by a simple monomial function during the transition stage, i.e., when $-t \leq s^{-1}$.

Therefore, we will call the background solution a ``solution with non-power-law behavior" if the background solution for the Hubble parameter \textit{cannot} be described by a monomial function of time even at the Genesis stage. Conversely, we will call the background solution a ``solution with  power-law behavior" if the background solution for the Hubble parameter \textit{can} be described by a monomial function of time during the Genesis stage.

The first set of parameters is:
\begin{align}
    \mu &= \frac{7}{10}, \;\delta = \frac{1}{10}, \; c = 10^{-4}, \; g = \frac{1}{77} \cdot 10^{-4}, \; s = 10^{-4}.
    \label{var4ParamSet}
\end{align}
This parameter set is consistent with the requirements in \eqref{boundsFor_a}.

Next, we calculate the value of $t_{nl}$ for our particular set of parameters in \eqref{var4ParamSet}:
\begin{align*}
    t_{nl} &\approx 10^{25} \gg \frac{1}{s} \approx 10^{4}\;.
\end{align*}
This indicates that $t_{nl}$ is much larger than $\frac{1}{s}$, suggesting that we will encounter non-power-law behavior well before the transition stage.

\begin{figure}[!ht]
\centering 
\includegraphics[width=7.2cm]{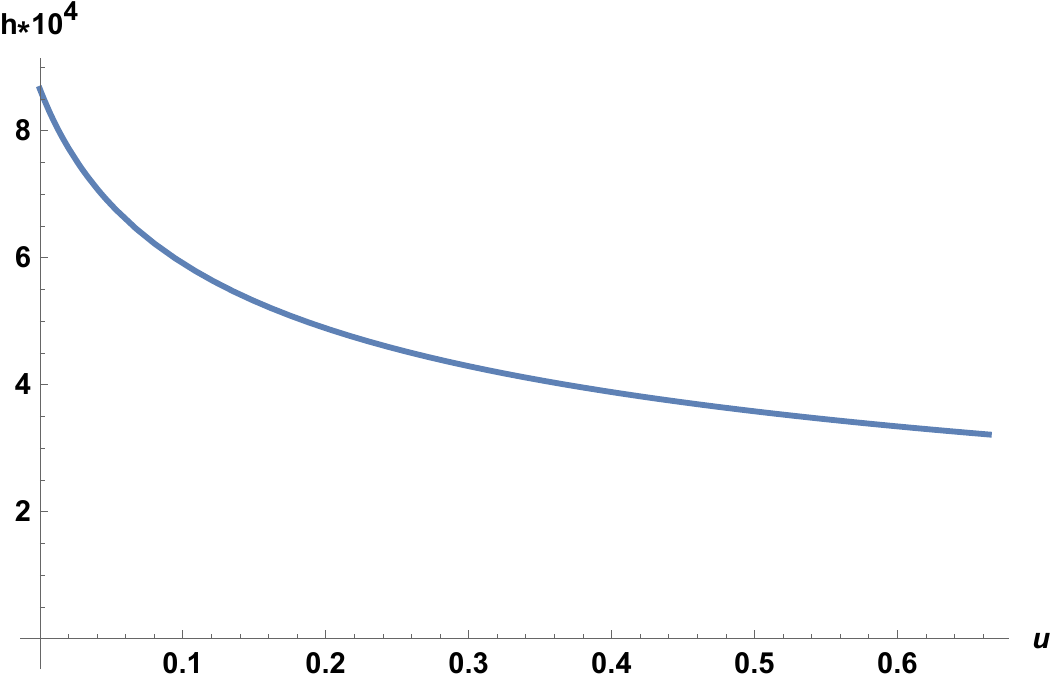}
\includegraphics[width=7.2cm]{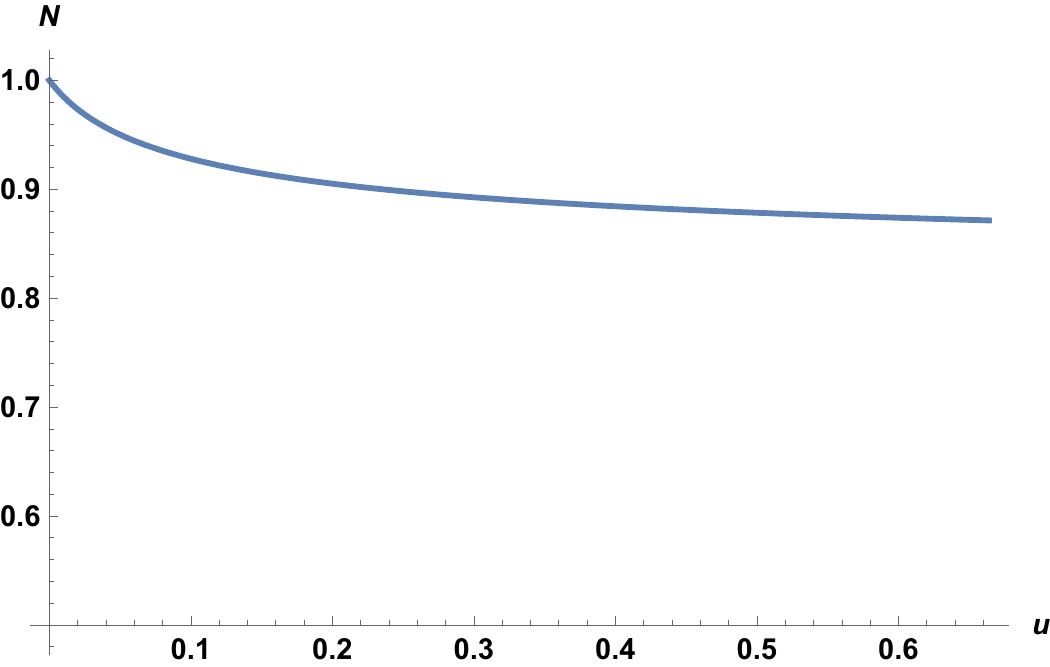}
 \caption{The $h(u)$ (left panel) and the lapse function $N(u)$ (right panel) during the Genesis stage.}
 \label{fig:Nu_Hu}
\end{figure}

We demonstrate the behavior of the Hubble parameter and the lapse function during the Genesis stage as functions of the variable $u$ in Fig.~\ref{fig:Nu_Hu}. Here, we note that small changes in the $u$ variable correspond to substantial changes in time. For example, $u = 10^{-3}$ corresponds to $t = -10^{34}$, while $u = 0.7$ corresponds to $t = -4 \cdot 10^{-5}$. Thus, the functions $h(u(t))$ and $N(u(t))$ in Fig.~\ref{fig:Nu_Hu} change extremely slowly with time.

\begin{figure}[!ht]
\centering 
\includegraphics[width=7.2cm]{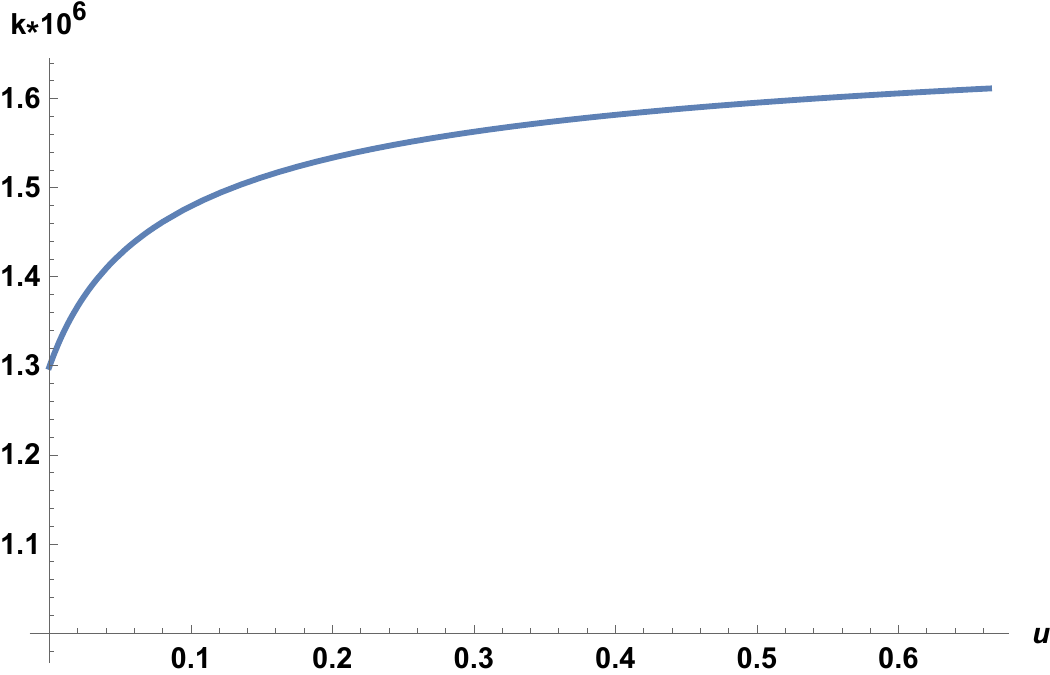}
\includegraphics[width=7.2cm]{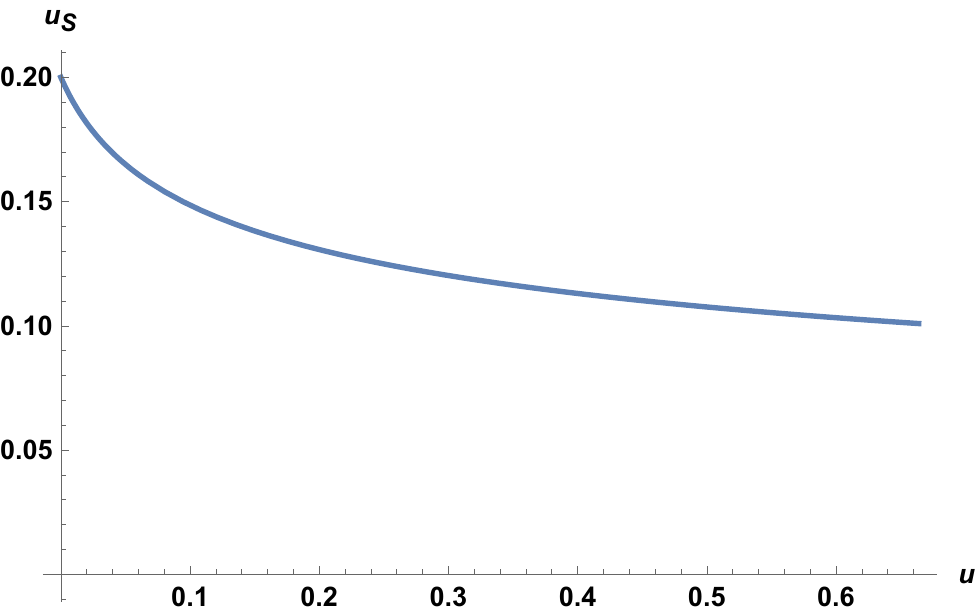}
 \caption{The $k(u)$ (left panel) and the scalar sound speed $u_S(u)$ (right panel) during the Genesis stage.}
 \label{fig:csu_kSu}
\end{figure}

The scalar sound speed and the combination $k \equiv \sqrt{2\mathcal{G}_S u h^2 N^2 (-c t)^{2\mu}}$ are shown in Fig.~\ref{fig:csu_kSu}. The combination $k$ is chosen because it provides a more convenient way to represent the results graphically, given that $\mathcal{G}_S$ can be very small. In terms of $\mathcal{G}_S$, the stability condition is $\mathcal{G}_S > 0$, while in terms of $k$, it requires $k \in \mathbb{R}$.

\begin{figure}[!ht]
\centering 
\includegraphics[width=7.2cm]{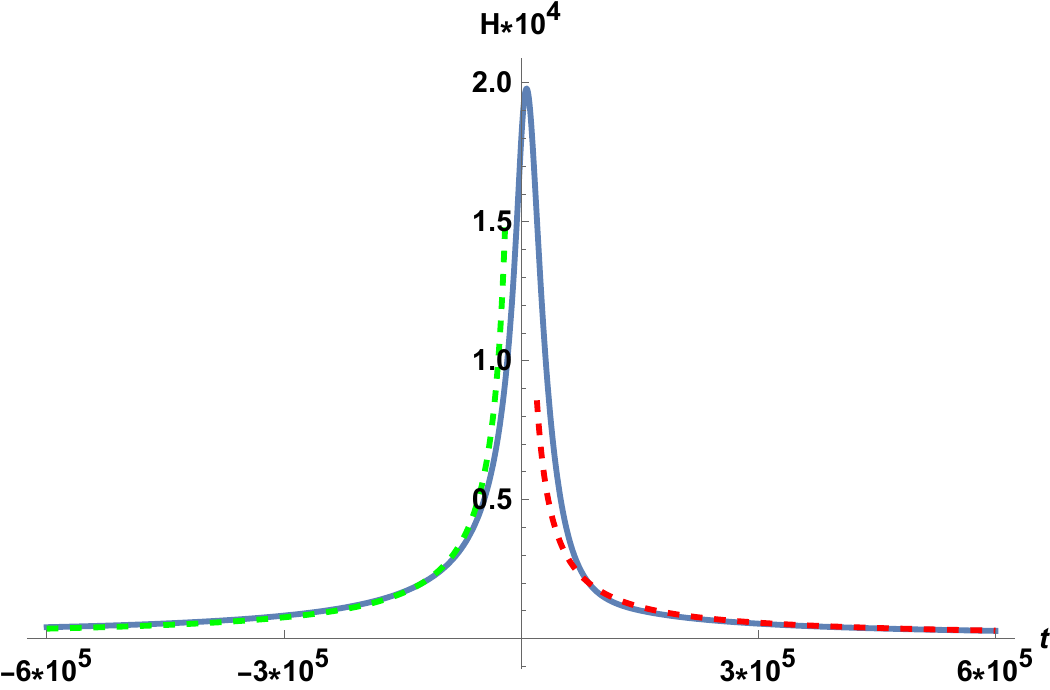}
\includegraphics[width=7.2cm]{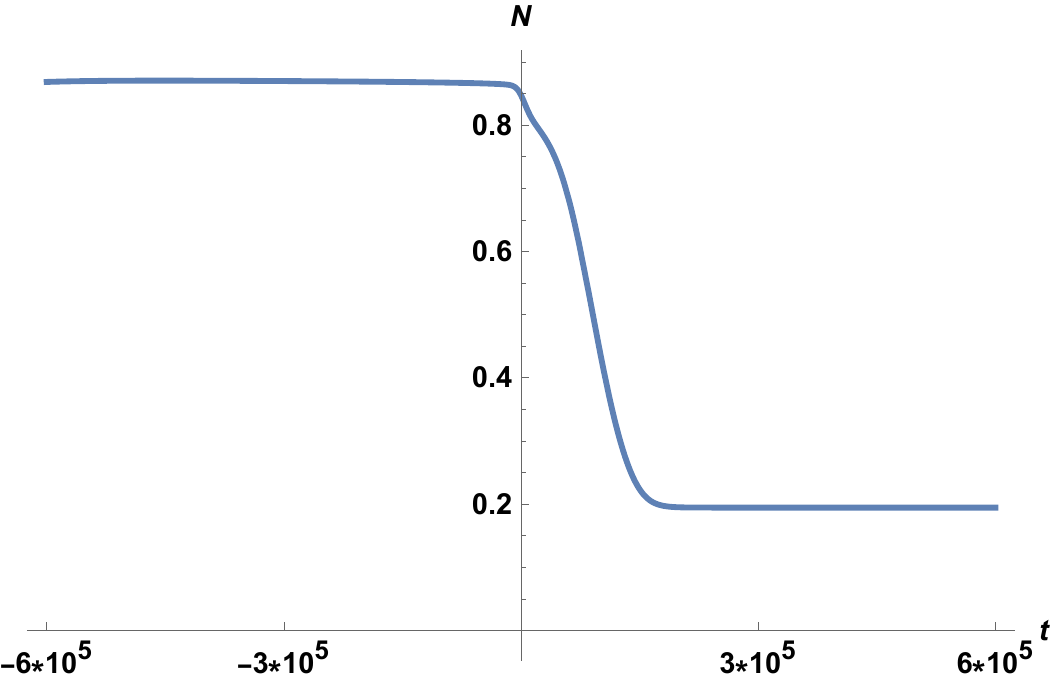}
 \caption{The $H(t)$ (left panel) and the lapse function $N(t)$ (right panel) during the late Genesis stage, transition stage, and kination stage, respectively. The red dashed line represents the Hubble parameter for the kination stage $H = \frac{1}{3t N_f}, \;N_f = N(t_0)$. The green dashed line indicates the Hubble parameter for the early Genesis stage $H = h_0 \cdot (-ct)^{-1 - \delta}$.}
 \label{fig:H_N}
\end{figure}

The Hubble parameter and lapse function for the late Genesis stage, transition stage, and kination stage are shown in Fig.~\ref{fig:H_N}. The red dashed line in Fig.~\ref{fig:H_N} represents the Hubble parameter for the kination stage, given by $H = \frac{1}{3t N_f}, \;N_f = N(t_0)$. We observe that our scenario indeed concludes with the kination epoch, as expected. The green dashed line indicates the Hubble parameter for the early Genesis stage, expressed as $H = h_0 \cdot (-ct)^{-1 - \delta}$.

\begin{figure}[!ht]
\centering 
\includegraphics[width=7.2cm]{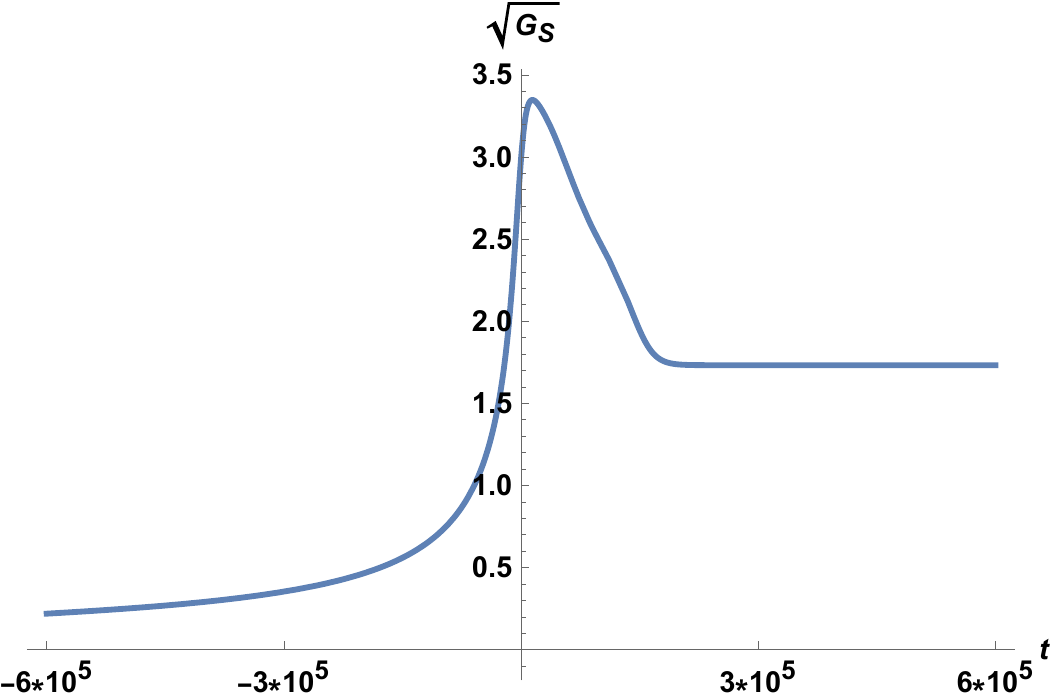}
\includegraphics[width=7.2cm]{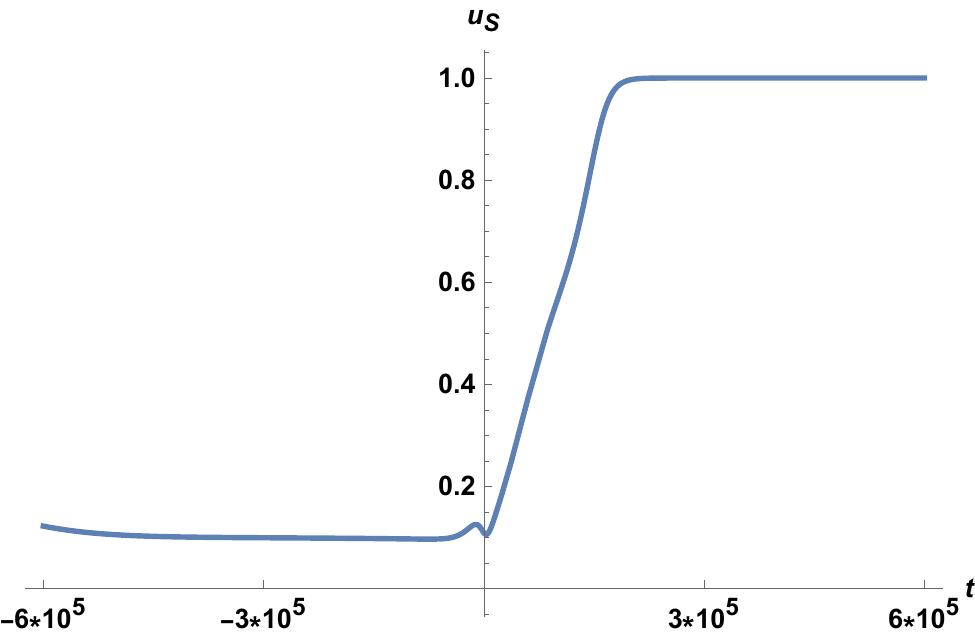}
 \caption{The $\sqrt{\mathcal{G}_S}$ (left panel) and the scalar sound speed $u_S(t)$ (right panel) for the late Genesis stage, transition stage, and kination stage, respectively.}
 \label{fig:Gs_cs}
\end{figure}

In Fig.~\ref{fig:Gs_cs}, we present $\sqrt{\mathcal{G}_S}$ and the scalar sound speed $u_S$ for the late Genesis stage, transition stage, and kination stage.

\newpage

\subsubsection{Solution with power-law behavior}

The second set of parameters is as follows:
\begin{align}
      \label{param. set no power law corrections}
      \mu &= \frac{7}{10}, \;\delta = \frac{1}{10}, \; c =  \frac{1}{6000}, \; g = \frac{35}{20} \cdot 10^{-7}, \; s = \frac{1}{450}\;.  
\end{align}
This parameter set is also consistent with the requirements in \eqref{healthy_region}, which address stability and the absence of strong coupling at early times.

\begin{figure}[!ht]
\centering 
\includegraphics[width=7.2cm]{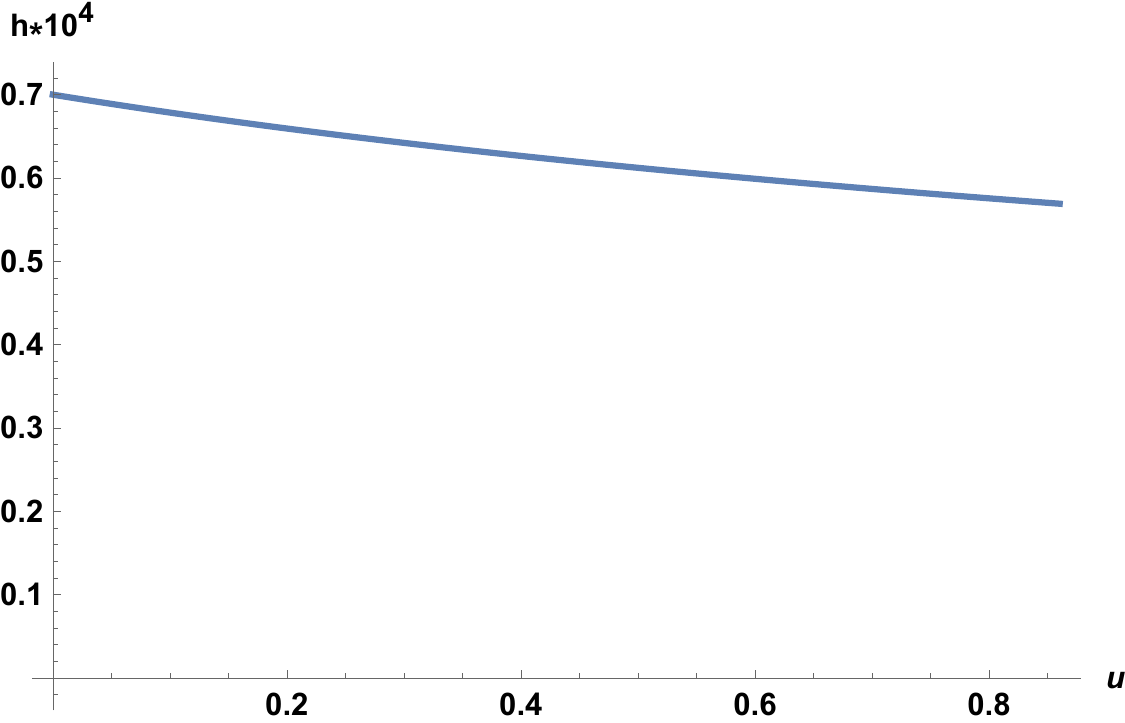}
\includegraphics[width=7.2cm]{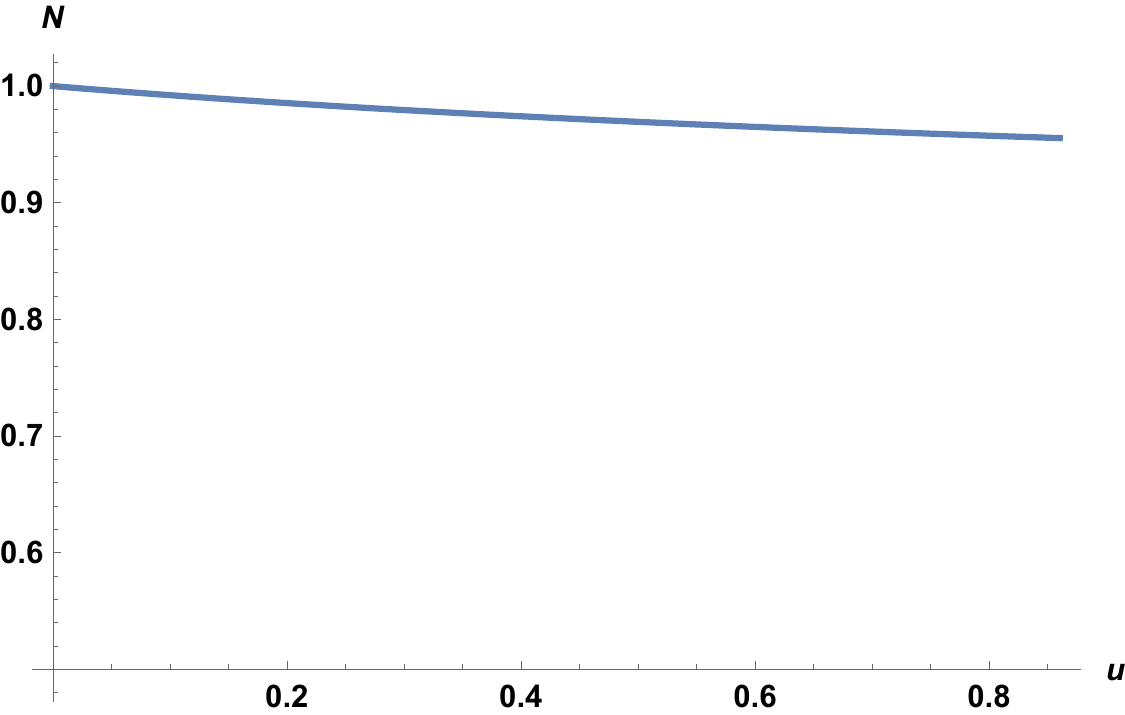}
 \caption{The $h(u)$ (left panel) and the lapse function $N(u)$ (right panel) during the Genesis stage for the variant with power-law behavior.} 
 \label{fig:hh_NN_Lin}
\end{figure}

\begin{figure}[!ht]
\centering 
\includegraphics[width=7.2cm]{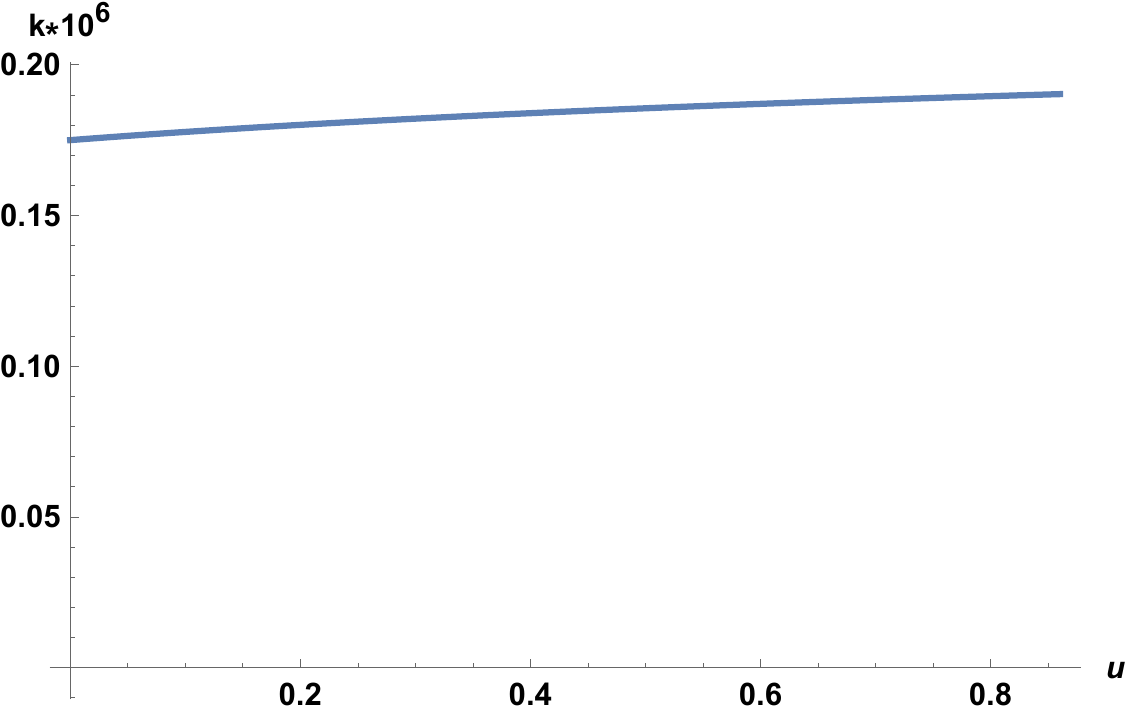}
\includegraphics[width=7.2cm]{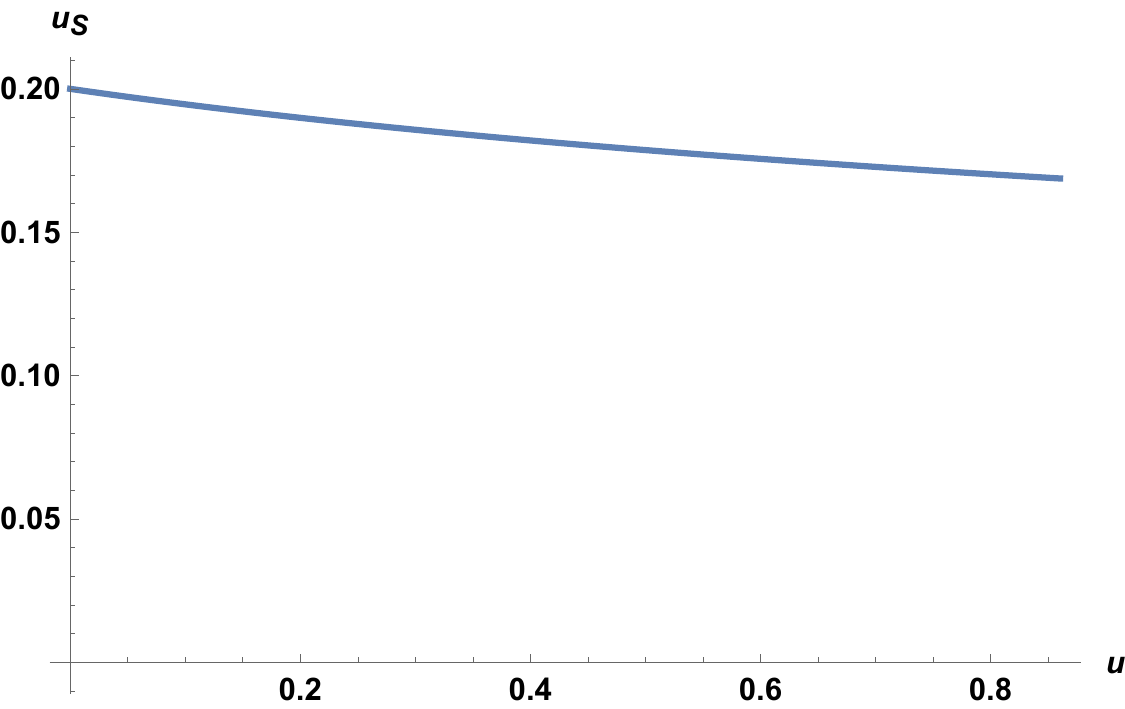}
 \caption{The $k(u)$ (left panel) and the scalar sound speed $u_S(u)$ (right panel) during the Genesis stage for the variant with power-law behavior.} 
 \label{fig:stab_Lin}
\end{figure}

In Fig.~\ref{fig:hh_NN_Lin}, we show the behavior of the functions $h(u)$ and $N(u)$. We note that the function $h(u)$ has slight deviations from a constant value, indicating that the solution has power-law behavior during the Genesis stage. In Fig.~\ref{fig:stab_Lin}, we display the function $k(u)$ and the scalar sound speed. We observe that the function $k(u)$ is real, which signifies the absence of ghosts, while the scalar sound speed $u_S$ is also real and does not exceed unity. Therefore, during the Genesis stage, the solution does not exhibit gradient instabilities and does not permit superluminal propagation for the perturbations.

\begin{figure}[!ht]
\centering 
\includegraphics[width=7.2cm]{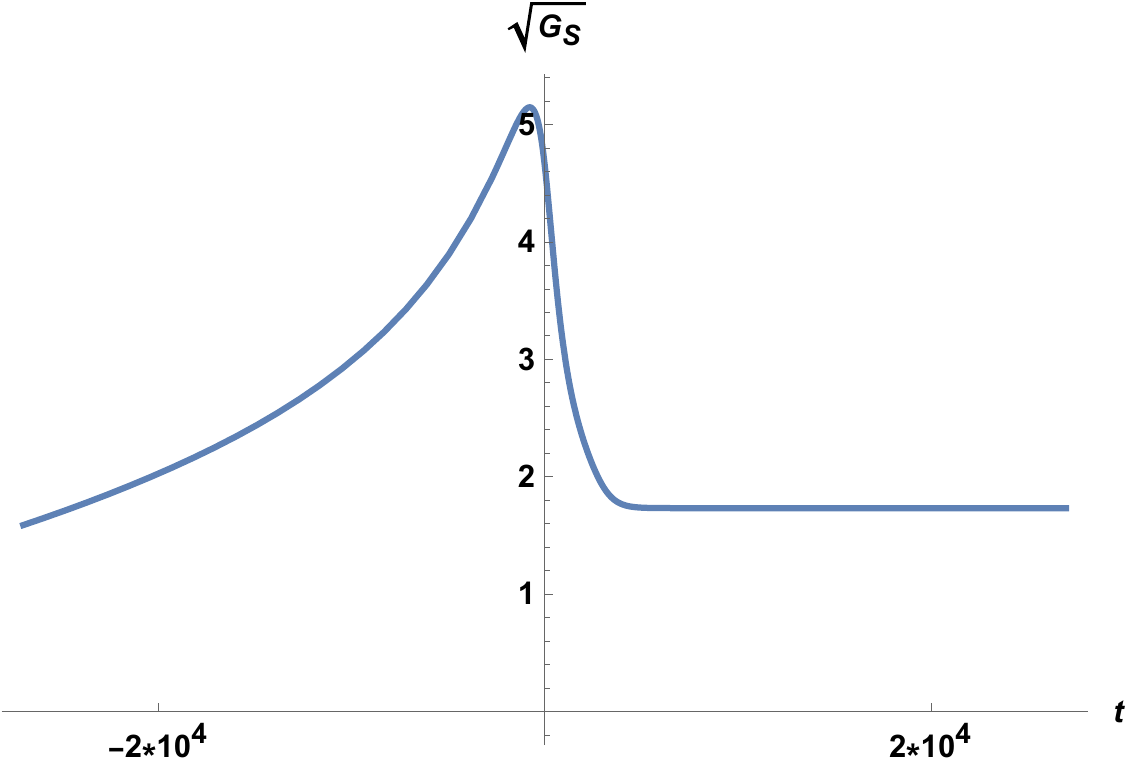}
\includegraphics[width=7.2cm]{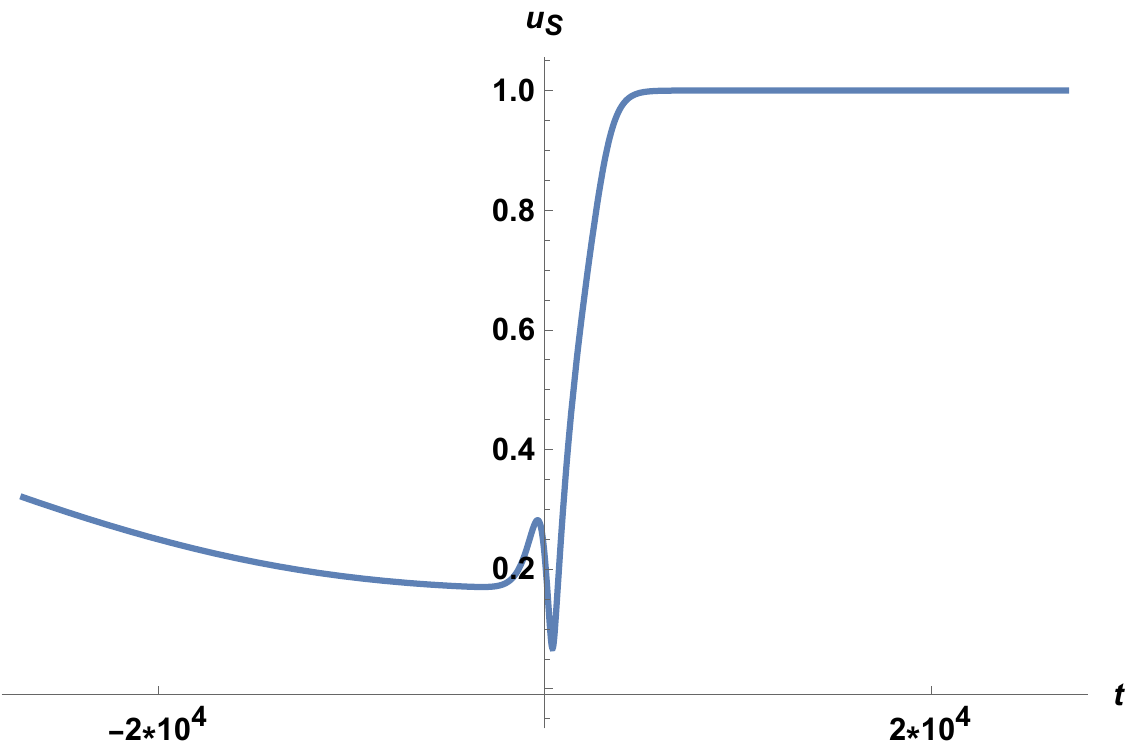}
 \caption{The $\sqrt{\mathcal{G}_S}$ (left panel) and the scalar sound speed $u_S$ (right panel) for the parameter set \eqref{param. set no power law corrections}.}
 \label{fig:Gs_cs_Lin}
\end{figure}

Below in Fig.~\ref{fig:H_N_Lin}, we show the Hubble parameter and the lapse function during the late Genesis and transition stages, respectively. At late times, the Hubble parameter approaches the red dashed line, indicating that the scenario enters the kination stage as expected. Next, in Fig.~\ref{fig:Gs_cs_Lin}, we display $\sqrt{\mathcal{G}_S}$ and $u_S$. We observe that the model avoids ghost and gradient instabilities during the transition stage. Additionally, the speed of scalar perturbations remains below unity at all times, ensuring no superluminal behavior arises during the transition stage.

\begin{figure}[!ht]
\centering 
\includegraphics[width=7.2cm]{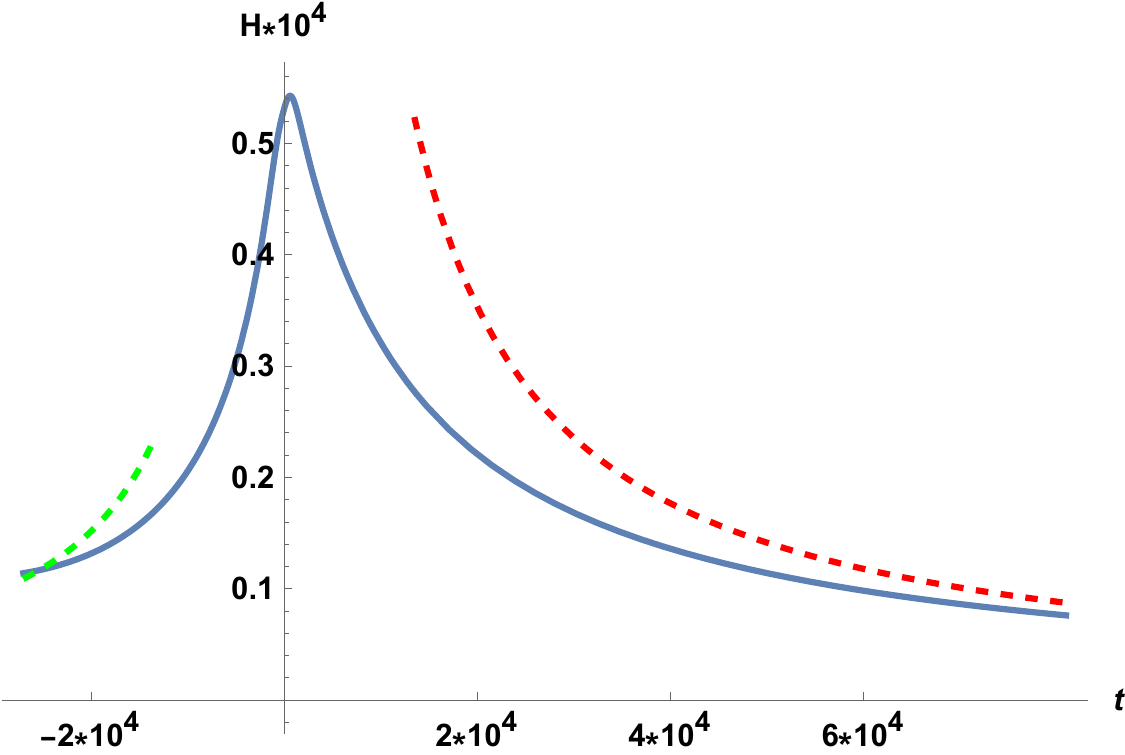}
\includegraphics[width=7.2cm]{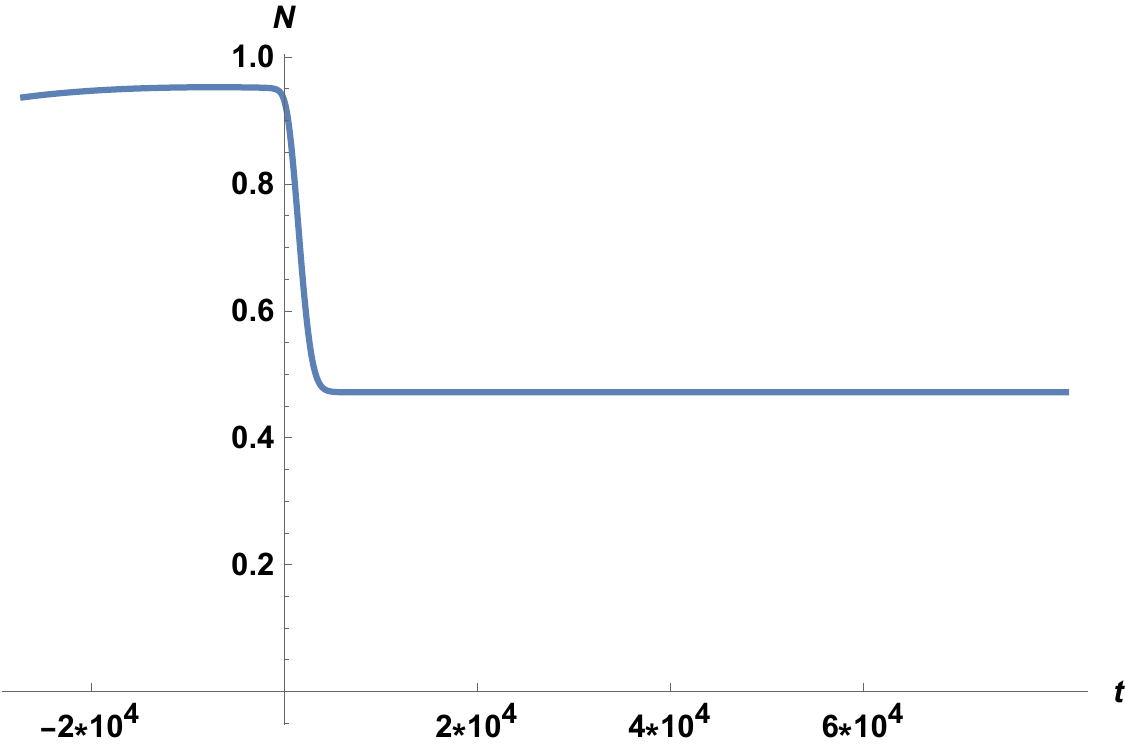}
 \caption{The left panel shows $H(t)$, while the right panel depicts the lapse function $N(t)$. The green dashed line represents the Hubble parameter for the early Genesis stage $H = h_0 \cdot (-ct)^{-1 - \delta}$, and the red dashed line corresponds to the Hubble parameter for the kination stage $H = (3tN_f)^{-1}$.}
 \label{fig:H_N_Lin}
\end{figure}

\newpage
From the analysis above, we conclude that both of our scenarios are completely stable and end in the kination stage, where the free massless scalar field governs the evolution. Additionally, we observe that the scalar sound speed approaches unity in the kination stage, as expected.

In the next section, we will analyze the power spectrum for scalar and tensor perturbations, respectively.


\section{Scalar primordial power spectrum}
\label{sec: Scalar primordial power spectrum}

Here, we present the calculation of the power spectrum for perturbations, starting with the primordial scalar spectrum. The quadratic action for scalar perturbations is given by Eq.~\eqref{quadraticActionScalar}. This action leads to the following equation for the mode $\zeta$:
\begin{equation}
    \ddot{\zeta} + \dot{\zeta} \cdot \frac{\theta^s}{t} + \vec{k}^2 \cdot \mathcal{B}^s \cdot \zeta = 0,
    \label{zetaModeEq}
\end{equation}
where we introduce
\begin{align*}
    \mathcal{A}^s &\equiv  \frac{\mathcal{G}_S a^3}{N}, \\
    \theta^s &\equiv t \cdot \frac{\dot{\mathcal{A}}^s}{\mathcal{A}^s}, \\
    \mathcal{B}^s &\equiv \frac{u_S^2 N^2}{a^2}.
\end{align*}
We note that in general, the coefficients $\mathcal{A}^s$, $\theta^s$, and $\mathcal{B}^s$ could be time-dependent. For the early Genesis stage described in Section~\ref{sec: The setup}: $\mathcal{A}^s$, $\theta^s$ and $\mathcal{B}^s$ are constants with the following values:
\begin{align*}
    \mathcal{A}_{g}^{s} &\equiv \frac{4 (-c t)^{\delta - 2\mu} \left( 2 a_2^{\prime}(1) + a_2^{\prime\prime}(1) \right)}{\left( 4 h_0 + a_3^{\prime}(1) \right)^2} \cdot a_{g}^{3}\;, \\
    \theta_{g}^{s} & \equiv t \cdot \frac{\dot{\mathcal{A}}_{g}^{s}}{\mathcal{A}_{g}^{s}} = \delta - 2\mu < 0\;, \\
    \mathcal{B}_{g}^{s} &=  \frac{c \left( 2 \mu - 1 - \delta \right) \left( a_3^{\prime}(1) + 4 h_0 \right)}{a_g^2 \cdot \left( a_2^{\prime\prime}(1) + 2 a_2^{\prime}(1) \right)} > 0\;.
\end{align*}
Here, we label the values of $A^s$, $\theta^s$ and $\mathcal{B}^s$ during the early Genesis stage as $A^s_g$, $\theta_g^{s}$ and $\mathcal{B}_g^{s}$, respectively.


Let us assume that the mode freezes at early times when the following approximation is valid:
\begin{equation}
    \mathcal{B}^{s} \approx \mathcal{B}_{g}^{s},\;\;\theta^{s} \approx \theta_{g}^{s}.
    \label{earlyTimeFreeze}
\end{equation}
In this particular case, it is legitimate to introduce the canonically normalized field $\psi$ via
\begin{equation*}
    \zeta \equiv \frac{\psi}{\left( 2\mathcal{A}_{g}^{s} \right)^{1 / 2}}\;,
\end{equation*}
so that the quadratic action is given by:
\begin{equation*}
    \mathcal{S}_{\psi \psi}^{(2)} = \int d^{3} x \, dt
    \left[\frac{1}{2} \dot{\psi}^{2} - \frac{\mathcal{B}_{g}^{s}}{2} \big(\vec{\nabla} \psi\big)^{2} + O\big(t^{-2}\big)\right]\;.
\end{equation*}
The last term in the integrand is negligible at early times when $t \to -\infty$, rendering the field $\psi$ a free field. Thus, the negative-frequency normalized solution is given by
\begin{equation}
    \psi_{-\infty} = \frac{1}{(2 \pi)^{3 / 2}} \frac{1}{\sqrt{2 \omega}} \cdot
    \mathrm{e}^{- i \int \omega dt}\;,  
    \label{earlyTimePsi}
\end{equation}
where
\begin{equation*}
    \omega \equiv \sqrt{\vec{k}^2 \mathcal{B}_{g}^{s}}.
\end{equation*}


Now, let us turn to the equation \eqref{zetaModeEq}. The solutions of this equation are given by:
\begin{equation*}
    \zeta_{1,2} = \mathcal{C} \cdot \left(-t\sqrt{\mathcal{B}_{g}^{s}}|\vec{k}|\right)^{\nu_s} \cdot H^{1,2}_{\nu_s}\left(-t\sqrt{\mathcal{B}_{g}^{s}}|\vec{k}|\right)\;,
\end{equation*}
where $\mathcal{C}$ is a normalization constant, and 
\begin{equation*}
    \nu_s \equiv \frac{1 - \theta_{g}^{s}}{2} = \mu  + \frac{1}{2} - \frac{\delta}{2}\;.
\end{equation*}
By matching the solution for $\zeta$ with the early-time asymptotic behavior described in \eqref{earlyTimePsi}, we arrive at:
\begin{equation*}
    \zeta = \frac{(|\vec{k}|\sqrt{\mathcal{B}_{g}^{s}})^{-\nu_s}}{2^3 \pi \sqrt{\mathscr{A}^{s}_g}} \cdot \big(-t\sqrt{\mathcal{B}_{g}^{s}}|\vec{k}|\big)^{\nu_s} \cdot H^{1,2}_{\nu_s}\big(-t\sqrt{\mathcal{B}_{g}^{s}}|\vec{k}|\big)\;,
\end{equation*}
Additionally, we define
\begin{equation*}
    \mathcal{A}^{s}_g \equiv \mathscr{A}^{s}_g \cdot (-t)^{\theta_{g}^{s}}.
\end{equation*}
At late times (i.e., $t \to 0$), this solution becomes a constant, given by:
\begin{equation*}
    \zeta(t)\Big|_{t\to 0} = \frac{\big(\sqrt{\mathcal{B}_{g}^{s}}|\vec{k}|\big)^{-\nu_s}\Gamma(\nu_s)}{2^{3 - \nu_s} \pi^2 \sqrt{\mathscr{A}_{g}^{s}}} \cdot (\text{phase factor})\;.
\end{equation*}


Afterward, it is relatively straightforward to find the scalar power spectrum:
\begin{equation*}
    \mathcal{P} = 4\pi |\vec{k}|^3 \cdot \big|\zeta(0)\big|^2\;.
\end{equation*}
This can be expressed as:
\begin{equation*}
    \mathcal{P} = A_{\zeta} \cdot \left(\frac{k}{k_*}\right)^{n_s - 1}\;,
\end{equation*}
where $k_* = 0.05 \; \text{Mpc}^{-1}$ is the pivot scale, and 
\begin{align}
    n_s &= 3 + \theta_{g}^s = 3 - 2\mu + \delta\;, \label{scalarSpectrumLin} \\
    A_\zeta &= \frac{\Gamma^{2}(\nu_s) (\mathcal{B}_{g}^s)^{-\nu_s} k_{*}^{3 - 2\nu_s}}{2^{4 - 2\nu_s} \pi^3 \mathscr{A}_{g}^s}\;.\nonumber
\end{align}
Here, $n_s$ and $A_\zeta$ represent the scalar spectral tilt and scalar amplitude, respectively. Additionally, we note that $A_\zeta$ does not depend on $a_g$ and/or $k_{*}$ separately:
\begin{equation*}
    A_{\zeta} \propto a_{g}^{2\nu_s} \cdot a_{g}^{-3} \cdot k_{*}^{3 - 2\nu_s} = \left(\frac{k_{*}}{a_{g}}\right)^{3 - 2\nu_s},
\end{equation*}
which makes sense, as only the combination $\frac{k_{*}}{a_{g}}$ carries physical significance. Thus, our result is truly self-consistent.


However, from the constraints in \eqref{healthy_region}, we immediately conclude that $n_s$ must be blue-tilted in our model. We illustrate this point in Fig.~\ref{fig:Mu_del} below.

\begin{figure}[!ht]
\centering 
\includegraphics[width=8cm]{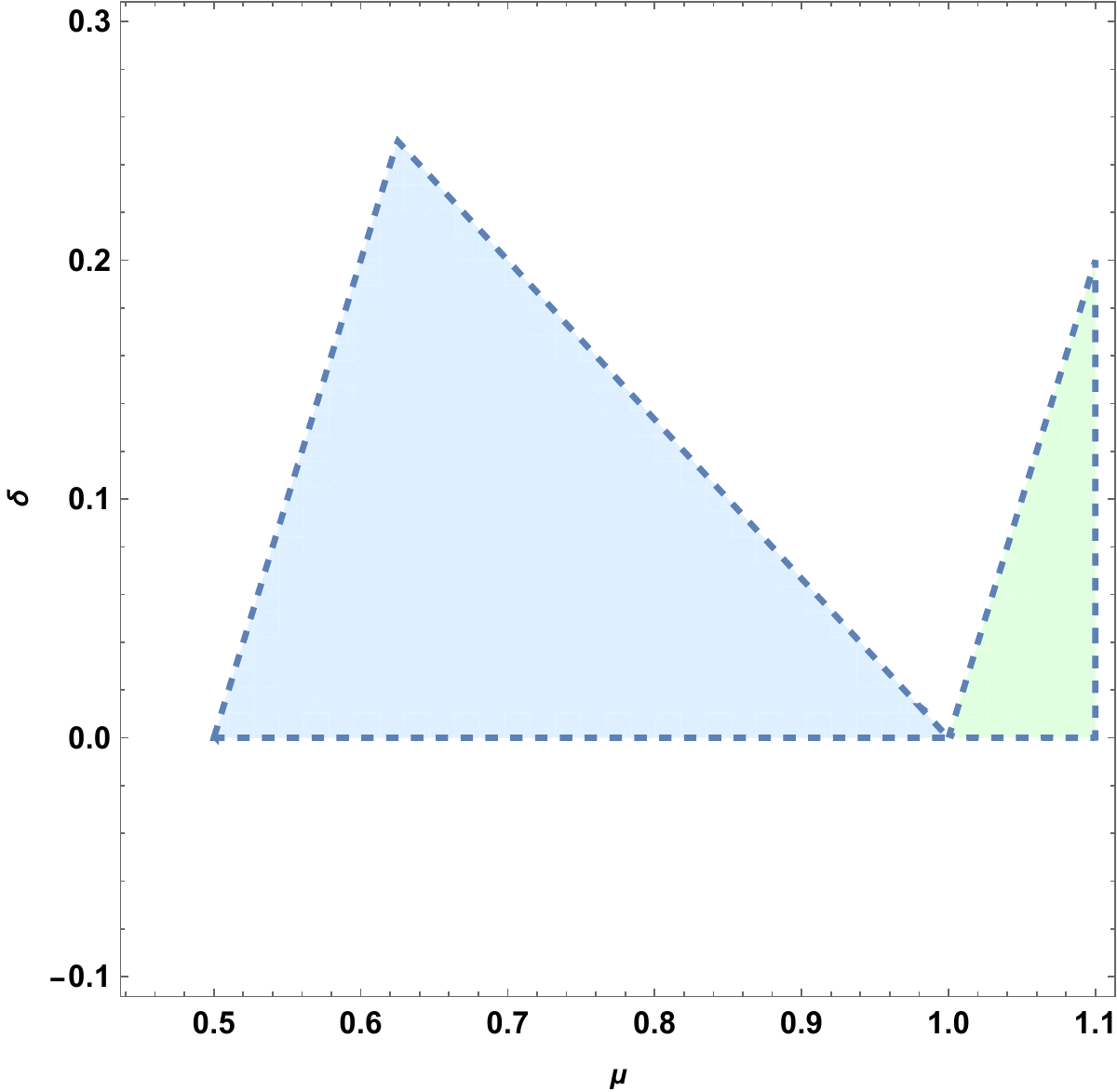}
 \caption{The range of parameters $(\mu,\;\delta)$. The blue area corresponds to the constraints in \eqref{healthy_region}, while the green area indicates the condition $n_s < 1$.}
 \label{fig:Mu_del}
\end{figure}

Thus, in the Genesis model, the validity of the classical field description leads to a blue spectrum. This situation is similar to bounce cosmology with strong gravity in the past, as discussed in Ref.~\cite{Ageeva:2022asq}. At first glance, this property makes the Genesis scenario seem unacceptable from an experimental standpoint since there are strong experimental constraints on the value of the scalar spectral tilt, $n_S = 0.9647 \pm 0.0043$ (see~\cite{Planck:2018nkj}). However, let us reiterate our assumptions. We assume that mode freezing occurs at an early time when the approximation \eqref{earlyTimeFreeze} holds. 

To check this approximation for the numerical model from Section \ref{sec:numerical_example_1}, we will make a rough estimation and take the following values for the parameters:
\begin{align*}
   T_{\text{reh}} &= 10^{-12} \; M_{\text{Pl}} \;\text{(reheating temperature)},\\
   T_{0} &= 10^{-32} \; M_{\text{Pl}} \;\text{(modern temperature)},\\  
   k_{*} &= 0.05 \; \text{Mpc}^{-1} = 2.6 \cdot 10^{-59} \; M_{\text{Pl}} \;\text{(pivot scale)}.
\end{align*}

Firstly, we estimate the initial value of the scale factor $a_g$, taking the modern value of the scale factor $a_0$ to be equal to one ($a_0 = 1$). We obtain
\begin{equation*}
    a_g = \exp\left[-\mathcal{N}_{\text{full}}\right],
\end{equation*}
where
\begin{equation*}
   \mathcal{N}_{\text{full}} = \frac{u_{\text{init}} h_0}{c\delta} + \int_{u_{\text{init}}}^{u_{0}} du\; \frac{h(u)}{c\delta} + \int_{t_0}^{-t_0} dt\; H(t) \cdot N(t) + \ln\left(\frac{T_{\text{reh}}}{T_{0}}\right),\; t_0 < 0.
\end{equation*}
Here, we suppose that reheating occurs instantly after the Genesis stage. For the numerical model from Section \ref{sec:numerical_example_1}, $\mathcal{N}_{\text{full}}$ and $a_g$ read
\begin{equation*}
    \mathcal{N}_{\text{full}} = 92,\; a_g = 9 \cdot 10^{-41}\;.
\end{equation*}


If we accept the approximation \eqref{earlyTimeFreeze}, the time of scalar mode freeze $t_{fr}^0$ can be estimated as
\begin{equation*}
    \frac{\theta_{g}^{s}}{t_{fr}^0} \sim \sqrt{\mathcal{B}_{g}^{s}} \cdot k_{*}\;.
\end{equation*}
On the other hand, if we do not accept the approximation \eqref{earlyTimeFreeze}, the time of freeze $t^{fr}$ can be obtained by numerically solving the following nonlinear equation:
\begin{equation*}
    \frac{\theta^{s}(t_{fr})}{t_{fr}} \sim \sqrt{\mathcal{B}^{s}(t_{fr})} \cdot k_{*}\;.    
\end{equation*}
For the current numerical model, we have
\begin{equation*}
    t^{0}_{fr} \approx -2.3\cdot 10^{19}, \quad t_{fr} \approx -2.5\cdot 10^{20}\;.
\end{equation*}

\begin{figure}[!ht]
\centering 
\includegraphics[width=8cm]{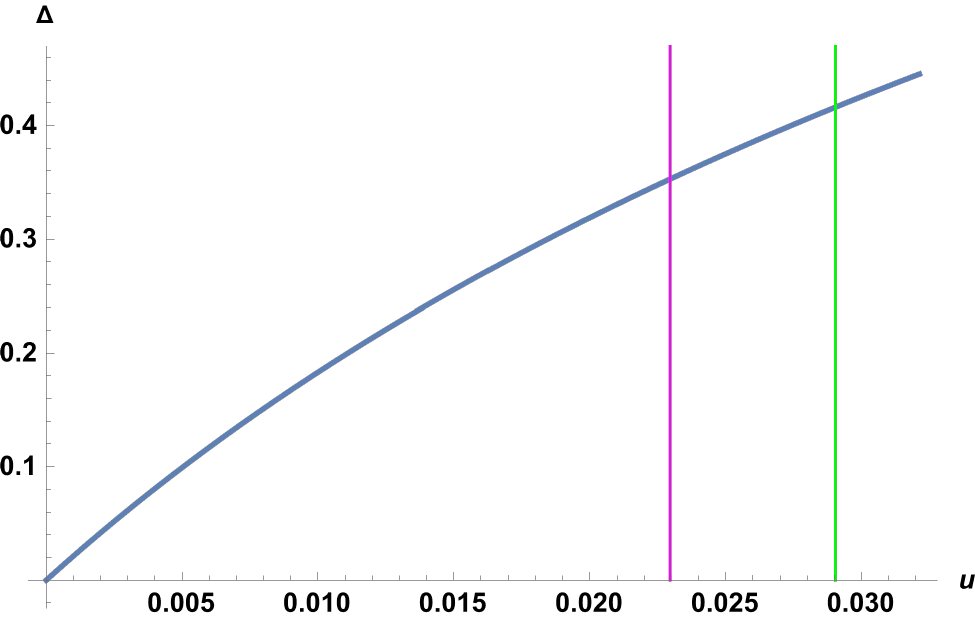}
 \caption{The $\Delta(u)$ in the vicinity of the freeze point. The magenta line represents $u(t^{fr})$, while the green line represents $u(t^{fr}_{0})$.}
 \label{fig:DeltaTh_Th0}
\end{figure}

Next, to check the validity of the approximation \eqref{earlyTimeFreeze}, let us plot the relative error $\Delta = \frac{2 \cdot |\theta_g^s - \theta^s|}{|\theta_g^s| + |\theta^s|}$ between the $\theta_g^s$ and $\theta^s$ values in the vicinity of the mode freeze point and compare the results. We choose $\theta^s$ because it governs the evolution of the friction term in equation \ref{zetaModeEq}, which has a significant impact on the value of $n_s$. The value of the relative error $\Delta$ is shown in Fig.~\ref{fig:DeltaTh_Th0}. We see that the relative error near the mode freeze point is around 30--40 percent, indicating that the approximation \eqref{earlyTimeFreeze}, along with the conclusion that the spectrum should be blue-tilted, is not acceptable. To find the correct value of $n_s$, we should proceed with a numerical analysis of the model. We will conduct such an analysis in Section~\ref{sec:TimeDependTermsScalarSpectrum}.

Additionally, we want to emphasize that there is no problem fitting the value of the scalar amplitude $A_\zeta$. To this end, we can introduce a new parameter $g_{1} > 0$ and deform the model as follows:
\begin{align*}
    \tilde{A}_2 &= \frac{g_1}{2} f^{-2\mu - 2 - \delta} \left( -\frac{g}{N^2} + \frac{g}{3 N^4} \right) (1 - U) + \frac{U}{3 N^2 \left( \frac{2f}{c} + t \right)^2}\;, \\
    \tilde{A}_3 &= 0\;, \\
    \tilde{A}_4 &= -\frac{g_1}{2} \left( f - 1 + g_{1}^{\frac{1}{2\mu}} \right)^{-2\mu}, \; f - 1 + g_{1}^{\frac{1}{2\mu}} > 0\;,
\end{align*}
where the value $g_1 = 1$ corresponds to the initial model. This deformation of the model does not alter the early-time behavior. Thus, during the Genesis stage, the values of the $r$ ratio, $n_s$, and $n_T$ remain unaffected, along with the stability conditions and the solutions for the Hubble parameter and lapse function. The only changes during the Genesis stage are in the absolute values of $\mathcal{F}_S$, $\mathcal{F}_T$, $\mathcal{G}_S$, and $\mathcal{G}_T$. This last fact modifies the absolute value of the scalar amplitude $A_\zeta$ by a factor of $g_1^{-1}$, which allows us to fit the scalar amplitude to the correct experimental value.

Moreover, this model deformation does not impact late-time behavior, since the new functions, namely $\tilde{A}_{2-4}$, share the same late-time asymptotics as those of the original functions in \eqref{fullGenesisLagrangian}. However, the transition stage could be affected. Thus, for each particular choice of $g_1$, one must check the stability conditions. Nevertheless, we argue that for a wide range of values of the parameter $g_1$, it is indeed likely that a stable transition between the Genesis and kination stages can be achieved. Indeed, as shown in Ref.~\cite{Ageeva:2021yik}, a similar class of models admits stable transitions between different cosmological stages in various scenarios. Therefore, one can tune the transition phase to be stable using the same techniques developed in Ref.~\cite{Ageeva:2021yik}.

\begin{figure}[!ht]
\centering 
\includegraphics[width=8cm]{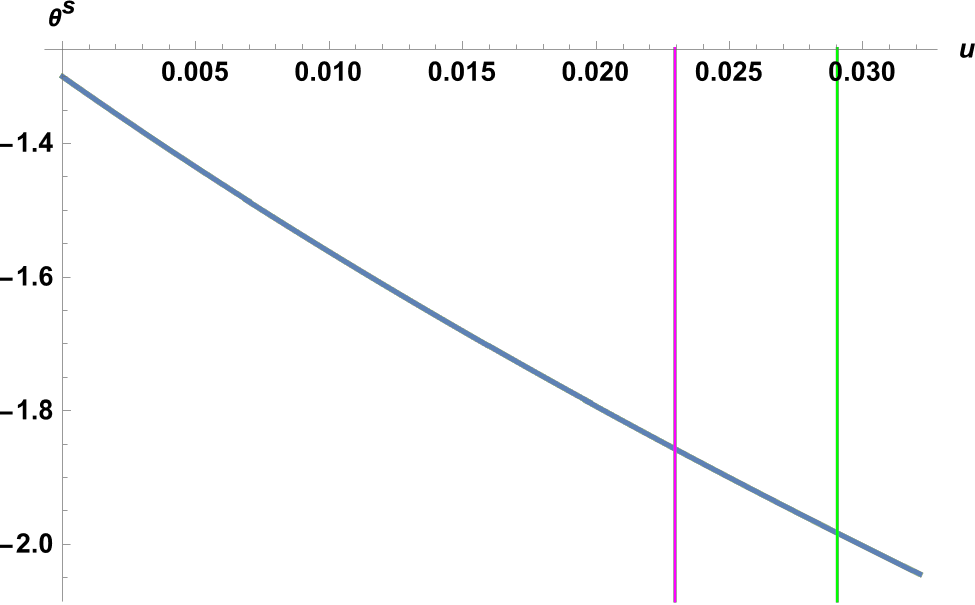}
 \caption{The $\theta^s(u)$ in the vicinity of the freeze point. The magenta line represents $u(t^{fr})$, while the green line corresponds to $u(t^{fr}_{0})$.}
 \label{fig:ThetaS}
\end{figure}

Below, in Fig.~\ref{fig:ThetaS}, we show the behavior of $\theta^{s}$ near the freeze point. From the plot, we conclude that the value of $\theta^s$ decreases for large $u$ and could be less than $-2$. Recalling the expression for $n_s = 3 + \theta^s$, we deduce that, in principle, our model could lead to a red-tilted scalar spectrum. However, to determine whether this statement is true, a more accurate analysis of equation \eqref{zetaModeEq} is necessary, accounting for the slow time dependencies of the functions $\theta^s(t)$ and $\mathcal{B}^s(t)$. We will conduct such an analysis in Section \ref{sec:TimeDependTermsScalarSpectrum}. In the next section, we will also explore the unitarity bounds on our model in more detail.

Finally, let us comment on how to calculate the tensor spectrum for this particular model. These calculations are completely analogous to those for the scalar spectrum, so we will describe them only briefly. To obtain the tensor spectrum, one should make the following replacements: $\mathcal{G}_S \rightarrow \mathcal{G}_T/8$, $u_s \rightarrow u_T = 1$, and then multiply the result by 2 to account for the two polarizations of the graviton. This procedure yields the following results:
\begin{align*}
    \mathcal{A}^T &\equiv  \frac{\mathcal{G}_T a^3}{8 N}, \\
    \theta^T &\equiv t \cdot \frac{\dot{\mathcal{A}}^T}{\mathcal{A}^T}, \\
    \mathcal{B}^T &\equiv \frac{N^2}{a^2}\;, \\
    \mathcal{A}^{T}_g &\equiv   \mathscr{A}^{T}_g \cdot (-t)^{\theta_{g}^{T}},
\end{align*}
where 
\begin{align*}
    \mathcal{A}_{g}^{T} &\equiv \frac{(-ct)^{-2\mu}}{8} \cdot a_{g}^{3}\;, \\
    \theta_{g}^{T} & \equiv t \cdot \frac{\dot{\mathcal{A}}_{g}^{s}}{\mathcal{A}_{g}^{s}} = -2\mu < 0\;, \\
    \mathcal{B}_{g}^{T} & = \frac{1}{a_{g}^{2}} > 0\;.
\end{align*}
Using the expressions above, the tensor power spectrum can be written as
\begin{align*}
    n_T &= 2 + \theta_{g}^{T} = 2 - 2\mu\;, \\
    A_T &= \frac{2\Gamma^{2}(\nu_T) (\mathcal{B}_{g}^{T})^{-\nu_T} k_{*}^{3 - 2\nu_T}}{2^{4 - 2\nu_T} \pi^3 \mathscr{A}_{g}^{T}}\;,
\end{align*}
where 
\begin{align*}
    \nu_T \equiv \frac{1 - \theta_{g}^{T}}{2} = \mu + \frac{1}{2}\;.
\end{align*}
We note that we have a blue spectrum for tensor perturbations when strong coupling is avoided, i.e., when the condition \eqref{healthy_region} is satisfied. This concludes the discussion on the tensor power spectrum.


\section{Unitarity bounds for the non-power-law backgrounds}
\label{sec: Unitarity Bounds for the late times}

The condition for the legitimacy of the classical description at early times is given by \eqref{earlyUnitarityBound}. The allowed range of parameters from Fig.~\ref{fig:Mu_del}, where we see that $\theta^{s}_{g}$ must be greater than minus two. Thus, the requirement for the absence of strong coupling leads to a blue-tilted scalar spectrum if the mode freezes at times when $\theta^{s} \approx \theta^{s}_{g}$. 

Conversely, from Fig.~\ref{fig:ThetaS}, we immediately note that $\theta^{s}$ can fall below $-2$. This behavior initially may suggests that the spectrum should be red-tilted (we will check this claim explicitly in the next section). However, a value of $\theta < -2$ may violate unitarity bounds, potentially invalidating the classical field description. Thus, we see that at least for a power-law solution, there is a direct connection between the value of $\theta^s$ and the legitimacy of the classical description. 

It then raises the question of whether this connection holds in cases where the solution is in a non-power-law regime. In this section, we will explicitly show that there is no connection between the values of $\theta^s$ and the absence of strong coupling. Moreover, we will demonstrate that one could indeed consider highly non-power-law backgrounds (i.e., backgrounds where $\theta^s < -2$) while remaining on safe ground—meaning that the background solution can still be described by classical field theory. This important observation suggests that our setup could remain valid even in quite extreme cases, when the solution exhibits non-power-law behavior during the Genesis stage! 


Now, let us shift our focus to the analysis of the strong-coupling condition. As time approaches minus infinity, the strictest constraints arise from the cubic scalar Lagrangian for perturbations. Therefore, we will restrict our analysis to the cubic scalar sector and proceed with a rough approximation. We assume that the strongest constraints at late times derive from the same terms that provide the most stringent unitarity constraints asymptotically as time approaches minus infinity. Additionally, we disregard numerical coefficients, the tensor structure of interaction terms, and any possible cancellations among the terms in the interaction Lagrangian. This last approximation can only make the conditions for the absence of strong coupling more stringent; thus, it is reasonably safe to proceed with this type of approximation.

The fact that we are disregarding numerical coefficients in the strong coupling scale $E_{\text{strong}}$ is also justifiable, since at the considered times, the ratio $E_{\text{strong}}/E_{\text{classical}}$ is enhanced by the power of time $(-t)^{\delta}$, making this enhancement far greater than any numerical coefficients in $E_{\text{strong}}$. Here, $E_{\text{classical}}$ is the classical energy scale, which is given by:
\begin{equation*}
    E_{\text{classical}} = \max\left\{ H, \frac{\dot{H}}{H}, \frac{\dot{\phi}}{\phi} \right\} \propto \frac{1}{|t|}\;.
\end{equation*}

On the other hand, $E_{\text{strong}}$, the strong coupling energy scale, can be found by imposing the condition of unitarity of the $S$ matrix and checking the optical theorem at tree level~\cite{deRham:2017aoj, Ageeva:2022nbw, Ageeva:2022asq, Cai:2022ori, Grojean:2007zz}. 

It remains an open question what impact loops will have during this analysis in Horndeski gravity; we leave this broad and interesting topic for future work. For a discussion of loop corrections to cosmological solutions, see Refs.~\cite{Leblond:2008gg, Senatore:2009cf}. 

Another way to estimate $E_{\text{strong}}$ is through dimensional analysis~\cite{Ageeva:2020gti, Ageeva:2020buc}. Although dimensional analysis is generally less accurate than the explicit conditions from the optical theorem, it has some advantages. For instance, dimensional analysis allows us to analyze unitarity at tree level for all orders of perturbation theory~\cite{Ageeva:2020buc}. However, a disadvantage of this approach is that dimensional analysis does not account for numerical coefficients, tensor structures, or possible cancellations between terms. Sometimes, these cancellations can be significant, as noted in Ref.~\cite{Ageeva:2022fyq}. 

In summary, if we do not take into account loop corrections (i.e., considering the optical theorem only at tree level), dimensional analysis will yield an actual unitarity bound (accurate up to a numerical factor) or a more stringent constraint on model parameters, again accurate up to a numerical factor.


If the classical energy scale $E_{\text{classical}}$ is below $E_{\text{strong}}$, the theory is valid as an effective field theory (EFT) and can be described by classical field theory and weakly coupled quantum field theory.   

At early times $(t \to -\infty)$, the strongest constraints on model parameters arise from the ($\Lambda_1,\;\Lambda_3,\;\Lambda_{7},$ $\Lambda_{10},\;\Lambda_{14},\;\Lambda_{16}\;$) terms in the cubic action for perturbations (see Refs.~\cite{Ageeva:2020gti, Ageeva:2020buc, Ageeva:2021yik}). The interaction terms given below produce the constraint $\mu + \frac{3\delta}{2} < 1$ on model parameters (see Ref.~\cite{Ageeva:2020gti}).   


\begin{align*}
    S^{(3)}_{0} \equiv \int N \, dt \, a^3 \, d^{3}x \Big[ &\Lambda_1 \frac{\dot{\zeta}^3}{N^3} 
        + \Lambda_3 \frac{\dot{\zeta}^2}{N^2 a^2} \partial^2 \zeta  
        + \Lambda_7 \frac{\dot{\zeta}}{N a^4} \left(\partial^2 \zeta\right)^2 + \Lambda_{10} \frac{\dot{\zeta}}{N a^4} \left(\partial_i \partial_j \zeta \right)^2 \\ 
        +  &\Lambda_{14} \frac{\dot{\zeta}}{N} \left(\partial_i \partial_j \psi \right)^2 
        + \Lambda_{16} \frac{\dot{\zeta}}{N a^2} \partial_i \partial_j \zeta \partial_i \partial_j \psi \Big]\;,
\end{align*}
where 
\begin{equation*}
    \psi \equiv \frac{1}{N} \partial^{-2} \dot{\zeta}\;.
\end{equation*}

The coefficients in the cubic action are:
\begin{align*}
\Lambda_1 &= -\frac{\mathcal{G}_T^3}{3\Theta^3}(\Sigma - N \Sigma_N + H \Xi) 
        + \frac{\mathcal{G}_T^2 \Xi}{\Theta^2} 
        - \frac{\mathcal{G}_T \mathcal{G}_S \Xi}{3\Theta^2} 
        + \frac{\Gamma \mathcal{G}_S^2}{2\Theta \mathcal{G}_T}  - \frac{2\Gamma \mathcal{G}_S}{\Theta} + \frac{3\Gamma \mathcal{G}_T}{\Theta}\;, \\
\Lambda_3 &= \frac{\mathcal{G}_T^3 \Xi}{3\Theta^3} 
        - \frac{\mathcal{G}_T \mathcal{G}_S \Gamma}{\Theta^2} + \frac{2\Gamma \mathcal{G}_T^2}{\Theta^2}\;,\\
\Lambda_7 &= \frac{\Gamma \mathcal{G}_T^3}{2\Theta^3}\;,\\
\Lambda_{10} &= -\frac{\Gamma \mathcal{G}_T^3}{2 \Theta^3}\;,\\
\Lambda_{14} &= -\frac{\Gamma \mathcal{G}_S^2}{2\Theta \mathcal{G}_T}\;,\\
\Lambda_{16} &= \frac{\mathcal{G}_T \mathcal{G}_S \Gamma}{\Theta^2}\;,
\end{align*}
where $\mathcal{G}_T$, $\mathcal{F}_T$, $\mathcal{G}_S$, $\mathcal{F}_S$, $\Theta$, and $\Sigma$ are given by equations~\eqref{FS_GS} -- \eqref{Sigma_Theta}, respectively, while $\Gamma$ and $\Xi$ are defined as
\begin{align*}
    \Xi &= -12HB_4, \;\; \Gamma = 2B_4\;.
\end{align*}
The expressions above are valid only when $A_3 \equiv 0$ and $A_4 \equiv A_4(t)$.


The full constrained cubic Lagrangian for all sectors can be found in Refs.~\cite{Gao:2011qe, DeFelice:2011uc, Gao:2012ib}, while the unconstrained cubic Lagrangian is provided in Ref.~\cite{Ageeva:2020gti, Gao:2012ib}. The action $S^{(3)}_{0}$ leads to the lowest strong-coupling energy scale $E_{\text{strong}}^{0}$. The sub-leading contributions to the strong coupling scale arise from the following interaction terms: ($\Lambda_2,\;\Lambda_4,$ $\Lambda_5,\;\Lambda_8,$
$\Lambda_9,\;\Lambda_{11},$
$\Lambda_{12},\;\Lambda_{13},$
$\Lambda_{15},\;\Lambda_{17}$). 

The terms below yield the constraint $\mu + \frac{\delta}{2} < 1$ on model parameters (see Ref.~\cite{Ageeva:2020gti}).
\begin{align*}
    S^{(3)}_{1} \equiv \int N dt a^3 d^{3}x \Big[&\Lambda_2 (\dot{\zeta}^2/N^2)\zeta
    + \big(a^{-2}\big)\Lambda_4 (\dot{\zeta}/N)\zeta \partial^2 \zeta
    +\big(a^{-2}\big) \Lambda_5 (\dot{\zeta}/N) \left(\partial_i \zeta \right)^2 \\ \nonumber
    &+ \big(a^{-4}\big) \Lambda_8\zeta \left(\partial^2 \zeta \right)^2
    + \big(a^{-4}\big) \Lambda_9 \partial^2 \zeta \left(\partial_i \zeta \right)^2 
    + \big(a^{-4}\big) \Lambda_{11} \zeta \big( \partial_i \partial_j \zeta \big)^2  \\&+ 
     \Lambda_{12}(\dot{\zeta}/N) \partial_i \zeta \partial^i \psi  + \big(a^{-2}\big)\Lambda_{13} \partial^2 \zeta \partial_i \zeta \partial^i \psi  +
     \Lambda_{15} \zeta \big( \partial_i \partial_j \psi \big)^2 
     \\
     &+ \big(a^{-2}\big)\Lambda_{17} \zeta \partial_i \partial_j \zeta \partial^i \partial^j \psi + d_1 \zeta \big(\dot{h}_{ij}/N\big)^2 
     + d_2\zeta h_{ij,k}h_{ij,k} 
     &
     \\&+ d_3 \psi_{,k}\big(\dot{h}_{ij}/N\big)h_{ij,k}\Bigg]\;,
\end{align*}    
where 
\begin{align*}
    \Lambda_2 &= \frac{3\mathcal{ G}_T^2 \Sigma}{\Theta^2} 
        + 9\mathcal{ G}_T - \frac{3\mathcal{ G}_S^2}{2\mathcal{ G}_T} \;,\\
    \Lambda_4 &= \frac{3\mathcal{ G}_T \mathcal{ G}_S}{\Theta} 
        - \frac{2\mathcal{ G}_T^2}{\Theta}\;,\\
    \Lambda_5 &= -\frac{\mathcal{ G}_T^2}{\Theta} + \frac{2\mathcal{ G}_T \mathcal{ G}_S}{\Theta}\;,\\
    \Lambda_8 &= -\frac{3\mathcal{ G}_T^3}{2\Theta^2}\;,\\
    \Lambda_9 &= -\frac{2\mathcal{ G}_T^3}{\Theta^2}\;,\\
    \Lambda_{11} &= \frac{3\mathcal{ G}_T^3}{2 \Theta^2}\;,\\
    \Lambda_{12} &= -\frac{2\mathcal{ G}_S^2}{\mathcal{ G}_T}\;,\\
    \Lambda_{13} &= \frac{2 \mathcal{ G}_T\mathcal{ G}_S}{ \Theta}\;,\\
    \Lambda_{15} &= \frac{3\mathcal{ G}_S^2}{2\mathcal{ G}_T}\;,\\
    \Lambda_{17} &= -\frac{3\mathcal{ G}_T\mathcal{ G}_S}{\Theta}\;,\\
    d_1 &= \frac{3\mathcal{ G}_T}{8}\left[1-\frac{H\mathcal{ G}_T^2}{\Theta\mathcal{ F}_T}
        +\frac{\mathcal{ G}_T}{3}\frac{d}{Nd t}\left(\frac{\mathcal{ G}_T}{\Theta\mathcal{ F}_T}\right)
        \right]\;,\\
    d_2 &= \frac{\mathcal{F}_{S}}{8}\;,\\
    d_3 &= -\frac{\mathcal{G}_S}{4}\;.
\end{align*}

Other terms from the interaction Lagrangian impose weaker constraints on the model parameters (see Refs.~\cite{Ageeva:2020gti, Ageeva:2020buc}). Here, we will estimate the first sub-leading correction to the strong coupling scale; thus, it is sufficient to consider only $S^{(3)}_{0}$ and $S^{(3)}_{1}$.

Now we are ready to analyze the validity of the classical description. First of all, we must estimate the matrix elements. 
During the Genesis stage, the scale factor is slowly varying and can be treated as nearly constant. As a result, the space-time is close to Minkowski spacetime, 
allowing us to straightforwardly define the in- and out-states and calculate the $2 \to 2$ matrix element. 
In this context, below we will work in the regime where $\frac{\dot{N}(t)}{N(t)}$ and $\frac{\dot{a}(t)}{a(t)}$ are small compared to the characteristic energy of the scattering particles -- $E_{\text{scatter}}$. Specifically, we assume that $E_{\text{strong}} \gg E_{\text{scatter}} \gg E_{\text{classical}} > H$. In this regime, the Hubble friction term in the equations of motion for perturbations is negligible compared to the gradient term. Moreover, we can treat the sound speed of perturbations and coefficients in the quadratic Lagrangian as approximately constant during the scattering process. Indeed, we note that the following statements can be easily proved using both the equations of motion~\eqref{solBackEoMInU} and expressions for coefficients in the quadratic actions~\eqref{FS_GS}:
\begin{align*}
    \mathcal{G}_{S}^{2} &\propto \frac{(-t)^{-2\mu}}{u} \cdot (const_{1} + O(u))\;, \\
    \mathcal{F}_{S}^{2} &\propto \frac{(-t)^{-2\mu}}{u} \cdot (const_{1} + O(u))\;, \\
    u_s^{2} &\propto const_{3} + O(u)\;, \\
    N &\propto 1 + O(u)\;, \\
    a &\propto a_g (1 + O(u))\;,
\end{align*}
where
\begin{equation*}
    u = (-c t)^{-\delta} \gg (-ct)^{-1}\;.
\end{equation*}
Thus, we see that  $\dot{\mathcal{F}_{S}}/\mathcal{F}_{S}$, $\dot{\mathcal{G}_{S}}/\mathcal{G}_{S}$, $\dot{N}/N$, and $\dot{a}/a$ are of the order of $E_{class} \propto (-t)^{-1}$, which is much smaller than the scattering energy $E_{scatter}$ under our assumptions. 

Next, in order to simplify the calculations, we proceed with the following variable redefinitions:
\begin{equation*}
    d\tilde{x} = a \cdot dx, \;\; d\tilde{t} = N \cdot dt~,
\end{equation*}
where $a$ and $N$ are treated as constants during the scattering process.
We also make the canonical normalization of the field $\zeta$ as follows:
\begin{equation*}
    \zeta_c \propto \sqrt{\mathcal{G}_S} \zeta\;,
\end{equation*}
where the field $\zeta_c$ has the following dispersion relation:
\begin{equation*}
    w^2 = u_s^{2} |\vec{k}|^2\;.
\end{equation*}


Afterward, we rewrite the cubic action components $(\Lambda_1, \; \Lambda_3, \; \Lambda_{7}, \; \Lambda_{10}, \; \Lambda_{14}, \; \Lambda_{16})$ for scalar perturbations as follows:
\begin{align}
\label{S^3_0_rewrite}
    S^{(3)}_{0} = \int d \tilde{t} \, d^{3} \tilde{x} \Big[ & \Lambda_1 \frac{\zeta_{c}^{\prime 3}}{\mathcal{G}_{S}^{3/2}} 
        + \Lambda_3 \frac{\zeta_{c}^{\prime 2}}{\mathcal{G}_{S}^{3/2}} \tilde{\partial}^2 \zeta_c  
        + \Lambda_7 \frac{\zeta_{c}^{\prime}}{\mathcal{G}_{S}^{3/2}} \left(\tilde{\partial}^2 \zeta_{c}\right)^2 + \Lambda_{10} \frac{\zeta_{c}^{\prime}}{\mathcal{G}_{S}^{3/2}} \left(\tilde{\partial}_i \tilde{\partial}_j \zeta_{c}\right)^2 \\ \nonumber 
        +  &\Lambda_{14} \frac{\zeta_{c}^{\prime}}{\mathcal{G}_{S}^{3/2}} \left(\tilde{\partial}_i \tilde{\partial}_j \tilde{\psi}\right)^2 
        + \Lambda_{16} \frac{\zeta_{c}^{\prime}}{\mathcal{G}_{S}^{3/2}} \tilde{\partial}_i \tilde{\partial}_j \zeta_{c} \tilde{\partial}_i \tilde{\partial}_j \tilde{\psi} \Big]\;,
\end{align}
where 
\begin{align*}
   \tilde{\partial}_i &\equiv  \frac{\partial}{\partial \tilde{x}^{i}}\;,\\
   \zeta_{c}^{\prime} &\equiv \partial_{\tilde{t}} \zeta_{c}\;,\\
   \tilde{\psi} &\equiv \tilde{\partial}^{-2} \zeta_{c}^{\prime}.
\end{align*}
Similarly, we also rewrite the expression for $S^{(3)}_{1}$. We do not quote this expression here, as it is the exact analogue of the expression in \eqref{S^3_0_rewrite}.

Now, we schematically write the matrix element as

\begin{equation*}
    |M| \propto \frac{|V (\mathcal{E})|^2}{\mathcal{E}^2}\;,
\end{equation*}
where $V(\mathcal{E})$ is the vertex corresponding to the interaction terms ($\Lambda_1,\;\Lambda_3,$ $\Lambda_{7},$ $\Lambda_{10},\;\Lambda_{14},$ $\Lambda_{16}\;$), ($\Lambda_2,\;\Lambda_4,$ $\Lambda_5,\;\Lambda_8,$
$\Lambda_9,\;\Lambda_{11},$
$\Lambda_{12},\;\Lambda_{13},$
$\Lambda_{15},\;\Lambda_{17}$). The quantity $\mathcal{E}$ is defined as 
\begin{equation*}
    \mathcal{E} \equiv \frac{E}{N}\;.
\end{equation*}

The unitarity bound is saturated when the absolute value of the tree matrix element is roughly equal to unity. To obtain the exact unitarity bound (at tree level), one needs to calculate the $s$, $u$, and $t$ channels for the tree-level $2 \to 2$ matrix element, then proceed to the partial wave amplitudes (PWAs) and use the optical theorem. This procedure can be found in Refs.~\cite{Grojean:2007zz, deRham:2017aoj, Ageeva:2022nbw, Ageeva:2022byg, Cai:2022ori}. 

Here, we are not interested in the exact numerical coefficients in $E_{\text{strong}}$; therefore, we do not distinguish between $s$, $t$, and $u$ channels. Thus, the strong energy scale can be obtained as follows. The PWAs for this particular case of non-unity sound speed for perturbations are given by
\begin{equation*}
    a_l \propto \int d(\cos x) P_l(\cos x) M,
\end{equation*}
where all numerical coefficients are omitted.


After that, we can write for $ l = 0 $:
\begin{equation*}
    a_0 \sim M.
\end{equation*}
Thus, the corresponding strong coupling energy scale can be estimated from the unitarity bound (which is a direct consequence of the optical theorem for PWA):
\begin{align*}
    \Big| \text{Re}\big[a_0\big] \Big| = \frac{1}{2}.
\end{align*}
Omitting all numerical factors, we arrive at
\begin{equation*}
    |M(\mathcal{E}_{\text{strong}})| \propto 1.
\end{equation*}

However, if scalar particles (with non-unity sound speed $u_s < 1$) are in the initial and final states, we obtain an enhancement factor of $u_s^3$ (see, for example, Ref.~\cite{Ageeva:2022nbw}). This factor strengthens the unitarity bounds, at least in the scalar sector. Thus, regardless of the initial and/or final states, we will consider the following condition for the absence of strong coupling:
\begin{equation*}
    |M(\mathcal{E}_{\text{strong}})| \propto u_s^3 < 1.
\end{equation*}
This condition can only reinforce the requirement for the absence of a strong coupling regime.

Now, let us assume that there exists $M_R$ such that the inequality
\begin{align*}
    |M(\mathcal{E})| \leq |M_R(\mathcal{E})|
\end{align*}
holds for every energy. We can then obtain two energy scales from $M$ and $M_R$, respectively:
\begin{align*}
    |M_{R}(\mathcal{E}_{\text{strong}}^{R})| &= u_s^3\;, \\
    |M(\mathcal{E}_{\text{strong}})| &= u_s^3\;.
\end{align*}
The fact that $ |M| < |M_{R}| $ for every value of $\mathcal{E}$ implies that $\mathcal{E}_{\text{strong}}^{R} \leq  \mathcal{E}_{\text{strong}}$. Therefore, the condition $ |M_{R}(\mathcal{E}_{\text{strong}}^{R})| = u_s^3 $ provides the actual unitarity constraints, or a more stringent one.Thus, the condition $\mathcal{E}_{\text{classical}} \ll \mathcal{E}_{\text{particle}} \ll \mathcal{E}_{\text{strong}}^{R}$ is sufficient for unitarity. We assume that, for the estimations above, all energy scales, including $\mathcal{E}_{\text{strong}}^{R}$, still tend to zero as time approaches minus infinity.

Now, we estimate $M_R$ as follows:
\begin{align*}
    |M| \leq \frac{1}{\mathcal{G}_{S}^{3} \cdot \mathcal{E}^2}\cdot \Big(& |\Lambda_1|\cdot \mathcal{E}^3 +  
|\Lambda_3|\cdot \mathcal{E}^4 u_{s}^{-2} + |\Lambda_7|\cdot \mathcal{E}^5 u_{s}^{-4}\\
&+|\Lambda_{10}|\cdot \mathcal{E}^5 u_{s}^{-4} + |\Lambda_{14}|\cdot \mathcal{E}^3 
 + |\Lambda_{16}|\cdot \mathcal{E}^4 u_{s}^{-2} 
 + |\Lambda_2|\cdot\mathcal{E}^2
 \\&+ |\Lambda_4|\cdot\mathcal{E}^3 u_{s}^{-2} 
 + |\Lambda_5|\cdot\mathcal{E}^3 u_{s}^{-2}
 +|\Lambda_8|\cdot\mathcal{E}^4 u_{s}^{-4}
 +|\Lambda_9|\cdot\mathcal{E}^4 u_{s}^{-2}
 +|\Lambda_{11}|\cdot\mathcal{E}^4 u_{s}^{-2}\\
 &+|\Lambda_{12}|\cdot\mathcal{E}^4 u_{s}^{-2}
 +|\Lambda_{13}|\cdot\mathcal{E}^3 u_{s}^{-2}
 +|\Lambda_{15}|\cdot\mathcal{E}^2 
 +|\Lambda_{17}|\cdot\mathcal{E}^3 u_{s}^{-2}
 +\frac{\mathcal{G}_S}{\mathcal{G}_T}|d_1|\mathcal{E}^2\\
 &+\frac{\mathcal{G}_S}{\mathcal{G}_T}|d_2|\mathcal{E}^2
 +\frac{\mathcal{G}_S}{\mathcal{G}_T}|d_3|\mathcal{E}^2 u_{s}^{1}
 \Big)^2 \equiv M_R\;.
\end{align*}

\begin{figure}[!ht]
\centering 
\includegraphics[width=8cm]{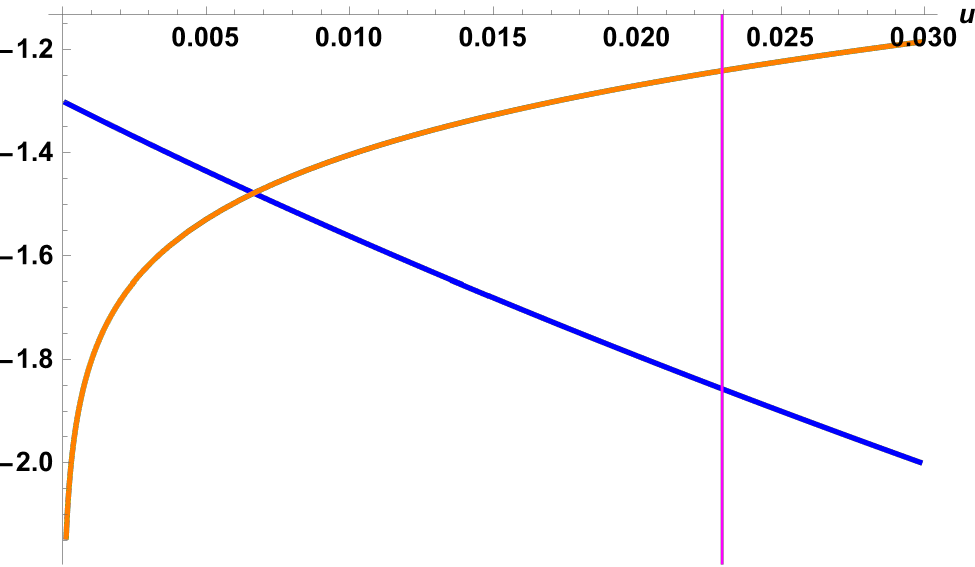}
 \caption{In this figure, we consider the parameter set \eqref{var4ParamSet}. The blue line represents the value of $\theta^s (u)$. The orange line depicts $-\log_{10} \left(\frac{\mathcal{E}_{\text{strong}}}{\mathcal{E}_{\text{classical}}}\right)$. The magenta line corresponds to $u(t^{fr})$.}
 \label{fig:EStrongVar4}
\end{figure}

In Fig.~\ref{fig:EStrongVar4}, the blue line represents the value of $\theta^{s}(u)$. We have also plotted the logarithm of the ratio $\frac{\mathcal{E}_{\text{strong}}}{\mathcal{E}_{\text{classical}}}$, denoted as $-\log_{10}\left(\frac{\mathcal{E}_{\text{strong}}}{\mathcal{E}_{\text{classical}}}\right)$ (orange line). At first glance, one might deduce that the value of $\theta^{s} \approx -2$ creates a small tension with the absence of the strong coupling condition. However, this is not the case. 

\begin{figure}[!ht]
\centering 
\includegraphics[width=8cm]{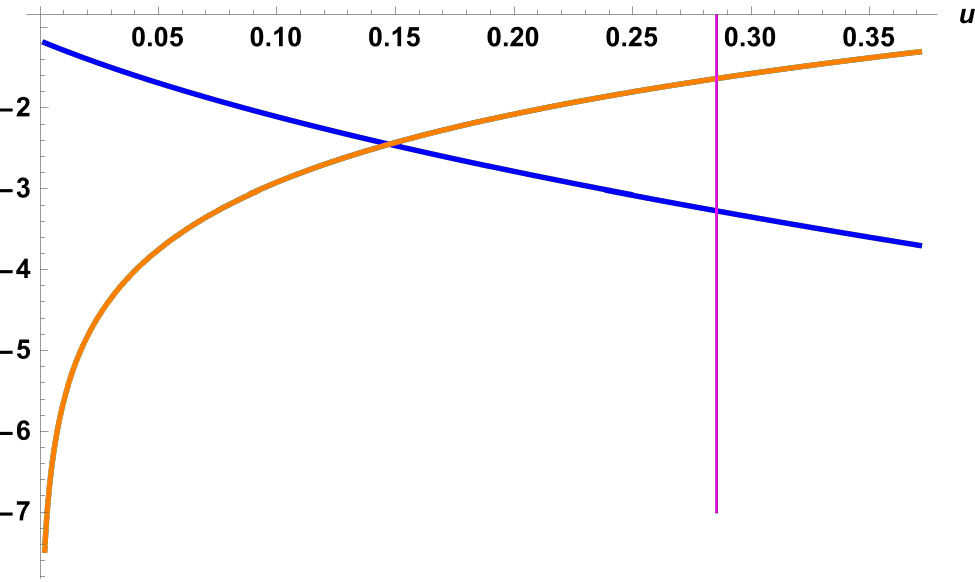}
 \caption{Here, we consider the parameter set \eqref{var9ParamSet}. The blue line represents $\theta^s (u)$, the orange line indicates $-\log_{10}\left(\frac{\mathcal{E}_{\text{strong}}}{\mathcal{E}_{\text{classical}}}\right)$, and the magenta line corresponds to $u(t^{fr})$.}
 \label{fig:EStrongVar9}
\end{figure}

To illustrate our point, let us choose a different set of model parameters:
\begin{align}
    \mu &= \frac{3}{5}, \;\delta = \frac{1}{30}, \; c = 10^{-4}, \; g = \frac{1}{19} \cdot 10^{-5}, \; s = 10^{-4}.
    \label{var9ParamSet}
\end{align}
This parameter set is fully consistent with the requirements in \eqref{boundsFor_a}. Additionally, this choice of parameters provides a stable cosmological background solution, as confirmed by numerical simulation. In Fig.~\ref{fig:EStrongVar9}, we show the logarithm of the ratio $\frac{\mathcal{E}_{\text{strong}}}{\mathcal{E}_{\text{classical}}}$ alongside the value of $\theta^s$.

One can see that the ratio $\frac{\mathcal{E}_{\text{strong}}}{\mathcal{E}_{\text{classical}}}$ is much greater than unity for every reasonable value of $u$. This indicates that for some ranges of parameters, the corrections due to the $u$ variable do not drastically alter the strong-coupling absence constraints, and the theory remains valid as an effective field theory even at late times. 

Moreover, when $\theta^{s}$ approaches the value of $-2$, the ratio $\frac{\mathcal{E}_{\text{strong}}}{\mathcal{E}_{\text{classical}}}$ is roughly equal to $10^3$, which is more than sufficient to establish that our theory is valid as an EFT, meaning that the evolution could be described by classical field theory and weakly coupled quantum field theory. Therefore, when the background solution has non-power-law behavior during the Genesis stage, there is no direct connection between the value of the friction term $\theta^s$ and the absence of strong coupling.

In addition, we would like to note that the expressions for both $\theta^S$ and the ratio $\frac{\mathcal{E}_{\text{strong}}}{\mathcal{E}_{\text{classical}}}$ are evaluated numerically, and we do not use Taylor series or any other approximations or simplifications.

Furthermore, we would like to highlight that when $\mathcal{E}_{\text{scatter}}$ is close to the energy $\mathcal{E}_*$ of the mode with characteristic momentum roughly equal to $k_*$, all the previous analysis becomes invalid in the vicinity of the freeze point. The reason is that near the freeze point, the expressions for the matrix elements at energies of order $\mathcal{E}_*$ are not valid because it is impossible to define in-states and out-states in a conventional way. Thus, the applicability of the unitarity constraints during the entire mode evolution remains an open question, particularly in cases where the mode approaches the horizon. We leave this compelling issue for future work.

To justify the method discussed above, let us add one more comment. The calculation of $2 \to 2$ scattering is valid only if the coefficients in the vertices $\Lambda_i$ change slowly in comparison with the characteristic timescale $t_{\text{scatter}}$ of scattering. This timescale can be estimated as 
$t_{\text{scatter}} \propto \frac{1}{E_{\text{scatter}}}$. Therefore, we require the following condition:
\begin{align*}
    \Big| \frac{\dot{\Lambda}}{\Lambda E_{\text{scatter}}} \Big| \ll 1.
\end{align*}

Let us check whether this condition holds numerically. To this end, we choose the parameter set in \eqref{var4ParamSet} and plot $\Big| \frac{\dot{\Lambda}}{\Lambda E_{\text{class}}} \Big|$. For this demonstration, we focus on $\Lambda_7$; however, the other $\Lambda_i$ exhibit very similar behavior. We choose $\Lambda_7$ because it imposes the most stringent constraints on the absence of strong coupling, while the other $\Lambda_i$ provide weaker or similar constraints.

\begin{figure}[!ht]
\centering 
\includegraphics[width=8cm]{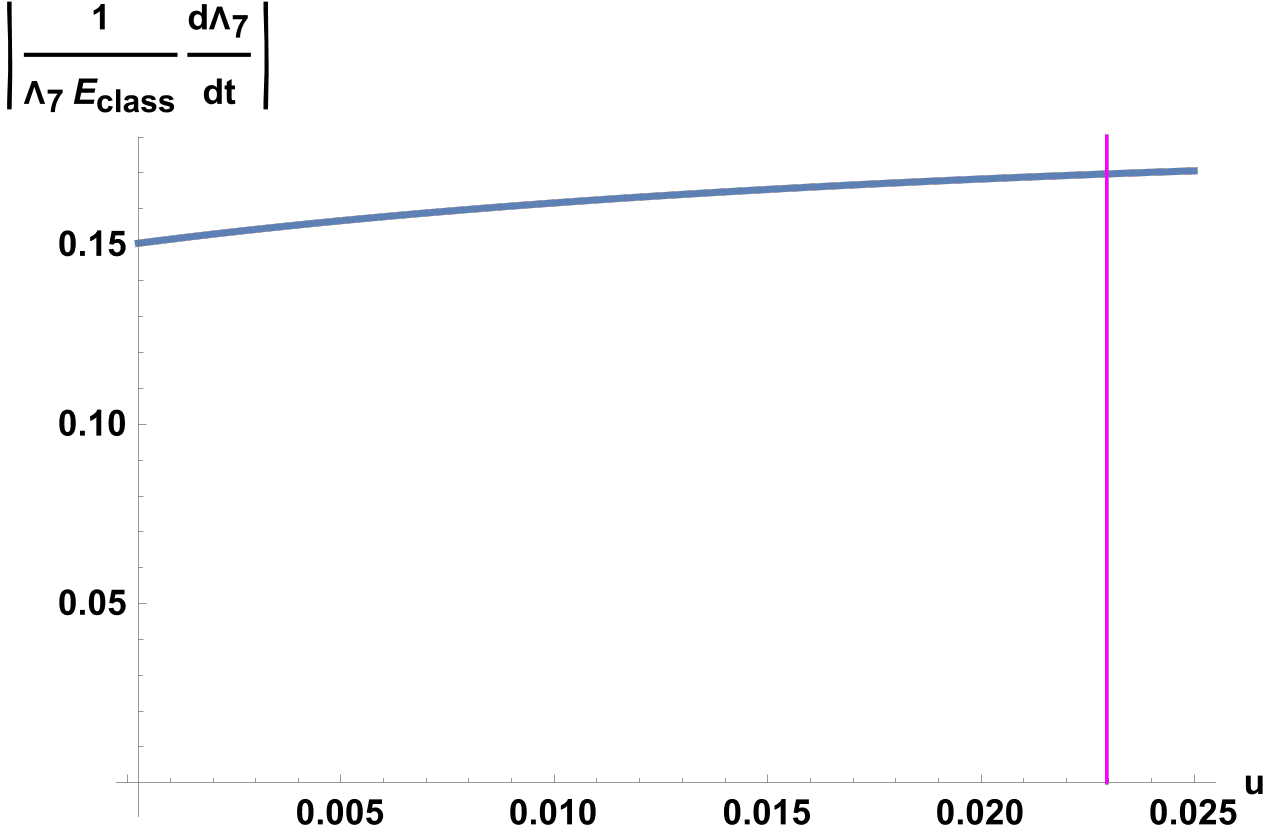}
 \caption{The blue line represents $\Big| \frac{\dot{\Lambda}_7}{\Lambda_7 E_{\text{class}}} \Big|$, while the magenta line shows $u(t^{fr})$. All plots are obtained with the parameter set~\eqref{var4ParamSet}.}
 \label{fig:DLog7LamOverEClass}
\end{figure}

From Fig.~\ref{fig:DLog7LamOverEClass}, we observe that $\Big| \frac{\dot{\Lambda}_7}{\Lambda_7} \Big|$ is marginally less than $E_{\text{class}}$. Given that $E_{\text{scatter}} \gg E_{\text{class}}$, it follows that $\Big| \frac{\dot{\Lambda}_7}{\Lambda_7} \Big| \ll E_{\text{scatter}}$. This observation fully justifies the calculation of the scattering matrix element and the subsequent analysis of the unitarity bounds.


\section{Numerical evaluation of scalar power spectral index}
\label{sec:TimeDependTermsScalarSpectrum}

The purpose of this section is to obtain the scalar spectral index in the case when Genesis stage may exhibits non-power-law behavior. To this end, we will use the numerical methods. The mode equation during the Genesis stage is described by Eq.~(\ref{zetaModeEq}). The logarithm of the amplitude, $\chi = \ln |\zeta|$, and the phase $\varphi$ of $\zeta$ follow:
\begin{equation}\label{chievol}
    \ddot{\chi} + \dot{\chi}^2 - \dot{\varphi}^2 + \frac{\theta^s}{t} \dot{\chi} + k^2 \mathcal{B}^s = 0 ~,
\end{equation}
\begin{equation} \label{phevol}
    \ddot{\varphi} + \left(2 \dot{\chi} + \frac{\theta^s}{t}\right) \dot{\varphi} = 0.
\end{equation}
The solution of Eq. \eqref{phevol} is given by
\begin{equation} \label{phsol}
    \dot{\varphi} = \dot{\varphi}_0 \exp[-2(\chi - \chi_0) - (\ln \mathcal{A}^s - \ln \mathcal{A}^s_0)] ~,
\end{equation}
where the subscript $0$ denotes the evaluation of the function at $t_0$, the initial time of the numerical mode evolution. We obtain numerical solutions for $\chi$ by solving Eq. \eqref{chievol} with $\dot{\varphi}$ replaced by the solution in Eq. \eqref{phsol}.

The initial conditions $\dot{\varphi}_0$, $\chi_0$, and $\dot{\chi}_0$ are estimated using the Eikonal approximation (see Appendix B), which accurately describes mode evolution before the freeze-out time $t_{fr}$. They are given by:
\begin{equation} \label{ini}
    \begin{split}
        &\dot{\varphi}_0 \simeq -k \sqrt{\mathcal{B}^s_0}\\
        &\chi_0 = 0\\
        &\dot{\chi}_0 = -\frac{1}{2}\frac{\theta^s_0}{t_0} - \frac{1}{4}\frac{\dot{\mathcal{B}}^s_0}{\mathcal{B}^s_0}
    \end{split}~.
\end{equation}
Note that we used the freedom in the value of $\chi_0$ to set $\chi_0=0$, which is equivalent to $|\zeta_0|=1$. After solving the equations of motion, $\zeta$ is rescaled to
\begin{equation*}
    |\zeta_0| \propto (k \mathcal{A}^s_0)^{-1/2} \mathcal{B}_0^{-1/4},
\end{equation*}
which matches with the Eikonal approximation and the normalization condition.

The initial time of mode evolution should be set far behind the freeze-out time $t_{fr}$ to ensure the accuracy of the initial conditions given in Eq. (\ref{ini}). To determine a reasonable initial time point $t_0$, we estimate the phase evolution before the freeze-out time by
\begin{equation*}
    \Delta \varphi \simeq \int_{t_0}^{t_{fr}} dt \, k \sqrt{\mathcal{B}^s}~.
\end{equation*}
We choose $t_0$ such that $\Delta \varphi$ is sufficiently large. We found that $\Delta \varphi = 20\pi$ is enough to achieve accurate mode evolution.

\begin{figure}[!ht]
    \centering
    \includegraphics[width=0.5\linewidth]{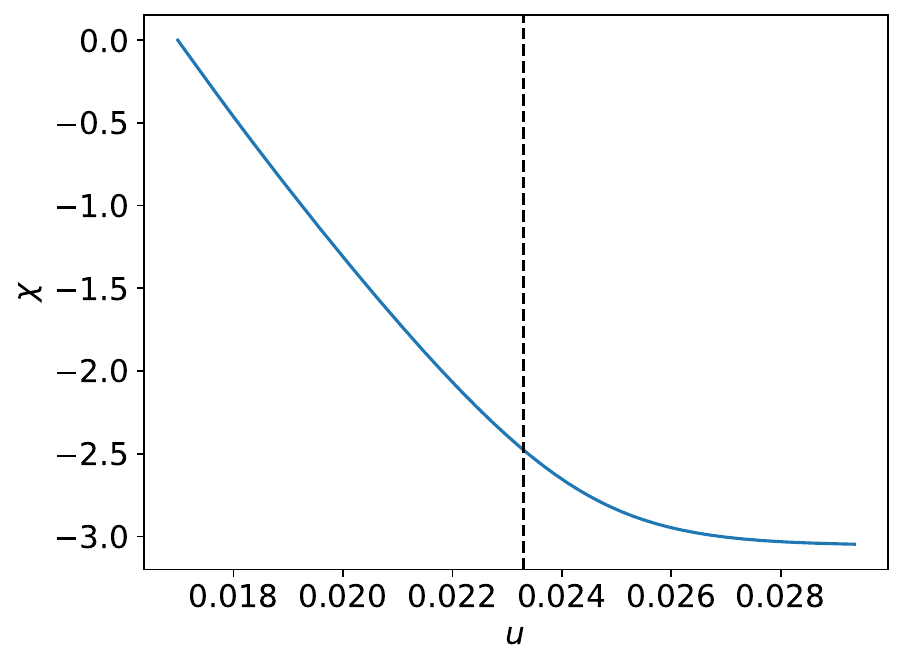}
    \caption{The numerical evolution example of the log-amplitude $\chi$. The horizontal axis is re-parameterized as $u = (-c t)^{-\delta}$, where increasing $u$ represents the direction of time evolution. The vertical dashed line denotes the freeze-out $u$ value, $u_{fr} = (-c t_{fr})^{-\delta}$. The model parameters are the same as in Eq. (\ref{var4ParamSet}).}
    \label{fig:chi}
\end{figure}

The variable $\chi$ decreases rapidly until it reaches $t_{fr}$, at which point it freezes to a constant value. An example of the numerical solution is shown in Fig. \ref{fig:chi}. We terminate the numerical evolution at $t_e = \frac{t_{fr}}{10}$ and take the corresponding $\chi$ value. The freeze-out value of $\zeta_{fr}$ can be obtained by
\begin{equation}
    \ln|\zeta|_{fr}(k) \simeq \chi(t_e) - \frac{1}{2}\ln k - \frac{1}{2}\ln \mathcal{A}^s_0 - \frac{1}{4}\ln \mathcal{B}^s_0 + \text{const.}~,
\end{equation}
where $\text{const.}$ is a $k$-independent constant. Note that $\mathcal{A}^s_0$ and $\mathcal{B}^s_0$ are $k$-dependent due to their dependence on $t_0$. To obtain the spectral tilt, we compute the finite difference
\begin{equation*}
    \frac{d\ln |\zeta|_{fr}(k)}{d \ln k} \simeq \frac{\ln \zeta_{fr}(k_2) - \ln \zeta_{fr}(k_1)}{\Delta \ln k}~,
\end{equation*}
where $k_1 = k\left(1 - \frac{\Delta \ln k}{2}\right)$ and $k_2 = k\left(1 + \frac{\Delta \ln k}{2}\right)$.

\begin{figure}[!ht]
    \centering
    \includegraphics[width=0.40\linewidth]{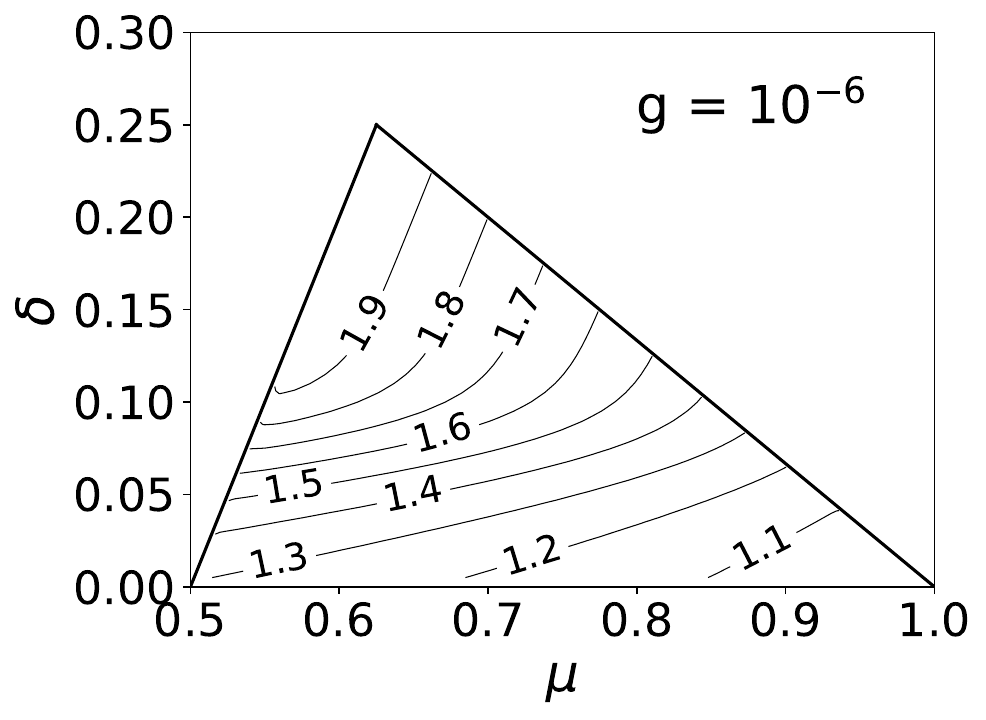}\includegraphics[width=0.40\linewidth]{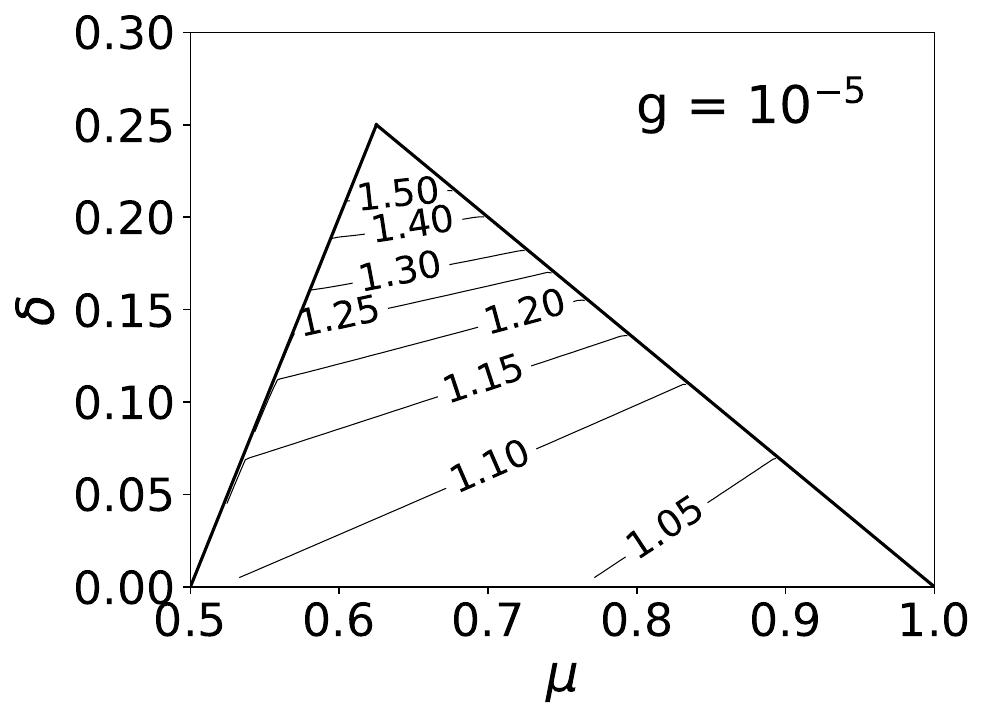}    
    \includegraphics[width=0.40\linewidth]{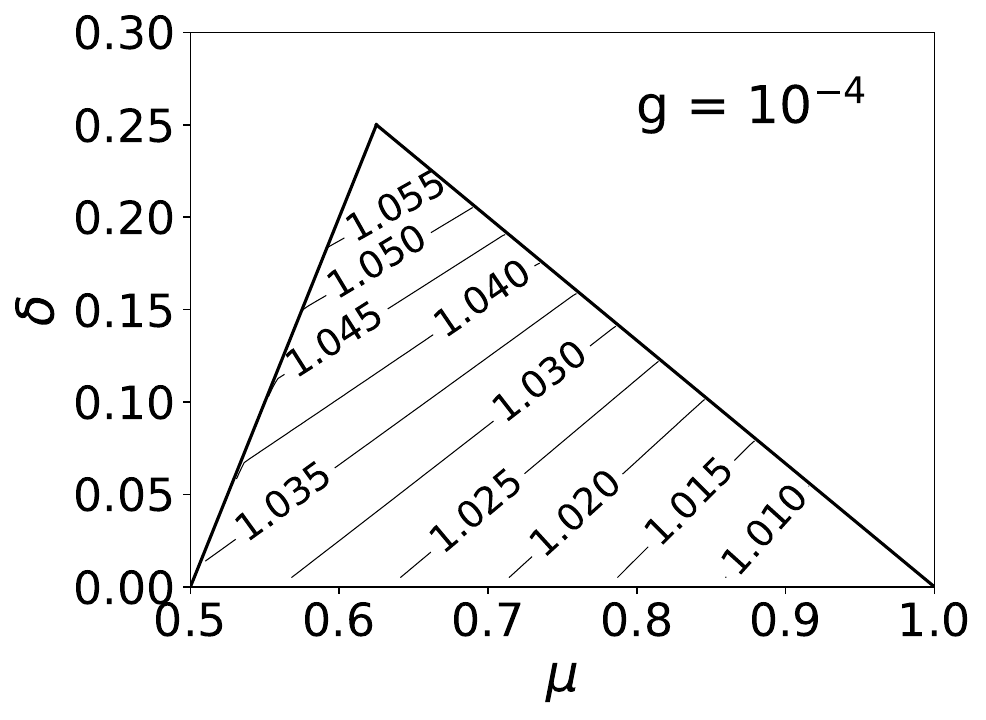}\includegraphics[width=0.40\linewidth]{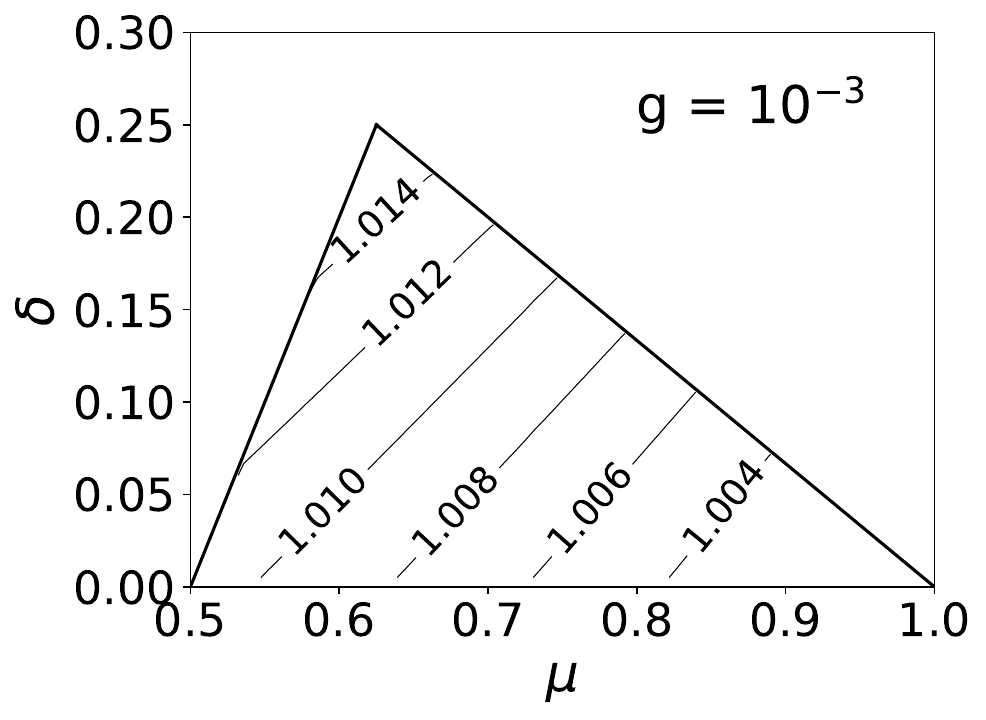}
    \caption{Numerical results of the spectral index $n_s$ in the $\mu-\delta$ plane, shown as level curves. $n_s$ values are sampled only when $\mu$ and $\delta$ satisfy Eq. (\ref{earlyUnitarityBound}) (inside the triangular region). Each panel assumes different $g$ values, displayed in the upper right corners. The values $c$ and $s$ are fixed at $10^{-4}$ for all panels.}
    \label{fig:nsnum}
\end{figure}

Equipped with the numerical method, we investigated the values of the spectral index $n_s$ over a broad range of model parameters. An example of the results is shown in Fig.~\ref{fig:nsnum}. When the model parameters $\mu$ and $\delta$ satisfy the no-go and unitarity bounds described in Eq. (\ref{earlyUnitarityBound}), the resulting $n_s$ values are more red-shifted than those obtained from the linear estimation $n_s = 3 - 2\mu + \delta$. However, we found that $n_s$ is always blue-tilted, at least in the parameter space we explored. For this reason, we introduced a spectator field into our Genesis scenario, which will be discussed in the next section. 

\newpage

\section{The spectator field}

\label{sec: The spectator field}

As we observed in the previous section, it is nearly impossible to obtain a red-tilted spectrum, even at the non-linear stage, when the background evolution includes significant corrections from the $u$ variable. This problem for Genesis is not new. Generally speaking, for classical Genesis without strong gravity in the past \cite{Creminelli:2010ba}, or for certain scenarios with early NEC violation \cite{Tahara:2020fmn}, it is impossible to generate a red-tilted scalar spectrum. The resolution to this problem is quite straightforward: the introduction of a spectator field. 

However, before introducing an additional spectator scalar field, let us discuss the symmetries of our theory. For $\mu = 1$ and $\delta = 0$, the Jordan frame Lagrangian is invariant under scale symmetry. Under the scaling transformation
\begin{align}
\label{scalingTransformation}
 \tilde{\phi} = \phi - \ln \lambda\;, \\
 \tilde{g}_{\mu\nu} = \lambda^2 g_{\mu\nu}\;,\nonumber
\end{align}
the Lagrangian \eqref{JordanFrameLagrangian} transforms as follows:
\begin{align*}
   \sqrt{-\tilde{g}}& \left[ \tilde{G}_2 - \tilde{G}_3 \tilde{\square} \tilde{\phi} + \tilde{G}_4 \frac{\tilde{R}}{2} \right] = \lambda^4 \sqrt{-g} \Bigg[ \frac{X \left(3 c^2 e^{2 \phi} \left(4 c^2 \ln \left(\frac{X}{\lambda^2 X_0}\right) - g\right) + 2g X\right)}{3 c^4 \lambda^4} \\
   &- \frac{e^{2 \phi} \left(\ln \left(\frac{X}{\lambda^2 X_0}\right) + 2\right)}{\lambda^4} \square \phi + \frac{e^{2 \phi}}{2 \lambda^4}\frac{R}{2} \Bigg] \Rightarrow \sqrt{-g}\left[G_2 - G_3 \square \phi + G_4 \frac{R}{2}\right]\;,
\end{align*}
where the arrow indicates integration by parts.

This symmetry ensures that for the case $\mu = 1$ and $\delta = 0$, the perturbation $\zeta$ of the field $\phi$ should have a flat power spectrum. This observation aligns perfectly with the formula~\eqref{scalarSpectrumLin} for the scalar spectral index $n_S$. Thus, we will introduce a spectator field in the spirit of Refs.~\cite{Creminelli:2010ba, Tahara:2020fmn, Libanov:2016kfc}, specifically in a manner where the spectator field is invariant under the scaling transformation~\eqref{scalingTransformation}
\begin{align}
    S_{\sigma} = \int dt \, d^3x \, \sqrt{-g} \, e^{2\phi} \left( -\frac{1}{2}(\partial \sigma)^2 \right).
    \label{spectatorAction}
\end{align}
Indeed, the action above transforms in the following way:
\begin{align*}
    \sqrt{-\tilde{g}} \, e^{2\tilde{\phi}} \left( -\frac{1}{2} \tilde{g}^{\nu\rho} \partial_{\nu} \sigma \partial_{\rho}\sigma \right) = \lambda^4 \sqrt{-g} \, e^{2\phi} \left( -\frac{1}{2\lambda^4} g^{\nu\rho} \partial_{\nu} \sigma \partial_{\rho} \sigma\right)\;.
\end{align*}
Thus, the effective scale factor for the spectator field $\sigma$ is expressed as
\begin{align*}
    a_{\text{eff}} = e^{\phi} \cdot a.
\end{align*}

Now, we introduce the conformal time as follows:
\begin{align*}
    \eta = \int \frac{dt}{a} \propto t < 0\;.
\end{align*}
Here, we remind the reader that for Genesis, the scale factor $a$ tends toward a constant in the asymptotic past.

Recalling the solution for the field $\phi$ \eqref{jordanFrameSolutionsPhiH}, we write the effective scale factor as
\begin{equation*}
    a_{eff} = -\frac{1}{H_{\sigma}\eta},
\end{equation*}
where $H_{\sigma}$ is a constant. This is precisely the scale factor for de Sitter space. Therefore, we immediately see that the field $\sigma$ is massless and experiences the effective de Sitter metric. Thus, we conclude that the power spectrum for perturbations generated by the additional scalar $\sigma$ is flat. By adding potential and/or massive terms to the action~\eqref{spectatorAction}, one can tilt the power spectrum in either direction and ultimately obtain a red-tilted spectrum that is consistent with the experimental data~\cite{Planck:2018nkj}. The conversion of fluctuations in $\sigma$ into adiabatic modes could occur through one of the mechanisms proposed in Refs.~\cite{Lyth:2001nq, Dvali:2003em, Dvali:2003ar}. This mechanism is model-dependent, and we leave a precise analysis of it for future work. 

Finally, we would like to comment on the ratio value. As proposed in Sec.~\ref{sec: Scalar primordial power spectrum}, the model suggests a natural way to deform it in order to achieve a definite value for the scalar amplitude. In the case of the spectator field, the scalar spectrum is generated by the new field $\sigma$, but the same mechanism of model deformation could change the amplitude of tensor perturbations, which provides us with an opportunity to obtain a subsequent small value for the ratio $r$.


\section{Conclusion}
\label{sec:conclusion}

In this article, we find the minimal setup within the framework of Horndeski gravity that can describe non-singular cosmology. In this setup, we construct the Genesis scenario. Our scenario begins with flat space and time, then expands and transitions to the kination stage, during which General Relativity is restored. This Genesis scenario circumvents the no-go theorem at the cost of encountering the danger of strong coupling in the past; that is, the effective Planck mass tends to zero in the asymptotic past. This implies a risk of violating unitarity at early times.

We demonstrate that the background solution remains stable throughout the entire evolution, and the speed of scalar perturbations does not exceed the speed of light, while the speed of tensor perturbations remains equal to unity. Moreover, in our model, there are two distinct regimes. In the first regime, the early-stage solution exhibits a roughly power-law behavior; thus, the unitarity bounds from~\eqref{healthy_region} fully apply to this first solution. We implicitly show that our first solution does not break unitarity at early times.

The second solution, however, deviates sufficiently from power-law behavior even at early times. Therefore, one cannot directly apply the unitarity bound~\eqref{healthy_region} to this solution. Nevertheless, we have shown that despite its highly non power-law behavior, there still exist parameters for which the background solution remains outside the strong-coupling regime. This last observation makes our setup more universal.

Next, we investigate the scalar and tensor power spectra. We find that if one assumes power-law behavior to the solution at the Genesis stage, it is impossible to simultaneously have a red-tilted scalar spectrum while maintaining unitarity in the theory. Additionally, we find that, in most cases, the tensor spectrum is blue-tilted.

Following that, we analyze solutions that exhibit non-power-law behavior at early times. This case cannot be analyzed analytically, so we performed numerical simulations. Unfortunately, we were unable to find a point in the parameter space that allows for a red-tilted power spectrum. Thus, to the best of our knowledge, it is also impossible to produce a correct spectrum of scalar perturbations, even in a highly non-power-law regime.

Therefore, we were compelled to look for another mechanism to produce a red-tilted spectrum for scalar perturbations. We adopt the spectator field mechanism. This additional spectator field couples in a conformally invariant way with the Horndeski field, providing a natural and straightforward mechanism for producing the red-tilted scalar spectrum. Additionally, we suggest a deformation of the model; this model deformation enables us to achieve sufficiently small values for the $r$ ratio.


\section*{ACKNOWLEDGMENTS}

We are grateful to Yulia Ageeva, Shingo Akama, Antonio De Felice, 
 Alexander Ganz, Dra\v{z}en Glavan, Mohammad Ali Gorji, Nils Albin Nilsson, Alexander Vikman, Yi Wang, and Christof Wetterich for useful discussion and comments. H.G. and P.P. are supported by IBS under the project code, IBS-R018-D3. M.Y. is supported by IBS under the project code, IBS-R018-D3, and by JSPS Grant-in-Aid for Scientific Research Number JP23K20843.

\newpage
\appendix

\section{From Jordan Frame to Einstein Frame}

\numberwithin{equation}{section}
\label{app:fromJToE}

Here, we obtain the general expressions for the Einstein frame functions $ G_{2}^{E} $ and $ G_{3}^{E} $ in the case where the function $ G_4(\phi) $ depends only on the field $ \phi $. 

Let us start with the following conformal transformation of the metric:
\begin{align*}
    \tilde{g}_{\mu\nu} = \Omega(\phi) g_{\mu\nu} = \text{e}^{2K(\phi)} g_{\mu\nu}, 
\end{align*}
where $ \tilde{g}_{\mu\nu} $ is the Einstein frame metric. The inverse metric is given by
\begin{align*}
    \tilde{g}^{\mu\nu} = \text{e}^{-2K(\phi)} g^{\mu\nu}.
\end{align*}
The Ricci scalar transforms as 
\begin{equation*}
    \tilde{R} = \text{e}^{-2K} \left( R - 6 g^{\mu\nu} \nabla_{\mu}\nabla_{\nu} K - 6 g^{\mu\nu} \partial_{\mu} K \partial_{\nu} K \right),
\end{equation*}
and
\begin{equation*}
    \sqrt{-\tilde{g}} = \text{e}^{4K} \sqrt{-g}.
\end{equation*}
Additionally, we have
\begin{equation*}
    \tilde{\Gamma}^{\lambda}_{\mu\nu} = \Gamma^{\lambda}_{\mu\nu} + \delta^{\lambda}_{\mu} \partial_{\nu} K + \delta^{\lambda}_{\nu} \partial_{\mu} K - g_{\mu\nu} g^{\lambda\rho} \partial_{\rho} K.
\end{equation*}
Also, one can write
\begin{align*}
    \tilde{\square}\phi &= \tilde{g}^{\mu\nu} \tilde{\nabla}_{\mu} \partial_{\nu} \phi = \text{e}^{-2K} g^{\mu\nu} \left[ \partial_{\mu} \partial_{\nu} \phi - \tilde{\Gamma}^{\lambda}_{\mu\nu} \partial_{\lambda} \phi \right] \nonumber \\
    &= \text{e}^{-2K} g^{\mu\nu} \left[ \partial_{\mu} \partial_{\nu} \phi - \Gamma^{\lambda}_{\mu\nu} \partial_{\lambda} \phi - \delta^{\lambda}_{\mu} \partial_{\nu} K \partial_{\lambda} \phi - \delta^{\lambda}_{\nu} \partial_{\mu} K \partial_{\lambda} \phi + g_{\mu\nu} g^{\lambda\rho} \partial_{\rho} K \partial_{\lambda} \phi \right] \nonumber \\
    &= \text{e}^{-2K} \left[ \square\phi - g^{\lambda\nu} \partial_{\nu} K \partial_{\lambda} \phi - g^{\lambda\mu} \partial_{\mu} K \partial_{\lambda} \phi + 4 g^{\lambda\rho} \partial_{\rho} K \partial_{\lambda} \phi \right] \nonumber \\
    &= \text{e}^{-2K} \left[ \square\phi + 2 g^{\mu\nu} \partial_{\mu} K \partial_{\nu} \phi \right].
\end{align*}
Meanwhile, the kinetic term $ X $ transforms as follows:
\begin{equation*}
    \tilde{X} = -\frac{1}{2} \tilde{g}^{\mu\nu} \partial_{\mu} \phi \partial_{\nu} \phi = \text{e}^{-2K} X.
\end{equation*}

Now, we are ready to apply the expressions above and substitute them into the action:
\begin{align*}
    S &= \int d^4 x \sqrt{-\tilde{g}} \Big[\tilde{G}_2(\phi,\tilde{X}) - \tilde{G}_3(\phi,\tilde{X})\tilde{\square}\phi + \frac{1}{2}\tilde{R}\Big] \nonumber\\
    &= \int d^4 x \; \text{e}^{4K}\sqrt{-g} \Big[\tilde{G}_2(\phi,\tilde{X}) - \tilde{G}_3(\phi,\tilde{X})\left(\text{e}^{-2K} \left[\square\phi + 2g^{\mu\nu} \partial_{\mu}K\partial_{\nu}\phi\right]\right) + \frac{1}{2}\tilde{R}\Big] \nonumber\\
    &= \int d^4 x \; \sqrt{-g}\Big[\text{e}^{4K}\tilde{G}_2(\phi,\tilde{X}) - \text{e}^{2K}\tilde{G}_3(\phi,\tilde{X}) \square\phi - 2\text{e}^{2K}\tilde{G}_3(\phi,\tilde{X}) g^{\mu\nu} \partial_{\mu}K\partial_{\nu}\phi \nonumber\\
    &+\frac{1}{2}\text{e}^{2K} \left(R - 6 g^{\mu\nu} \nabla_{\mu}\nabla_{\nu}K - 6 g^{\mu\nu}\partial_{\mu}K\partial_{\nu}K\right)\Big] \nonumber\\
    &= \int d^4 x \; \sqrt{-g}\Big[\text{e}^{4K}\tilde{G}_2(\phi,\tilde{X}) - \text{e}^{2K}\tilde{G}_3(\phi,\tilde{X}) \square\phi - 2 \text{e}^{2K} \tilde{G}_3(\phi,\tilde{X}) g^{\mu\nu} \partial_{\mu}K\partial_{\nu}\phi \nonumber\\
    &+\frac{1}{2}\left(\text{e}^{2K} R + 12\text{e}^{2K} g^{\mu\nu} \partial_{\mu}K\partial_{\nu}K - 6\text{e}^{2K} g^{\mu\nu}\partial_{\mu}K\partial_{\nu}K\right)\Big] \nonumber\\
    &= \int d^4 x \; \sqrt{-g}\Big[\text{e}^{4K}\tilde{G}_2(\phi,\tilde{X}) - \text{e}^{2K}\tilde{G}_3(\phi,\tilde{X}) \square\phi - 2 \text{e}^{2K} \tilde{G}_3(\phi,\tilde{X}) K_{\phi} g^{\mu\nu} \partial_{\mu}\phi\partial_{\nu}\phi \nonumber\\
    &+\frac{1}{2}(\text{e}^{2K} R + 6 \text{e}^{2K} K_{\phi}^2 g^{\mu\nu} \partial_{\mu}\phi\partial_{\nu}\phi) \Big] \\
    &= \int d^4 x \; \sqrt{-g} \left[ G_{2}(\phi,X) - G_{3}(\phi,X) \square \phi + G_{4}(\phi) R \right],
\end{align*}
thus
\begin{align*}
    G_{2}(\phi,X) &= \text{e}^{4K}\tilde{G}_{2}(\phi,\text{e}^{-2K}X) + 4\text{e}^{2K}\tilde{G}_{3}(\phi,\text{e}^{-2K}X) K_{\phi} X - 6 \text{e}^{2K} K_{\phi}^2 X, \\
    G_{3}(\phi,X) &= \text{e}^{2K}\tilde{G}_{3}(\phi,\text{e}^{-2K}X),\\
    G_{4}(\phi) &= \frac{1}{2} \text{e}^{2K}.
\end{align*}
In order to transition to the Einstein frame, one should choose $ e^{2K} = 2G_{4}(\phi) $; then, the Einstein frame Lagrangian is given by the expressions below:
\begin{align*}
    G_{2}^{E}(\phi,X^{E}) &= e^{-4K} G_{2}(\phi,X^{E} e^{2K}) - 4 e^{-4K} G_{3}(\phi,X^{E} e^{2K}) K_{\phi} X + 6 K_{\phi}^{2} X^{E} \;,\\  
    G_{3}^{E}(\phi,X^{E}) &= e^{-2K} G_{3}(\phi,X^{E} e^{2K})\;,\\ 
    G_{4}^{E} &= \frac{1}{2}\;,
\end{align*}
where, in the Einstein frame, instead of ``$ \tilde{\ldots} $", we write ``$ E $".


\section{Eikonal approximation}

\label{app: Eikonal approximation}

Here, we discuss how we can solve the mode equation Eq. (\ref{zetaModeEq}) when its coefficients have more complicated time dependence. We found that the evolution of $ \zeta $ before freeze-out can be well described by the Eikonal approximation. Let us consider the $ k (= |\vec{k}|) $ mode equation of the form
\begin{equation}
    \ddot{\zeta} + \frac{\theta^s}{t} \dot{\zeta} + k^2 \mathcal{B}^s \zeta = 0~,
\end{equation}
where the definitions of $ \theta^s $ and $ \mathcal{B}^s $ can be found in Sec.~\ref{sec: Scalar primordial power spectrum}. Decomposing $ \zeta $ into its log-amplitude $ \chi = \ln |\zeta| $ and its phase $ \varphi $, we find that the equation of motion for $ \chi $ is given by
\begin{equation}
    \ddot{\chi} + \dot{\chi}^2 - \dot{\varphi}^2 + \frac{\theta^s}{t} \dot{\chi} + k^2 \mathcal{B}^s = 0 ~,
\end{equation}
where
\begin{equation} 
    \dot{\varphi} = \dot{\varphi}_0 \exp[-2(\chi - \chi_0) - (\ln \mathcal{A}^s - \ln \mathcal{A}^s_0)] ~.
\end{equation}
See Sec.~\ref{sec:TimeDependTermsScalarSpectrum} for more detailed descriptions.

The equation of motion for $ \chi $ allows a simple approximate solution under the condition $ \theta^s/t \ll k\sqrt{\mathcal{B}^s} $. Physically, this means that the oscillation timescale of the mode is much shorter than the timescale of background evolution, which allows us to use the Eikonal approximation. In this formalism, we assume that $ (k \sqrt{\mathcal{B}^s})^{-1} $ and $ \dot{\varphi}^{-1} $ are small quantities of order $ \epsilon $, and we solve the equation of motion perturbatively with $ \chi = \chi^{(0)} + \epsilon \chi^{(1)} + \epsilon^2 \chi^{(2)} + \cdots $. The leading-order term of the equation of motion is
\begin{equation}
    -\dot{\varphi}_0^2 \exp[-4(\chi^{(0)} - \chi_0) - 2(\ln \mathcal{A}^s - \ln \mathcal{A}^s_0)] + k^2 \mathcal{B}^s = 0~.
\end{equation}
Evaluating the equation at $ t = t_0 $, we find that
\begin{equation}
    \chi^{(0)}(t_0) = \chi_0~
\end{equation}
and
\begin{equation}
    \dot{\varphi}_0^2 = k^2 \mathcal{B}^s_0~,
\end{equation}
where $ \mathcal{B}^s_0 = \mathcal{B}^s(t_0) $. Solving the equation for $ \chi^{(0)} $ leads to
\begin{equation}
    \chi^{(0)}(t) = \chi_0 - \frac{1}{2}(\ln \mathcal{A}^s(t) - \ln \mathcal{A}^s_0) - \frac{1}{4}(\ln \mathcal{B}^s(t) - \ln \mathcal{B}^s_0)~.
\end{equation}
Its time derivative is given by
\begin{equation}
    \dot{\chi}^{(0)}(t) = -\frac{1}{2}\frac{\theta^s}{t} - \frac{1}{4}\frac{\dot{\mathcal{B}}^s}{\mathcal{B}^s}~.
\end{equation}
Here, we used $ \theta^s/t = \dot{\mathcal{A}}^s/\mathcal{A}^s $. These results are used to determine the approximate initial conditions for the numerical evolution of the mode function in Sec. \ref{sec:TimeDependTermsScalarSpectrum}.


\newpage
\bibliographystyle{JHEPmod}
\bibliography{ref.bib}

\providecommand{\href}[2]{#2}\begingroup\raggedright\begin{thebibliography}{10}

\bibitem{Borde:1996pt}
A.~Borde and A.~Vilenkin, {\it {Singularities in inflationary cosmology: A Review}}, \href{https://doi.org/10.1142/S0218271896000497}{Int. J. Mod. Phys. D {\bfseries 5} (1996) 813} [\href{http://arxiv.org/abs/gr-qc/9612036}{{\ttfamily arXiv:gr-qc/9612036}}].

\bibitem{Penrose:1964wq}
R.~Penrose, {\it {Gravitational collapse and space-time singularities}}, \href{https://doi.org/10.1103/PhysRevLett.14.57}{Phys. Rev. Lett. {\bfseries 14} (1965) 57}.

\bibitem{Rubakov:2014jja}
V.~A. Rubakov, {\it {The Null Energy Condition and its violation}}, \href{https://doi.org/10.3367/UFNe.0184.201402b.0137}{Phys. Usp. {\bfseries 57} (2014) 128} [\href{http://arxiv.org/abs/1401.4024}{{\ttfamily arXiv:1401.4024}}].

\bibitem{Kobayashi:2019hrl}
T.~Kobayashi, {\it {Horndeski theory and beyond: a review}}, \href{https://doi.org/10.1088/1361-6633/ab2429}{Rept. Prog. Phys. {\bfseries 82} (2019) 086901} [\href{http://arxiv.org/abs/1901.07183}{{\ttfamily arXiv:1901.07183}}].

\bibitem{Horndeski:1974wa}
G.~W. Horndeski, {\it {Second-order scalar-tensor field equations in a four-dimensional space}}, \href{https://doi.org/10.1007/BF01807638}{Int. J. Theor. Phys. {\bfseries 10} (1974) 363}.

\bibitem{Zumalacarregui:2013pma}
M.~Zumalac\'arregui and J.~Garc\'\i{}a-Bellido, {\it {Transforming gravity: from derivative couplings to matter to second-order scalar-tensor theories beyond the Horndeski Lagrangian}}, \href{https://doi.org/10.1103/PhysRevD.89.064046}{Phys. Rev. D {\bfseries 89} (2014) 064046} [\href{http://arxiv.org/abs/1308.4685}{{\ttfamily arXiv:1308.4685}}].

\bibitem{Gleyzes:2014dya}
J.~Gleyzes, D.~Langlois, F.~Piazza and F.~Vernizzi, {\it {Healthy theories beyond Horndeski}}, \href{https://doi.org/10.1103/PhysRevLett.114.211101}{Phys. Rev. Lett. {\bfseries 114} (2015) 211101} [\href{http://arxiv.org/abs/1404.6495}{{\ttfamily arXiv:1404.6495}}].

\bibitem{Langlois:2015cwa}
D.~Langlois and K.~Noui, {\it {Degenerate higher derivative theories beyond Horndeski: evading the Ostrogradski instability}}, \href{https://doi.org/10.1088/1475-7516/2016/02/034}{JCAP {\bfseries 02} (2016) 034} [\href{http://arxiv.org/abs/1510.06930}{{\ttfamily arXiv:1510.06930}}].

\bibitem{Creminelli:2010ba}
P.~Creminelli, A.~Nicolis and E.~Trincherini, {\it {Galilean Genesis: An Alternative to inflation}}, \href{https://doi.org/10.1088/1475-7516/2010/11/021}{JCAP {\bfseries 11} (2010) 021} [\href{http://arxiv.org/abs/1007.0027}{{\ttfamily arXiv:1007.0027}}].

\bibitem{Creminelli:2012my}
P.~Creminelli, K.~Hinterbichler, J.~Khoury, A.~Nicolis and E.~Trincherini, {\it {Subluminal Galilean Genesis}}, \href{https://doi.org/10.1007/JHEP02(2013)006}{JHEP {\bfseries 02} (2013) 006} [\href{http://arxiv.org/abs/1209.3768}{{\ttfamily arXiv:1209.3768}}].

\bibitem{Hinterbichler:2012fr}
K.~Hinterbichler, A.~Joyce, J.~Khoury and G.~E.~J. Miller, {\it {DBI Realizations of the Pseudo-Conformal Universe and Galilean Genesis Scenarios}}, \href{https://doi.org/10.1088/1475-7516/2012/12/030}{JCAP {\bfseries 12} (2012) 030} [\href{http://arxiv.org/abs/1209.5742}{{\ttfamily arXiv:1209.5742}}].

\bibitem{Elder:2013gya}
B.~Elder, A.~Joyce and J.~Khoury, {\it {From Satisfying to Violating the Null Energy Condition}}, \href{https://doi.org/10.1103/PhysRevD.89.044027}{Phys. Rev. D {\bfseries 89} (2014) 044027} [\href{http://arxiv.org/abs/1311.5889}{{\ttfamily arXiv:1311.5889}}].

\bibitem{Pirtskhalava:2014esa}
D.~Pirtskhalava, L.~Santoni, E.~Trincherini and P.~Uttayarat, {\it {Inflation from Minkowski Space}}, \href{https://doi.org/10.1007/JHEP12(2014)151}{JHEP {\bfseries 12} (2014) 151} [\href{http://arxiv.org/abs/1410.0882}{{\ttfamily arXiv:1410.0882}}].

\bibitem{Nishi:2015pta}
S.~Nishi and T.~Kobayashi, {\it {Generalized Galilean Genesis}}, \href{https://doi.org/10.1088/1475-7516/2015/03/057}{JCAP {\bfseries 03} (2015) 057} [\href{http://arxiv.org/abs/1501.02553}{{\ttfamily arXiv:1501.02553}}].

\bibitem{Kobayashi:2015gga}
T.~Kobayashi, M.~Yamaguchi and J.~Yokoyama, {\it {Galilean Creation of the Inflationary Universe}}, \href{https://doi.org/10.1088/1475-7516/2015/07/017}{JCAP {\bfseries 07} (2015) 017} [\href{http://arxiv.org/abs/1504.05710}{{\ttfamily arXiv:1504.05710}}].

\bibitem{Qiu:2011cy}
T.~Qiu, J.~Evslin, Y.-F. Cai, M.~Li and X.~Zhang, {\it {Bouncing Galileon Cosmologies}}, \href{https://doi.org/10.1088/1475-7516/2011/10/036}{JCAP {\bfseries 10} (2011) 036} [\href{http://arxiv.org/abs/1108.0593}{{\ttfamily arXiv:1108.0593}}].

\bibitem{Easson:2011zy}
D.~A. Easson, I.~Sawicki and A.~Vikman, {\it {G-Bounce}}, \href{https://doi.org/10.1088/1475-7516/2011/11/021}{JCAP {\bfseries 11} (2011) 021} [\href{http://arxiv.org/abs/1109.1047}{{\ttfamily arXiv:1109.1047}}].

\bibitem{Battarra:2014tga}
L.~Battarra, M.~Koehn, J.-L. Lehners and B.~A. Ovrut, {\it {Cosmological Perturbations Through a Non-Singular Ghost-Condensate/Galileon Bounce}}, \href{https://doi.org/10.1088/1475-7516/2014/07/007}{JCAP {\bfseries 07} (2014) 007} [\href{http://arxiv.org/abs/1404.5067}{{\ttfamily arXiv:1404.5067}}].

\bibitem{Ijjas:2016tpn}
A.~Ijjas and P.~J. Steinhardt, {\it {Classically stable nonsingular cosmological bounces}}, \href{https://doi.org/10.1103/PhysRevLett.117.121304}{Phys. Rev. Lett. {\bfseries 117} (2016) 121304} [\href{http://arxiv.org/abs/1606.08880}{{\ttfamily arXiv:1606.08880}}].

\bibitem{Kobayashi:2016xpl}
T.~Kobayashi, {\it {Generic instabilities of nonsingular cosmologies in Horndeski theory: A no-go theorem}}, \href{https://doi.org/10.1103/PhysRevD.94.043511}{Phys. Rev. D {\bfseries 94} (2016) 043511} [\href{http://arxiv.org/abs/1606.05831}{{\ttfamily arXiv:1606.05831}}].

\bibitem{Libanov:2016kfc}
M.~Libanov, S.~Mironov and V.~Rubakov, {\it {Generalized Galileons: instabilities of bouncing and Genesis cosmologies and modified Genesis}}, \href{https://doi.org/10.1088/1475-7516/2016/08/037}{JCAP {\bfseries 08} (2016) 037} [\href{http://arxiv.org/abs/1605.05992}{{\ttfamily arXiv:1605.05992}}].

\bibitem{Ageeva:2018lko}
Y.~A. Ageeva, O.~A. Evseev, O.~I. Melichev and V.~A. Rubakov, {\it {Horndeski Genesis: strong coupling and absence thereof}}, \href{https://doi.org/10.1051/epjconf/201819107010}{EPJ Web Conf. {\bfseries 191} (2018) 07010} [\href{http://arxiv.org/abs/1810.00465}{{\ttfamily arXiv:1810.00465}}].

\bibitem{Ageeva:2020gti}
Y.~Ageeva, O.~Evseev, O.~Melichev and V.~Rubakov, {\it {Toward evading the strong coupling problem in Horndeski genesis}}, \href{https://doi.org/10.1103/PhysRevD.102.023519}{Phys. Rev. D {\bfseries 102} (2020) 023519} [\href{http://arxiv.org/abs/2003.01202}{{\ttfamily arXiv:2003.01202}}].

\bibitem{Ageeva:2022fyq}
Y.~Ageeva and P.~Petrov, {\it {On the strong coupling problem in cosmologies with \textquotedblleft{}strong gravity in the past\textquotedblright{}}}, \href{https://doi.org/10.1142/S0217732322501711}{Mod. Phys. Lett. A {\bfseries 37} (2022) 2250171} [\href{http://arxiv.org/abs/2206.10646}{{\ttfamily arXiv:2206.10646}}].

\bibitem{Cai:2022ori}
Y.~Cai, J.~Xu, S.~Zhao and S.~Zhou, {\it {Perturbative unitarity and NEC violation in genesis cosmology}}, \href{https://doi.org/10.1007/JHEP10(2022)140}{JHEP {\bfseries 10} (2022) 140} [\href{http://arxiv.org/abs/2207.11772}{{\ttfamily arXiv:2207.11772}}].

\bibitem{Ageeva:2022asq}
Y.~Ageeva, P.~Petrov and V.~Rubakov, {\it {Generating cosmological perturbations in non-singular Horndeski cosmologies}}, \href{https://doi.org/10.1007/JHEP01(2023)026}{JHEP {\bfseries 01} (2023) 026} [\href{http://arxiv.org/abs/2207.04071}{{\ttfamily arXiv:2207.04071}}].

\bibitem{Ageeva:2020buc}
Y.~Ageeva, P.~Petrov and V.~Rubakov, {\it {Horndeski genesis: consistency of classical theory}}, \href{https://doi.org/10.1007/JHEP12(2020)107}{JHEP {\bfseries 12} (2020) 107} [\href{http://arxiv.org/abs/2009.05071}{{\ttfamily arXiv:2009.05071}}].

\bibitem{Creminelli:2016zwa}
P.~Creminelli, D.~Pirtskhalava, L.~Santoni and E.~Trincherini, {\it {Stability of Geodesically Complete Cosmologies}}, \href{https://doi.org/10.1088/1475-7516/2016/11/047}{JCAP {\bfseries 11} (2016) 047} [\href{http://arxiv.org/abs/1610.04207}{{\ttfamily arXiv:1610.04207}}].

\bibitem{Rubakov:2022fqk}
V.~A. Rubakov and C.~Wetterich, {\it {Geodesic (in) Completeness in General Metric Frames}}, \href{https://doi.org/10.3390/sym14122557}{Symmetry {\bfseries 14} (2022) 2557} [\href{http://arxiv.org/abs/2210.11198}{{\ttfamily arXiv:2210.11198}}].

\bibitem{Wetterich:2024ung}
C.~Wetterich, {\it {Physical time for the beginning universe}}, \href{https://doi.org/10.1142/S0217751X24450052}{Int. J. Mod. Phys. A {\bfseries 39} (2024) 2445005} [\href{http://arxiv.org/abs/2408.01524}{{\ttfamily arXiv:2408.01524}}].

\bibitem{Wetterich:2014zta}
C.~Wetterich, {\it {Eternal Universe}}, \href{https://doi.org/10.1103/PhysRevD.90.043520}{Phys. Rev. D {\bfseries 90} (2014) 043520} [\href{http://arxiv.org/abs/1404.0535}{{\ttfamily arXiv:1404.0535}}].

\bibitem{Mironov:2024pjt}
S.~Mironov and V.~Volkova, {\it {Non-singular cosmological scenarios in scalar-tensor theories and their stability: a review}},  \href{http://arxiv.org/abs/2409.16108}{{\ttfamily arXiv:2409.16108}}.

\bibitem{Kobayashi:2011nu}
T.~Kobayashi, M.~Yamaguchi and J.~Yokoyama, {\it {Generalized G-inflation: Inflation with the most general second-order field equations}}, \href{https://doi.org/10.1143/PTP.126.511}{Prog. Theor. Phys. {\bfseries 126} (2011) 511} [\href{http://arxiv.org/abs/1105.5723}{{\ttfamily arXiv:1105.5723}}].

\bibitem{Deffayet:2011gz}
C.~Deffayet, X.~Gao, D.~A. Steer and G.~Zahariade, {\it {From k-essence to generalised Galileons}}, \href{https://doi.org/10.1103/PhysRevD.84.064039}{Phys. Rev. D {\bfseries 84} (2011) 064039} [\href{http://arxiv.org/abs/1103.3260}{{\ttfamily arXiv:1103.3260}}].

\bibitem{Ageeva:2021yik}
Y.~Ageeva, P.~Petrov and V.~Rubakov, {\it {Nonsingular cosmological models with strong gravity in the past}}, \href{https://doi.org/10.1103/PhysRevD.104.063530}{Phys. Rev. D {\bfseries 104} (2021) 063530} [\href{http://arxiv.org/abs/2104.13412}{{\ttfamily arXiv:2104.13412}}].

\bibitem{Akama:2022usl}
S.~Akama and S.~Hirano, {\it {Primordial non-Gaussianity from Galilean Genesis without strong coupling problem}}, \href{https://doi.org/10.1103/PhysRevD.107.063504}{Phys. Rev. D {\bfseries 107} (2023) 063504} [\href{http://arxiv.org/abs/2211.00388}{{\ttfamily arXiv:2211.00388}}].

\bibitem{Gleyzes:2013ooa}
J.~Gleyzes, D.~Langlois, F.~Piazza and F.~Vernizzi, {\it {Essential Building Blocks of Dark Energy}}, \href{https://doi.org/10.1088/1475-7516/2013/08/025}{JCAP {\bfseries 08} (2013) 025} [\href{http://arxiv.org/abs/1304.4840}{{\ttfamily arXiv:1304.4840}}].

\bibitem{Fasiello:2014aqa}
M.~Fasiello and S.~Renaux-Petel, {\it {Non-Gaussian inflationary shapes in $G^3$ theories beyond Horndeski}}, \href{https://doi.org/10.1088/1475-7516/2014/10/037}{JCAP {\bfseries 10} (2014) 037} [\href{http://arxiv.org/abs/1407.7280}{{\ttfamily arXiv:1407.7280}}].

\bibitem{Adams:2006sv}
A.~Adams, N.~Arkani-Hamed, S.~Dubovsky, A.~Nicolis and R.~Rattazzi, {\it {Causality, analyticity and an IR obstruction to UV completion}}, \href{https://doi.org/10.1088/1126-6708/2006/10/014}{JHEP {\bfseries 10} (2006) 014} [\href{http://arxiv.org/abs/hep-th/0602178}{{\ttfamily arXiv:hep-th/0602178}}].

\bibitem{deRham:2013hsa}
C.~de~Rham, M.~Fasiello and A.~J. Tolley, {\it {Galileon Duality}}, \href{https://doi.org/10.1016/j.physletb.2014.03.061}{Phys. Lett. B {\bfseries 733} (2014) 46} [\href{http://arxiv.org/abs/1308.2702}{{\ttfamily arXiv:1308.2702}}].

\bibitem{Mironov:2024idn}
S.~Mironov, A.~Shtennikova and M.~Valencia-Villegas, {\it {Reviving Horndeski after GW170817 by Kaluza-Klein compactifications}}, \href{https://doi.org/10.1016/j.physletb.2024.139058}{Phys. Lett. B {\bfseries 858} (2024) 139058} [\href{http://arxiv.org/abs/2405.02281}{{\ttfamily arXiv:2405.02281}}].

\bibitem{Mironov:2024zzk}
S.~Mironov, M.~Sharov and V.~Volkova, {\it {Time-dependent, spherically symmetric background in Kaluza-Klein compactified Horndeski theory and the speed of gravity waves}},  \href{http://arxiv.org/abs/2408.06329}{{\ttfamily arXiv:2408.06329}}.

\bibitem{Kamenshchik:2016gcy}
A.~Y. Kamenshchik, E.~O. Pozdeeva, S.~Y. Vernov, A.~Tronconi and G.~Venturi, {\it {Transformations between Jordan and Einstein frames: Bounces, antigravity, and crossing singularities}}, \href{https://doi.org/10.1103/PhysRevD.94.063510}{Phys. Rev. D {\bfseries 94} (2016) 063510} [\href{http://arxiv.org/abs/1602.07192}{{\ttfamily arXiv:1602.07192}}].

\bibitem{Wetterich:2020oyy}
C.~Wetterich, {\it {Crossing the Big Bang singularity}}, \href{https://doi.org/10.1016/j.dark.2021.100866}{Phys. Dark Univ. {\bfseries 33} (2021) 100866} [\href{http://arxiv.org/abs/2004.04506}{{\ttfamily arXiv:2004.04506}}].

\bibitem{Mironov:2019qjt}
S.~Mironov, V.~Rubakov and V.~Volkova, {\it {Genesis with general relativity asymptotics in beyond Horndeski theory}}, \href{https://doi.org/10.1103/PhysRevD.100.083521}{Phys. Rev. D {\bfseries 100} (2019) 083521} [\href{http://arxiv.org/abs/1905.06249}{{\ttfamily arXiv:1905.06249}}].

\bibitem{Ilyas:2020zcb}
A.~Ilyas, M.~Zhu, Y.~Zheng and Y.-F. Cai, {\it {Emergent Universe and Genesis from the DHOST Cosmology}}, \href{https://doi.org/10.1007/JHEP01(2021)141}{JHEP {\bfseries 01} (2021) 141} [\href{http://arxiv.org/abs/2009.10351}{{\ttfamily arXiv:2009.10351}}].

\bibitem{Zhu:2021ggm}
M.~Zhu and Y.~Zheng, {\it {Improved DHOST Genesis}}, \href{https://doi.org/10.1007/JHEP11(2021)163}{JHEP {\bfseries 11} (2021) 163} [\href{http://arxiv.org/abs/2109.05277}{{\ttfamily arXiv:2109.05277}}].

\bibitem{Armendariz-Picon:1999hyi}
C.~Armendariz-Picon, T.~Damour and V.~F. Mukhanov, {\it {k - inflation}}, \href{https://doi.org/10.1016/S0370-2693(99)00603-6}{Phys. Lett. B {\bfseries 458} (1999) 209} [\href{http://arxiv.org/abs/hep-th/9904075}{{\ttfamily arXiv:hep-th/9904075}}].

\bibitem{BazrafshanMoghaddam:2016tdk}
H.~Bazrafshan~Moghaddam, R.~Brandenberger and J.~Yokoyama, {\it {Note on Reheating in G-inflation}}, \href{https://doi.org/10.1103/PhysRevD.95.063529}{Phys. Rev. D {\bfseries 95} (2017) 063529} [\href{http://arxiv.org/abs/1612.00998}{{\ttfamily arXiv:1612.00998}}].

\bibitem{Planck:2018nkj}
{\scshape Planck} collaboration, N.~Aghanim et~al., {\it {Planck 2018 results. I. Overview and the cosmological legacy of Planck}}, \href{https://doi.org/10.1051/0004-6361/201833880}{Astron. Astrophys. {\bfseries 641} (2020) A1} [\href{http://arxiv.org/abs/1807.06205}{{\ttfamily arXiv:1807.06205}}].

\bibitem{deRham:2017aoj}
C.~de~Rham and S.~Melville, {\it {Unitary null energy condition violation in P(X) cosmologies}}, \href{https://doi.org/10.1103/PhysRevD.95.123523}{Phys. Rev. D {\bfseries 95} (2017) 123523} [\href{http://arxiv.org/abs/1703.00025}{{\ttfamily arXiv:1703.00025}}].

\bibitem{Ageeva:2022nbw}
Y.~A. Ageeva and P.~K. Petrov, {\it {Unitarity relation and unitarity bounds for scalars with different sound speeds}}, \href{https://doi.org/10.3367/UFNe.2022.11.039259}{Phys. Usp. {\bfseries 66} (2023) 1134} [\href{http://arxiv.org/abs/2206.03516}{{\ttfamily arXiv:2206.03516}}].

\bibitem{Grojean:2007zz}
C.~Grojean, {\it {New approaches to electroweak symmetry breaking}}, \href{https://doi.org/10.1070/PU2007v050n01ABEH006157}{Phys. Usp. {\bfseries 50} (2007) 1}.

\bibitem{Leblond:2008gg}
L.~Leblond and S.~Shandera, {\it {Simple Bounds from the Perturbative Regime of Inflation}}, \href{https://doi.org/10.1088/1475-7516/2008/08/007}{JCAP {\bfseries 08} (2008) 007} [\href{http://arxiv.org/abs/0802.2290}{{\ttfamily arXiv:0802.2290}}].

\bibitem{Senatore:2009cf}
L.~Senatore and M.~Zaldarriaga, {\it {On Loops in Inflation}}, \href{https://doi.org/10.1007/JHEP12(2010)008}{JHEP {\bfseries 12} (2010) 008} [\href{http://arxiv.org/abs/0912.2734}{{\ttfamily arXiv:0912.2734}}].

\bibitem{Gao:2011qe}
X.~Gao and D.~A. Steer, {\it {Inflation and primordial non-Gaussianities of 'generalized Galileons'}}, \href{https://doi.org/10.1088/1475-7516/2011/12/019}{JCAP {\bfseries 12} (2011) 019} [\href{http://arxiv.org/abs/1107.2642}{{\ttfamily arXiv:1107.2642}}].

\bibitem{DeFelice:2011uc}
A.~De~Felice and S.~Tsujikawa, {\it {Inflationary non-Gaussianities in the most general second-order scalar-tensor theories}}, \href{https://doi.org/10.1103/PhysRevD.84.083504}{Phys. Rev. D {\bfseries 84} (2011) 083504} [\href{http://arxiv.org/abs/1107.3917}{{\ttfamily arXiv:1107.3917}}].

\bibitem{Gao:2012ib}
X.~Gao, T.~Kobayashi, M.~Shiraishi, M.~Yamaguchi, J.~Yokoyama and S.~Yokoyama, {\it {Full bispectra from primordial scalar and tensor perturbations in the most general single-field inflation model}}, \href{https://doi.org/10.1093/ptep/ptt031}{PTEP {\bfseries 2013} (2013) 053E03} [\href{http://arxiv.org/abs/1207.0588}{{\ttfamily arXiv:1207.0588}}].

\bibitem{Ageeva:2022byg}
Y.~A. Ageeva and P.~K. Petrov, {\it {Unitarity relation and unitarity bounds for scalars with different sound speeds}}, \href{https://doi.org/10.3367/UFNe.2022.11.039259}{Phys. Usp. {\bfseries 66} (2023) 1134} [\href{http://arxiv.org/abs/2206.03516}{{\ttfamily arXiv:2206.03516}}].

\bibitem{Tahara:2020fmn}
H.~W.~H. Tahara and T.~Kobayashi, {\it {Nanohertz gravitational waves from a null-energy-condition violation in the early universe}}, \href{https://doi.org/10.1103/PhysRevD.102.123533}{Phys. Rev. D {\bfseries 102} (2020) 123533} [\href{http://arxiv.org/abs/2011.01605}{{\ttfamily arXiv:2011.01605}}].

\bibitem{Lyth:2001nq}
D.~H. Lyth and D.~Wands, {\it {Generating the curvature perturbation without an inflaton}}, \href{https://doi.org/10.1016/S0370-2693(01)01366-1}{Phys. Lett. B {\bfseries 524} (2002) 5} [\href{http://arxiv.org/abs/hep-ph/0110002}{{\ttfamily arXiv:hep-ph/0110002}}].

\bibitem{Dvali:2003em}
G.~Dvali, A.~Gruzinov and M.~Zaldarriaga, {\it {A new mechanism for generating density perturbations from inflation}}, \href{https://doi.org/10.1103/PhysRevD.69.023505}{Phys. Rev. D {\bfseries 69} (2004) 023505} [\href{http://arxiv.org/abs/astro-ph/0303591}{{\ttfamily arXiv:astro-ph/0303591}}].

\bibitem{Dvali:2003ar}
G.~Dvali, A.~Gruzinov and M.~Zaldarriaga, {\it {Cosmological perturbations from inhomogeneous reheating, freezeout, and mass domination}}, \href{https://doi.org/10.1103/PhysRevD.69.083505}{Phys. Rev. D {\bfseries 69} (2004) 083505} [\href{http://arxiv.org/abs/astro-ph/0305548}{{\ttfamily arXiv:astro-ph/0305548}}].

\end{thebibliography}\endgroup

\end{document}